\documentclass[openany, 11pt]{book}

\usepackage[text={15.5cm,23cm},centering]{geometry}
\usepackage{mathrsfs}
\usepackage{mathptmx}
\usepackage{amsmath}
\usepackage{amssymb}
\usepackage{amsfonts}
\usepackage{graphicx}
\usepackage{subfigure}
\usepackage{dcolumn}
\usepackage{bm}
\usepackage{epsf}
\usepackage{color}
\usepackage[colorlinks=true,citecolor=blue,linkcolor=blue,urlcolor=blue]{hyperref}
\usepackage{wrapfig}
\usepackage[framemethod=TikZ]{mdframed}
\usepackage{hhline}
\usepackage{amssymb}
\usepackage{cite}

\usepackage{chapterbib} 

\usepackage{tikz}
\usetikzlibrary{arrows,shapes,calc,matrix}

  \expandafter\let\csname equation*\endcsname\relax
  \expandafter\let\csname endequation*\endcsname\relax

\usepackage{fancyhdr}
\makeatletter
\newcommand\footnoteref[1]{\protected@xdef\@thefnmark{\ref{#1}}\@footnotemark}
\makeatother

\newcommand{\mean}[1]{\langle #1 \rangle}

\newcommand{\bra}[1]{\left\langle #1\right|}
\newcommand{\ket}[1]{\left|#1\right\rangle}
\newcommand{\braket}[2]{\left\langle #1|#2\right\rangle}
\newcommand{\ketbrad}[1]{|#1\rangle\!\langle #1|}
\newcommand{\tr}[1]{\mathrm{tr}\left\{#1\right\}}
\newcommand{\ptr}[2]{\mathrm{tr_{#1}}\left\{#2\right\}}

\newcommand{\la}{\left\langle}
\newcommand{\ra}{\right\rangle}
\newcommand{\pd}{\partial}
\newcommand{\de}[1]{\delta\left(#1\right)}
\newcommand{\td}{\mathrm{d}}

\newcommand{\etal}{\textit{et al. }}
\newcommand{\e}[1]{\exp{\left(#1\right)}}
\newcommand{\loge}[1]{\ln{\left(#1\right)}}
\newcommand{\lo}[1]{\ln{\left(#1\right)}}
\newcommand{\id}{\mathbb{I}}
\newcommand{\com}[2]{\left[#1,\,#2\right]}

\newcommand{\ch}[1]{\cosh{\left(#1\right)}}

\newcommand{\bla}{bla\\bla\\bla\\bla\\bla}

\newcommand{\mb}[1]{\mbox{\boldmath$#1$}}
\newcommand{\mc}[1]{\mathcal{#1}}
\newcommand{\mbb}[1]{\mathbb{#1}}
\newcommand{\mf}[1]{\mathfrak{#1}}
\newcommand{\mrm}[1]{\mathrm{#1}}

\newcommand{\dbar}{d\hspace*{-0.08em}\bar{}\hspace*{0.1em}}

\DeclareMathOperator*{\sumint}{%
\mathchoice%
  {\ooalign{$\displaystyle\sum$\cr\hidewidth$\displaystyle\int$\hidewidth\cr}}
  {\ooalign{\raisebox{.14\height}{\scalebox{.7}{$\textstyle\sum$}}\cr\hidewidth$\textstyle\int$\hidewidth\cr}}
  {\ooalign{\raisebox{.2\height}{\scalebox{.6}{$\scriptstyle\sum$}}\cr$\scriptstyle\int$\cr}}
  {\ooalign{\raisebox{.2\height}{\scalebox{.6}{$\scriptstyle\sum$}}\cr$\scriptstyle\int$\cr}}
}

\newcommand{\draftmode}{1}    
\newcommand{\notetoself}[1]{\ifnum \draftmode=1 {\color[rgb]{0,0,0.8} [#1]} \fi}  
\newcommand{\cuttext}[1]{\ifnum \draftmode=1 {\color[rgb]{0,0.5,0} [#1]} \fi}  
\newcommand{\warntext}[1]{\ifnum \draftmode=1 {\color[rgb]{0.9,0.6,0} #1} \else {#1} \color{black} \fi}

\title{Quantum Thermodynamics\\An introduction to the thermodynamics of quantum information} 
\author{Sebastian Deffner and Steve Campbell}

\begin{document}
\frontmatter
\pagenumbering{Roman}
\maketitle


\chapter{Abstract}

This book provides an introduction to the emerging field of quantum thermodynamics, with particular focus on its relation to quantum information and its implications for quantum computers and next generation quantum technologies. The text, aimed at graduate level physics students with a working knowledge of quantum mechanics and statistical physics, provides a brief overview of the development of classical thermodynamics and its quantum formulation in Chapter~\ref{chap:termo}. Chapter~\ref{chap:devices} then explores typical thermodynamic settings, such as cycles and work extraction protocols, when the working material is genuinely quantum. Finally, Chapter~\ref{chap:info} explores the thermodynamics of quantum information processing and introduces the reader to some more state-of-the-art topics in this exciting and rapidly developing research field.

\newpage

\quad\\\vspace{.45\textwidth}
\begin{flushright}
 \textit{Quidquid praecipies, esto brevis.}\\ \vspace{0.3em}\footnotesize(Horaz, Ars poetica 335)
\end{flushright}


\chapter{About the Authors}

\section*{Sebastian Deffner}
\begin{figure}[h!]
\begin{minipage}[l]{.4\textwidth}
\begin{center}
\includegraphics[width=7cm]{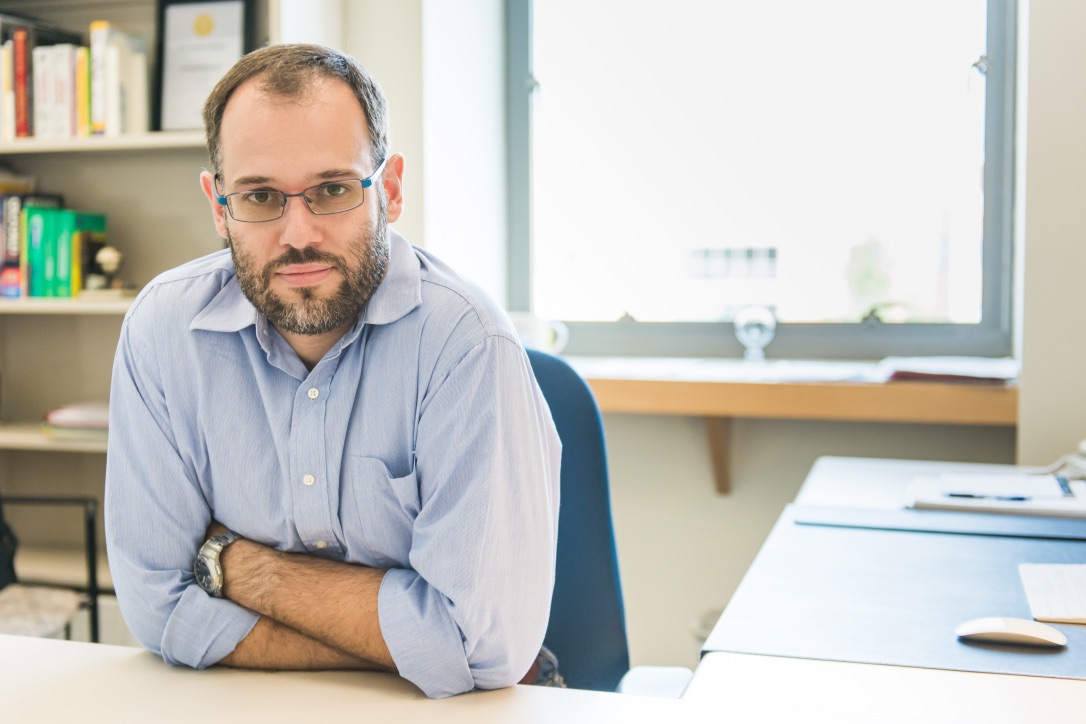}
\end{center}
\end{minipage}
\hfill
\begin{minipage}[r]{.51\textwidth}
Dr. Sebastian Deffner received his doctorate from the University of Augsburg in 2011 under the supervision of Eric Lutz. From 2011 to 2014 he was a Research Associate in the group of Chris Jarzynski at the University of Maryland, College Park and from 2011 to 2016 he was a Director's Funded Postdoctoral Fellow with Wojciech H. Zurek at the Los Alamos National Laboratory. Since 2016 he has been on the faculty of the Department of Physics at the University of Maryland, Baltimore County (UMBC), where he leads the quantum thermodynamics group.
\end{minipage}
\end{figure}

Dr. Deffner's contributions to quantum thermodynamics have been recognized through the Early Career Award 2016 from IOP's New Journal of Physics, and he was also awarded the Leon Heller Postdoctoral Publication Prize from the Los Alamos National Laboratory in 2016.

To date, Dr. Deffner has been reviewing for more than ten international funding agencies and more than thirty high-ranking journals. For these efforts he has been named Oustanding Reviewer for New Journal of Physics in 2016, Outstanding Reviewer for Annals of Physics in 2016, and in 2017 he was named APS Outstanding Referee. Since 2017 Dr. Deffner has been a member of the international editorial board for IOP's Journal of Physics Communications, and since 2019 he has been on the editorial advisory board of Journal of Nonequilibrium Thermodynamics.

As a theoretical physicist,  Dr. Deffner employs tools from statistical physics, open quantum dynamics, quantum information theory, quantum optics, quantum field theory, condensed matter theory, and optimal control theory to investigate the nonequilibrium properties of nanosystems operating far from thermal equilibrium.

\clearpage

\qquad \\
\qquad \\
\qquad \\
\qquad \\
\qquad \\
\qquad \\
\qquad \\

\section*{Steve Campbell}
\begin{figure}[h!]
\begin{minipage}[l]{.4\textwidth}
\begin{center}
\includegraphics[height=6.5cm]{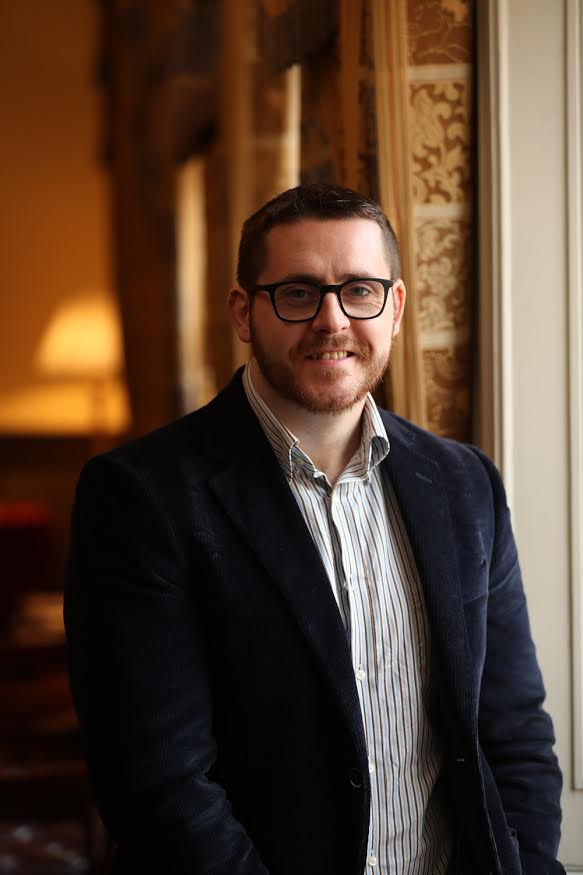}
\end{center}
\end{minipage}
\hfill
\begin{minipage}[r]{.58\textwidth}
After a PhD in Queen’s University Belfast in 2011 under the supervision of Mauro Paternostro, Dr. Steve Campbell moved to University College Cork to work with Thomas Busch in 2012. He spent 2013 at the Okinawa Institute of Science and Technology Graduate University in Japan. Returning to Belfast, he spent 2014 through to 2016 at his alma mater Queen’s University. In 2017 he was awarded a fellowship from the INFN Sezione di Milano and worked with Bassano Vacchini. From February 2019 he has been appointed as Senior Research Fellow at Trinity College Dublin through the award of a Science Foundation Ireland Starting Investigators Research Grant.
\end{minipage}
\end{figure}

Dr. Campbell is interested in exploring the role which fundamental bounds, such as the quantum speed limit, play in characterizing and designing thermodynamically efficient control protocols for complex quantum systems. He works on a variety of topics including open quantum systems, critical spin systems and phase transitions, metrology, and coherent control. 

\tableofcontents
\cleardoublepage


\mainmatter
\addcontentsline{toc}{chapter}{Prologue}
\fancyhead[LO]{ }
\fancyhead[RE]{ }

\chapter*{Prologue}

What is physics? According to standard definitions in encyclopedias \emph{physics is a science that deals with matter and energy and their interactions}\footnote{This and similar definitions can be found, for instance, in Merriam-Webster.}. However, as physicists what is that we actually do? At the most basic level, we formulate predictions for how inanimate objects behave in their natural surroundings. These predictions are based on our expectation that we extrapolate from observations of the \emph{typical behavior}. If typical behavior is universally exhibited by many systems of the same ``family'', then this typical behavior is phrased as a \emph{law}.

Take for instance the infamous example of an apple falling from a tree. The same behavior is observed for any kind of fruit and any kind of tree -- the fruit ``always'' falls from the tree to the ground. Well, actually the same behavior is observed for any object that is let loose above the ground, namely everything will  eventually  fall towards the ground. It is this observation of \emph{universal falling}  that is encoded in the \emph{law of gravity}. 

Most theories in physics then seek to understand the nitty-gritty details, for which finer and more accurate observations are essential. Generally, we end up with more and more fine-grained descriptions of nature that are packed into more and more sophisticated laws. For instance, from classical mechanics over quantum mechanics to quantum field theory we obtain an ever more detailed prediction for how smaller and smaller systems behave.

Realizing this typical mindset of physical theories, it does not come as a big surprise that many students have such a hard time wrapping their minds around \emph{thermodynamics}: 
\begin{quote}
\emph{Thermodynamics is a phenomenological theory to describe the average behavior of heat and work.}
\end{quote}
As a phenomenological theory, thermodynamics does not seek to formulate detailed predictions for the microscopic behavior of \emph{some} physical systems, but rather it aims to provide the most universal framework to describe the typical behavior of \emph{all} physical systems.

\paragraph{``Reflections on the Motive Power of Fire''.}

The origins of thermodynamics trace back to the beginnings of the industrial revolution \cite{Kondepudi1998}. For the first time, mankind started developing artificial devices that contained so many moving parts that it became practically impossible to describe their behavior in full detail. Nevertheless, already the first devices, steam engines, proved to be remarkably useful and dramatically increased the effectiveness of productive efforts.

The founding father of thermodynamics is undoubtedly Sadi Carnot. After Napoleon had been exiled, France started importing advanced steam engines from Britain, which made Carnot realize how far France had fallen behind its adversary from across the channel. Quite remarkably, a small number of British engineers, who totally lacked any formal scientific education, had started to collect reliable data about the efficiency of many types of steam engines. However, it was not at all clear whether there was an optimal design and what the highest efficiency would be.

Carnot had been trained in the latest developments in physics and chemistry, and it was he who recognized that steam engines need to be understood in terms of their energy balance. Thus, optimizing steam engines was not only a matter of improving the expansion and compression of steam, but actually needed an understanding of the relationship between work and heat \cite{Carnot1824}.
\begin{figure}
\begin{mdframed}[roundcorner=10pt]
\begin{minipage}[l]{.4\textwidth}
\begin{center}
\includegraphics[height=5cm]{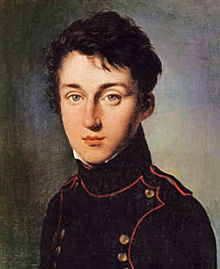}
\end{center}
\end{minipage}
\hfill
\begin{minipage}[r]{.58\textwidth}
Nicolas L\'{e}onard Sadi Carnot:\\ \emph{Everyone knows that heat can produce motion \cite{Carnot1824}.}
\end{minipage}
\end{mdframed}
\end{figure}

Sadly, Carnot's work \cite{Carnot1824} was largely ignored by the scientific community until the railroad engineer \'{E}mile Clapeyron quoted and generalized Carnot's results. Eventually 30 years later, it was Rudolph Clausius, who put Carnot's insight into a solid mathematical framework \cite{Clausius1854}, which is the same mathematical theory that we still use today -- thermodynamics.

Thus, thermodynamics is not only unique among the theories in physics with respect to its mindset, but also with respect to its beginnings. No other theory is so intimately connected with someone never holding an academic position -- Sadi Carnot. Formulating the original ideas was thus largely motivated by practical questions and not purely by scientific curiosity. This might explain why more than any other theory thermodynamics is a framework to describe the typical and universal behavior of any physical system.

\paragraph{Quantum computing -- Feynman's dream come true.}

A remarkable quote from Carnot's work \cite{Carnot1824} is the following:
\begin{quote}
\emph{The study of these engines is of the greatest interest, their importance is enormous, their use is continually increasing, and they seem destined to produce a great revolution in the civilized world.}
\end{quote}
If we replaced the word ``engines'' with ``quantum computers'', Carnot's sentence would fit nicely into the announcements of the various ``quantum initiatives'' around the globe \cite{Sanders2017}.

Ever since Feyman's proposal in the early 1980s \cite{Feynman1982} quantum computing as been a promise that could initiate a technological revolution. Over the last couple of years big corporations, such as Microsoft, IBM and Google, as well as smaller start-ups, such as D-Wave or Rigetti, have started to present more and more intricate technologies that promise to eventually lead to the development of a practically useful quantum computer. 
\begin{figure}
\begin{mdframed}[roundcorner=10pt]
\begin{minipage}[l]{.4\textwidth}
\begin{center}
\includegraphics[height=5cm]{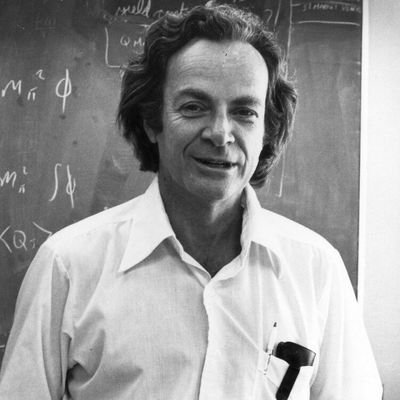}
\end{center}
\end{minipage}
\hfill
\begin{minipage}[r]{.58\textwidth}
Richard P. Feynman:\\ \emph{Nature isn't classical, dammit, and if you want to make a simulation of nature, you'd better make it quantum mechanical, and by golly it's a wonderful problem, because it doesn't look so easy \cite{Feynman1982}.}
\end{minipage}
\end{mdframed}
\end{figure}

Rather curiously, we are in a very similar situation that Carnot found in the beginning of the 19th century. Novel technologies are being developed by crafty engineers that are much too complicated to be described in full microscopic detail. Nevertheless, the question that we are really after is how to operate these technologies optimally in the sense that the least amount of resources, such as work and information, are wasted into the environment.

As physicists we know exactly which theory will prevail in the attempt to describe what is going on, since it is the only theory that is universal enough to be useful when faced with new challenges -- thermodynamics. However, this time the natural variables can no longer be volume, temperature, and pressure, which are characteristic for steam engines. Rather, in \emph{\textbf{Quantum Thermodynamics}} the first task has to be to identify the new \emph{canonical variables}, and then write the dictionary for how to translate between the universal thermodynamic framework and practically useful statements for the optimization of quantum technologies.

\paragraph{Purpose and target audience of this book.}

The purpose of this book is to provide a concise introduction to the conceptual building blocks of \emph{\textbf{Quantum Thermodynamics}} and their application in the description of quantum systems that process information. Large parts of this book arose from our lecture notes that we had put together for graduate classes in statistical physics or for workshops and summer schools dedicated to Quantum Thermodynamics. When teaching the various topics of Quantum Thermodynamics we always felt a bit unsatisfied as no single book contained a comprehensive overview of all the topics we deemed essential. Earlier monographs have become a bit outdated, such as \emph{Quantum Thermodynamics} by our colleagues Gemmer, Michel, and Mahler \cite{Gemmer2009}, or are simply not written as a textbook  suited for teaching, such as \emph{Thermodynamics in the Quantum Regime} which was edited by Binder \etal \cite{Binder2018}.

Thus, we took it upon ourselves to write a text that we will be using for advanced special topics classes in our graduate program. Considering graduate statistical physics and quantum mechanics as prerequisites the topics of the present book can be covered over the course of a semester. However, like always when designing a new course it is simply not possible to cover everything that would be interesting. Thus, we needed to make some tough choices and we hope that our colleagues will forgive us if they feel their work should have been a more prominent part of this text.

\vspace{1em}
\begin{flushright}
 \textit{Longum iter est per praecepta, breve et efficax per exempla.}\\ \vspace{0.3em}\footnotesize(Seneca Junior, 6th letter)
\end{flushright}

\vspace{1em}
\noindent
Baltimore, USA\hfill Sebastian Deffner\\
Dublin, Ireland\hfill Steve Campbell\\
\\
\today

\bibliographystyle{plain}
\bibliography{book}

\clearpage


\fancyhead[LO]{\rightmark}
\fancyhead[RE]{\leftmark}

\chapter{\label{chap:termo}The principles of modern thermodynamics}

Thermodynamics is a phenomenological theory to describe the average behavior of heat and work. Its theoretical framework is built upon five axioms, which are commonly called the \emph{laws of thermodynamics}. Thus, as an axiomatic theory, thermodynamic can never be wrong as long as it basic assumptions are fulfilled. 

Despite thermodynamics' unrivaled success, versatility, and universality it is plagued with three major shortcomings: (i) thermodynamics contains no microscopic information, nor does thermodynamics know how to relate its phenomenological framework to microscopic information; (ii) as an equilibrium theory, thermodynamics cannot characterize non-equilibrium states, and in particular only infinitely slow, quasistatic processes are fully describable; and (iii) as a classical theory the original mathematical framework is ill-equipped to be directly applied to quantum systems.

In the following we will briefly summarize the major building blocks of thermodynamics in Sec.~\ref{sec:thermo}, and its extension to Stochastic Thermodynamics in Sec.~\ref{sec:fluc}. We will then see how equilibrium states can be fully characterized from a quantum information theoretic point of view in Sec.~\ref{sec:envariance}, which we will use as a motivation to outline the framework of Quantum Thermodynamics in Sec.~\ref{sec:quwork}.

\section{\label{sec:thermo} A phenomenological theory of heat and work}

Thermodynamics was originally invented to describe and optimize the working principles of steam engines. Therefore, its natural quantities are work and heat. During the operation of such engines, work is understood as the useful part of the energy, whereas heat quantifies the waste into the environment.

\begin{figure}
\centering
\includegraphics[width=.48\textwidth]{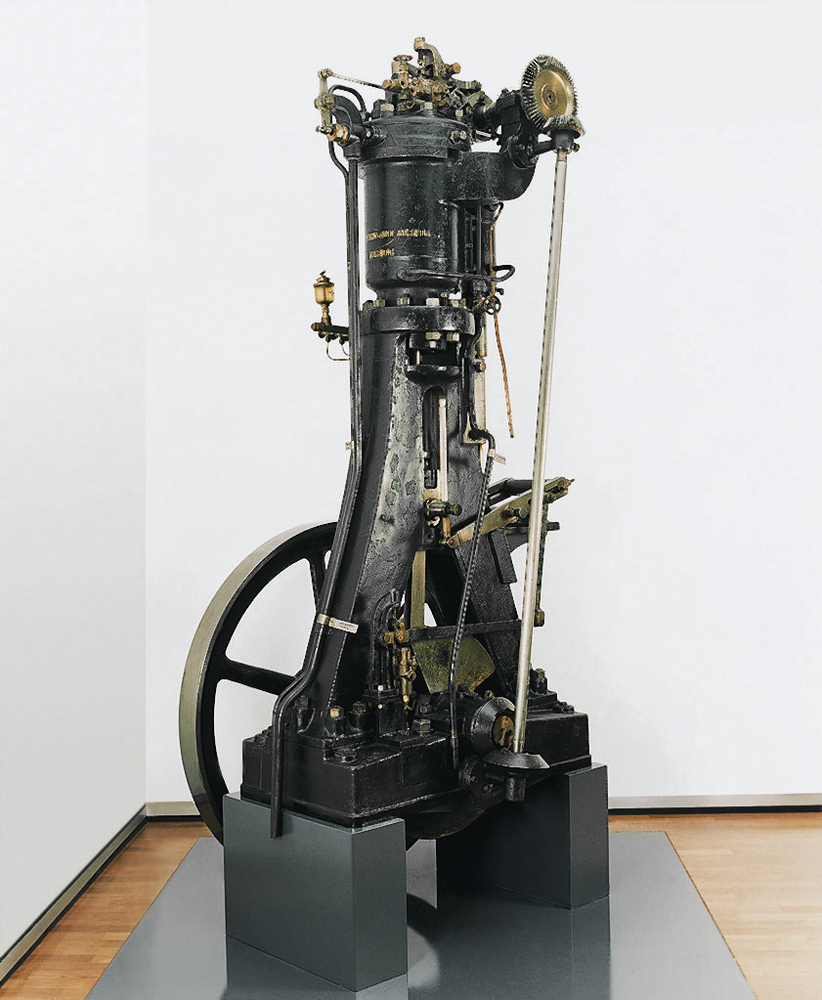}
\caption{\label{diesel} Paradigmatic thermodynamic engine: first operational Diesel test engine (M.A.N. museum in Augsburg, Germany)}
\end{figure}

In reality, steam engines are messy, stinky, and huge [cf Fig.~\ref{diesel}], which makes any attempt of describing their properties from a microscopic theory futile. Thermodynamics takes a very different perspective: rather than trying to understand all the nitty-gritty details, let's focus on the overall, average behavior once the engine is running smoothly -- once it has reached its stationary state of operation.

\subsection{The five laws of thermodynamics}

The framework of thermodynamics is built upon five \emph{laws}, which axiomatically paraphrase ordinary experience and observation of nature. The central notion is \emph{equilibrium}, and the central focus is on transformations of systems from one state of equilibrium to another.

\paragraph{Zeroth Law of Thermodynamics.}

The Zeroth Law of Thermodynamics defines a state of \emph{equilibrium} of a system relative to its environment. In its most common formulation it can be expressed as:
\begin{quote}
\textit{If two systems are in thermal equilibrium with a third system, then they are in thermal equilibrium with each other.}
\end{quote}

States of equilibrium are uniquely characterized by an \emph{equation of state}, which relates the experimentally accessible parameters. For a steam engine these parameters are naturally given by volume $V$, pressure $P$,  and temperature $T$. A sometimes under-appreciated postulate is then that all equilibria can be fully characterized by only \emph{three} accessible parameters, of which only \emph{two} are independent. The equation of state determines how these parameters are related to each other,
\begin{equation}
\label{eq:eos}
f(V,P,T)=0\,,
\end{equation}
where the function $f$ is characteristic for the system. For instance for an ideal gas Eq.~\eqref{eq:eos} becomes the famous $P V= N k_B T$, where $N$ is the number of particles and $k_B$ is Boltzmann's constant.

\paragraph{Thermodynamic manifolds and reversible processes.}

Mathematically speaking the equation of state \eqref{eq:eos} defines 2-1 maps, which allow to write one of the parameters as function of the other two, $V(P,T)$ or $P(V,T)$ or $T(V,P)$. Except under very special circumstances we regard $f$ as a continuous differentiable function\footnote{At loci where $f$ is not continuous differentiable, we have a so-called phase transitions.}. Thus, the equation of state can be represented as a smooth surface in three-dimensional space.

\begin{figure}
\centering
\includegraphics[width=.48\textwidth,trim={5cm 0 4cm 0},clip]{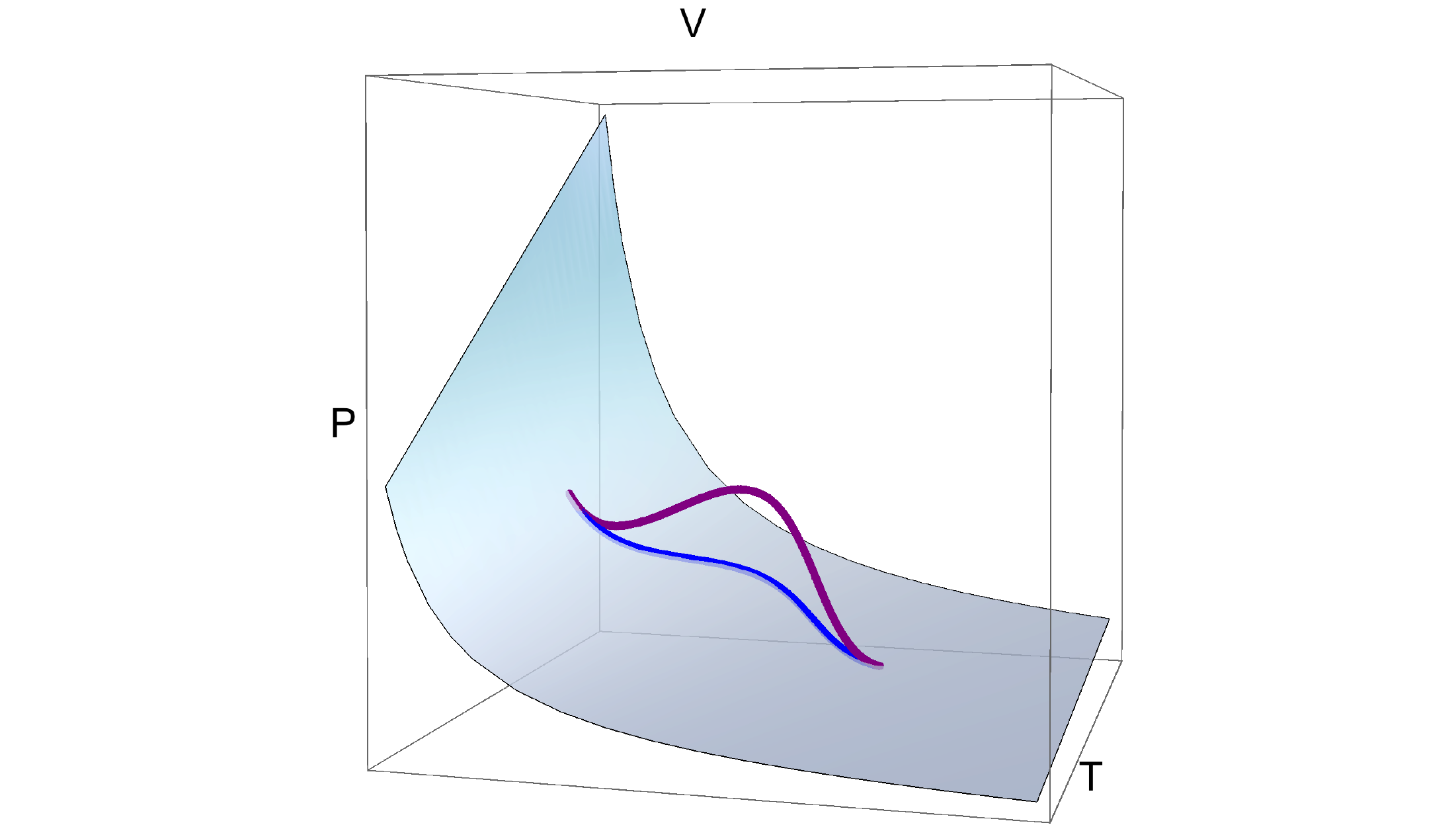}
\caption{Thermodynamic manifold for an ideal gas with $P V=N k_B T$, and a reversible state transformation (blue) and an irreversible process with the same end points (purple).\label{fig:path}}
\end{figure}

All equilibrium states for a specific substance are points on this surface. All \emph{thermodynamic transformations} are processes that take the system from one point on the surface to another, cf. Fig.~\ref{fig:path}.

In what follows we will see that only quasistatic processes are fully describable by means of thermodynamics. Quasistatic processes are so slow that the driven systems almost instantaneously relax back to equilibrium. Thus, such processes can be regarded as successions of equilibrium states, which correspond to paths on the surface spanned by the equation of state. Since the surface is smooth, i.e., continuous differentiable, the path cannot have any distinct directionality and this is why we call quasistatic processes that lie entirely in the thermodynamic manifold \emph{reversible}.

All real processes happen in finite time and at finite rates. Such processes necessarily comprise of nonequilibrium states, and paths corresponding to such processes have to leave the thermodynamic surface. Our goal has to be to quantify this irreversibility, which is the starting point of Stochastic Thermodynamics, see Sec.~\ref{sec:fluc} 

\paragraph{First Law of Thermodynamics.}

Before we move on to extensions of thermodynamics, however, we need to establish a few more concepts and notions. In classical mechanics the central concept is the energy of the system, since the complete dynamical behavior can be derived from it. We also know from classical mechanics that in isolated systems the energy is conserved, and that transformations of energy can depend on the path taken by the system -- think for instance of friction.

This leads naturally to the insight that
\begin{equation}
\label{eq:1st}
dE=\dbar W+\dbar Q\,,
\end{equation}
where $E$ is the internal energy, $W$ the work, and $Q$ denotes the heat. In Eq.~\eqref{eq:1st} work, $\dbar W$, is identified with the contribution to the change in internal energy that can be \emph{controlled}, whereas $\dbar Q$ denotes the amount of energy that is exchanged with a potentially vast bath. Moreover, $dE$ is an exact differential, which means that changes of the internal energy do not depend on which path is taken on the thermodynamic manifold. This makes sense, since we would expect energy to be only dependent on the state of the system, and not how the system has reached a state. In other words, $E$ is a \emph{state function}.

Already in classical mechanics, work is a very different concept. Loosely speaking work is given by a force along a trajectory, which clearly depends on the path a systems takes and which explains why $\dbar W$ is a non-exact differential. We can further identify infinitesimal changes in work as
\begin{equation}
\label{eq:work}
\dbar W=-P\,dV\,,
\end{equation}
which is fully analogous to classical mechanics. The other quantity, the one that quantifies the \emph{useless} change of internal energy, the part that is typically wasted into the environment, the heat $Q$ has no equivalent in classical mechanics. It is rather characterized and specified by the second law of thermodynamics.

\paragraph{Second Law of Thermodynamics.}

Let us inspect the first law of thermodynamics as expressed in Eq.~\eqref{eq:1st}. If $dE$ is an exact differential, and $\dbar W$ is a non-exact differential, then $\dbar Q$ also has to be non-exact. However, it is relatively simple to understand from its definition how $\dbar W$ can be written in terms of an exact differential. It is the force that depends in the path taken, yet the path length has to be an exact differential -- if you walk a closed loop you return to your point of origin with certainty. 

Finding the corresponding exact differential, i.e., the line element for $\dbar Q$ was a rather challenging task. A first account goes back to Clausius who realized \cite{Clausius1854} that
\begin{equation}
\label{eq:clausius}
\oint \frac{\dbar Q}{T}\leq 0
\end{equation}
where $T$ is the temperature of the substance undergoing the cyclic, thermodynamic transformation. Moreover, the inequality in Eq.~\eqref{eq:clausius} becomes an \emph{equality} for quasistatic processes. Thus, it seems natural to define a new state function, $S$, for reversible processes through 
\begin{equation}
\label{eq:ent_0}
dS\equiv \frac{\dbar Q}{T}\,.
\end{equation}
and that is known as \emph{thermodynamic entropy}.

\begin{figure}
\begin{mdframed}[roundcorner=10pt]
\begin{minipage}[l]{.4\textwidth}
\begin{center}
\includegraphics[height=5cm]{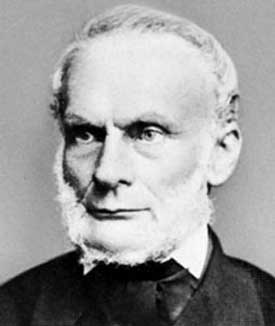}
\end{center}
\end{minipage}
\hfill
\begin{minipage}[r]{.58\textwidth}
Rudolf J. E. Clausius:\\ \emph{... as I hold it better to borrow terms for important magnitudes from the ancient languages, so that they may be adopted unchanged in all modern languages, I propose to call [it] the entropy of the body, from the Greek word ``trope” for ``transformation'' I have intentionally formed the word ``entropy” to be as similar as possible to the word ``energy"; for the two magnitudes to be denoted by these words are so nearly allied in their physical meanings, that a certain similarity in designation appears to be desirable \cite{Clausius1854}.}
\end{minipage}
\end{mdframed}
\end{figure}

To get a better understanding of this quantity consider a thermodynamic process that takes a system from a point $A$ on the thermodynamic manifold to a point $B$. Now imagine that the system is taken from $A$ to $B$ along a reversible path, and it returns from $B$ to $A$ along an irreversible path. For such a cycle, the latter two equations give combined,
\begin{equation}
\label{eq:clausius_1}
\Delta S_{A\rightarrow B} \geq \int_A^B \frac{\dbar Q}{T}\,,
\end{equation}
which is known as \emph{Clausius inequality}.

The Clausius inequality \eqref{eq:clausius_1} is an expression of the second law of thermodynamics. More generally, the second law is a collection of statements that at their core express that the entropy of the Universe is a non-decreasing function of time,
\begin{equation}
\label{eq:ent_1}
\Delta S_\mrm{Universe}\geq0\,.
\end{equation}
The most prominent, and also the oldest expressions of the second law of thermodynamics are formulated in terms of cyclic processes. The Kelvin-Planck statement asserts that
\begin{quote}
\textit{no process is possible whose sole result is the extraction of energy from a heat bath, and the conversion of all that energy into work.}
\end{quote}
The Clausius statement reads,
\begin{quote}
\textit{no process is possible whose sole result is the transfer of heat from a body of lower temperature to a body of higher temperature.}
\end{quote}
Finally, the Carnot statement declares that
\begin{quote}
\textit{no engine operating between two heat reservoirs can be more efficient than a Carnot engine operating between those same reservoirs.}
\end{quote}
These formulations refer to processes involving the exchange of energy among idealized subsystems: one or more heat reservoirs; a work source -- for example, a mass that can be raised or lowered against gravity; and a device that operates in cycles and affects the transfer of energy among the other subsystems. All three statements follow from simple entropy-balance analyzes and offer useful, logically transparent reference points as one navigates the application of the laws of thermodynamics to real systems.

\paragraph{Third Law of Thermodynamics.}

The Third Law of Thermodynamics or the Nernst Theorem paraphrases that in \emph{classical} systems the entropy vanishes in the limit of $T\to 0$. A little more precisely, the Nernst theorem states that as absolute zero of the temperature is approached, the entropy change $\Delta S$ for a chemical or physical transformation approaches 0,
\begin{equation}
\label{eq:nernst}
\lim _{T\to 0}\Delta S=0
\end{equation}
It is interesting to note that this equation is a modern statement of the theorem. Nernst often used a form that avoided the concept of entropy, since, e.g., for quantum mechanical systems the validity of Eq.~\eqref{eq:nernst} is somewhat questionable.

\paragraph{Fourth Law of Thermodynamics.}

The fourth law of thermodynamics takes the first step away from a mere equilibrium theory. In reality, few systems can ever be found in isotropic and homogeneous states of equilibrium. Rather, physical properties vary as functions of space $\vec{r}$ and time $t$.

Nevertheless, it is frequently not such a bad approximation to assume that a thermodynamic system is in a state of \emph{local equilibrium}. This means that for any point in space and time, the system appears to be in equilibrium, yet thermodynamic properties vary weakly on macroscopic scales. In such situations we can introduce the local temperature, $T(\vec{r},t)$, the local density, $n(\vec{r},t)$, and the local energy density, $e(\vec{r},t)$. The question now is, what general and universal statements can be made about the resulting transport driven by local gradients of the thermodynamic variables. 

The clearest picture arises if we look at the dynamics of the local entropy, $s(\vec{r},t)$. We can write
\begin{equation}
\label{eq:ent_prod}
\frac{ds}{dt}=\sum_k \frac{\pd s}{\pd X_k} \frac{d X_k}{d t},
\end{equation}
where $\{X_k\}_k$ is a set of extensive parameters that vary as a function of time. The time-derivative of these $X_k$ define the \emph{thermodynamic fluxes}
\begin{equation}
\label{eq:flux}
J_k\equiv\frac{d X_k}{d t}
\end{equation}
and the variation of the entropy as a function of the $X_k$ are the \emph{thermodynamic forces} or affinities, $F_k$. In short, we have
\begin{equation}
\label{eq:ent_prod_1}
\frac{ds}{dt}=\sum_k F_k J_k\,.
\end{equation}
This means that the rate of entropy production is the sum of products of each flux with its associated affinity. 

It should not come as a surprise that Eq.~\eqref{eq:ent_prod_1} is conceptually interesting, but practically of rather limited applicability. The problem is that generally the fluxes are complicated functions of all forces and local gradients, $J_k(F_0, F_1,\dots)$. A simplifying case is purely resistive systems, for which by definition the local flux only depends on the instantaneous local affinities. For small affinities, i.e., if the systems is in local equilibrium, $J_k$ can be expanded in $F_k$. In leading order we have,
\begin{equation}
\label{eq:flux_1}
J_k=\sum_j L_{j,k} F_j\,,
\end{equation}
where the kinetic coefficients $L_{j,k}$ are given by
\begin{equation}
\label{eq:L}
L_{j,k}\equiv\frac{\pd J_k}{\pd F_j}\bigg|_{F_j=0}\,,
\end{equation}
with $F_j=0$ in equilibrium.

The Onsager theorem \cite{Onsager1931}, which is also known as the Fourth Law of Thermodynamics, now states
\begin{equation}
\label{eq:onsager}
L_{j,k}=L_{k,j}\,.
\end{equation}
This means that the matrix of kinetic coefficients is symmetric. Therefore, to a certain degree Eq.~\eqref{eq:onsager} is a thermodynamic equivalent of Newton's third law. This analogy becomes even clearer if we interpret Eq.~\eqref{eq:flux_1} as a thermodynamic equivalent of Newton's second law.

\begin{figure}
\begin{mdframed}[roundcorner=10pt]
\begin{minipage}[l]{.4\textwidth}
\begin{center}
\includegraphics[height=5cm]{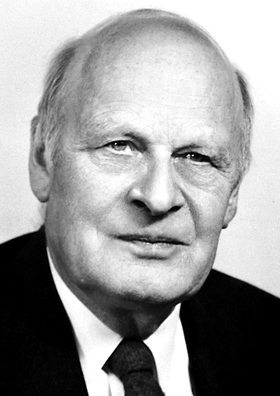}
\end{center}
\end{minipage}
\hfill
\begin{minipage}[r]{.58\textwidth}
Lars Onsager:\\ \emph{Now if we look at the condition of detailed balancing from the thermodynamic point of view, it is quite analogous to the principle of least dissipation \cite{OnsagerNobel}. }
\end{minipage}
\end{mdframed}
\end{figure}

It is interesting to consider when the above considerations break down. Throughout this little exercise we have explicitly assumed that the considered system is in a state of local equilibrium. This is justified as long as the flux and affinities are small. Consider, for instance, a system with a temperature gradient. For small temperature differences the flow is laminar, and the Onsager theorem \eqref{eq:onsager} is expected hold. For large temperature differences the flow becomes turbulent, and the fluxes can no longer be balanced.

\subsection{Finite-time thermodynamics and endoreversibility}
\label{sec:finite_time_thermo}

A standard exercise in thermodynamics is to compute the efficiency of cycles, i.e., to determine the relative work output for devices undergoing cyclic transformations on the thermodynamic manifold. However, all standard cycles, such as the Carnot, Otto, Diesel, etc. cycles have in common that they are comprised of only quasistatic state transformations, and hence their power output is strictly zero. 

This insight led Curzon and Ahlborn to ask a slightly different, yet a lot more practical question \cite{Curzon1975}: ``What is the efficiency of a Carnot engine at maximal power output?'' Obviously such a cycle can no longer be reversible, but we still would like to be able to use the methods and notions from thermodynamics. This is possible if one takes the aforementioned idea of \emph{local equilibrium} one step further.

Imagine a device, whose working medium is in thermal equilibrium at temperature $T_w$, but there is a temperature gradient over its boundaries to the environment at temperature $T$. A typical example is a not perfectly insulating thermo-can. Now let us now imagine that the device is slowly driven through a cycle, where slow means that the working medium remains in a local equilibrium state at all instants. However, we will also assume that the cycle operates too fast for the working medium to ever equilibrate with the environment, and thus from the point of view of the environment the device undergoes an irreversible cycle. Such state transformations are called \emph{endoreversible}, which means that locally the transformation is reversible, but globally irreversible. 

This idea can then be applied to the Carnot cycle, and we can determine its endoreversible efficiency. The standard Carnot cycle consists of two isothermal processes during which the systems absorbs/exhausts heat and two thermodynamically adiabatic, i.e., isentropic strokes. Since the working medium is not in equilibrium with the environment, we will have to modify the treatment of the isothermal strokes. The adiabatic strokes constitute \emph{no exchange of heat}, and thus they do not need to be re-considered.

During the hot isotherm the working medium is assumed to be a little cooler than the environment. Thus, during the whole stroke the system absorbs the heat
\begin{equation}
\label{eq:heat_hot}
Q_h=\kappa_h \tau_h \left(T_h-T_{hw}\right)\,,
\end{equation}
where $\tau_h$ is the time the isotherm needs to complete and $\kappa_h$ is a constant depending on thickness and thermal conductivity of the boundary separating working medium and environment. Note that Eq.~\eqref{eq:heat_hot} is nothing else but a discretized version of Fourier's law for heat conduction.

Similarly, during the cold isotherm the system is a little warmer than the cold reservoir. Hence the exhausted heat becomes
\begin{equation}
\label{eq:heat_cold}
Q_c=\kappa_c \tau_c \left(T_{cw}-T_{c}\right)
\end{equation}
where $\kappa_c$ is the heat transfer coefficient for the cold reservoir. 

As mentioned above, the adiabatic strokes are unmodified, but we note that the cycle is taken to be reversible with respect to the \emph{local temperatures} of the working medium. Hence, we can write
\begin{equation}
\label{eq:entropy}
\Delta S_h=-\Delta S_c\quad\mrm{and\,\, thus}\quad\frac{Q_h}{T_{hw}}=\frac{Q_c}{T_{cw}}\,.
\end{equation}
The latter will be useful to relate the stroke times $\tau_h$ and $\tau_c$ to the heat transfer coefficients $\kappa_h$ and $\kappa_c$.

We are now interested in determining the efficiency at maximal power. To this end, we write the power output of the cycle as
\begin{equation}
\label{eq:power}
P(\delta T_h,\delta T_c)=\frac{Q_h-Q_c}{\zeta (\tau_h+\tau_c)}
\end{equation}
where $\delta T_h=T_h-T_{hw}$ and $\delta T_c=T_{cw}-T_{c}$. In Eq.~\eqref{eq:power} we introduced the total cycle time $\zeta (\tau_h+\tau_c)$. This means we suppress any explicit dependence of the analysis on the lengths of the adiabatic strokes and exclusively focus on the isotherms, i.e, on the temperature difference between working medium and the hot and cold reservoirs.

It is then a simple exercise to find the maximum of $P(\delta T_h,\delta T_c)$ as a function of $\delta T_h$ and $\delta T_c$. After a few lines of algebra one obtains \cite{Curzon1975}
\begin{equation}
\label{eq:power_max}
P_\mrm{max}=\frac{\kappa_h\kappa_c}{\zeta}\left(\frac{\sqrt{T_h}-\sqrt{T_c}}{\sqrt{\kappa_h}+\sqrt{\kappa_c}}\right)^2\,,
\end{equation}
where the maximum is assumed for
\begin{equation}
\label{eq:deltaT_max}
\frac{\delta T_h}{T_h}=\frac{1-\sqrt{T_c/T_h}}{1+\sqrt{\kappa_h/\kappa_c}}\quad\mrm{and}\quad \frac{\delta T_c}{T_c}=\frac{\sqrt{T_h/T_c}-1}{1+\sqrt{\kappa_c/\kappa_h}}
\end{equation}
From these expressions we can now compute the efficiency. We have,
\begin{equation}
\label{eq:efficency}
\eta=\frac{Q_h-Q_c}{Q_h}=1-\frac{T_{cw}}{T_{hw}}=1-\frac{T_c+\delta T_c}{T_h-\delta T_h}
\end{equation}
where we used Eq.~\eqref{eq:entropy}. Thus, the efficiency of an endoreversible Carnot cycle at maximal power output is given by
\begin{equation}
\label{eq:efficency_CA}
\eta_{CA}=1-\sqrt{\frac{T_c}{T_h}}\,,
\end{equation}
which only depends on the temperatures of the hot and cold reservoirs.

The Curzon-Ahlborn efficiency is one of the first results that illustrate that (i) thermodynamics can be extended to treat nonequilibrium systems, and that (ii) also far from thermal equilibrium universal and mathematically simple relations govern the thermodynamic behavior. In the following we will analyze this observation a little more closely and see how universal statements arise from the nature of fluctuations.

\section{\label{sec:fluc} The advent of Stochastic Thermodynamics}

Relatively recently, Evans and co-workers \cite{Evans1993} discovered an unexpected symmetry in the simulation of sheared fluids. In small systems the dynamics is governed by thermal fluctuations and, thus, also thermodynamic quantities such as heat and work fluctuate.  Remarkably, single fluctuations can be at variance with the macroscopic statements of the second law. For instance, the change of entropy can be negative, or the performed work amounts to less than the free energy difference. Nevertheless, the probability distribution for the thermodynamic observables fulfills a symmetry relation, which has become known as \emph{fluctuation theorem}.

In its most general form the fluctuation theorem relates the probability to find a negative entropy production $\Sigma$ with the probability of the positive value,
\begin{equation}
\label{eq:FT}
\frac{\mc{P}\left(\Sigma=-A\right)}{\mc{P}\left(\Sigma=A\right)}=\e{-A}\,.
\end{equation}
Using Jensen's inequality for exponentials, $\e{-\la x\ra}\geq \la \e{-x}\ra$, Eq.~\eqref{eq:FT}, immediately implies that
\begin{equation}
\label{eq:sigma}
\la \Sigma\ra\geq0\,,
\end{equation}
which is a variation of the Clausius inequality Eq.~\eqref{eq:clausius_1}. Therefore, the fluctuation theorem can be interpreted as a generalization of the second law to systems far from equilibrium. For the average entropy production we retrieve the ``old'' statements. However, we also have that negative fluctuations of the entropy production do occur -- they are just exponentially unlikely.

The first rigorous proof of the fluctuation theorem was published by Gallavotti and Cohen in 1995 \cite{Gallavotti1995}, which was quickly generalized to Langevin dynamics \cite{Kurchan1998} and general Markov processes \cite{Lebowitz1999}.

The discovery of the fluctuation theorems has effectively opened a new area of thermodynamics, which adopted the name \emph{\textbf{Stochastic Thermodynamics}}. Rather than focusing on describing  macroscopic systems in equilibrium, Stochastic Thermodynamics is interested in the thermodynamic behavior of small systems that operate far from thermal equilibrium and whose dynamics are governed by fluctuations. Since quantum systems obviously fall into this class, we will briefly summarize the major achievements for classical systems that laid the ground work for what we will eventually be interested in -- the thermodynamics of quantum systems.

\subsection{Microscopic dynamics}

To fully understand and appreciate the fluctuation theorem Eq.~\eqref{eq:FT} we continue by briefly outlining the most important descriptions of random motion. Generally there are two distinct approaches: (i) explicitly modeling the dynamics of a stochastic observable, or (ii) describing the dynamics of the probability density function of a stochastic variable. Among the many variations of these two approaches the conceptually simplest notions are the \emph{Langevin equation} and the \emph{Klein-Kramers equation}.

\paragraph{Langevin equation.}

In 1908  Paul Langevin, a French physicist, proposed a powerful description of Brownian motion \cite{Langevin1908,Lemmons1997}. The Langevin equation is a Newtonian equation of motion  for a single Brownian particle driven by a stochastic force modeling the random kicks from the environment,
\begin{equation}
\label{eq:Langevin}
m\, \ddot x+m\gamma\, \dot x+V'(x)=\xi(t)\,.
\end{equation}
Here, $m$ denotes the mass of the particle, $\gamma$ is the damping coefficient and $V'(x)=\pd_x\, V(x)$ is a conservative force from a confining potential. The stochastic force, $\xi(t)$ describes the randomness in a small, but open system due to thermal fluctuations.  In the simplest case, $\xi(t)$ is assumed to be Gaussian white noise, which is characterized by,
\begin{equation}
\label{eq:correlation_noise}
\begin{split}
 \la \xi(t) \ra=0\quad\mrm{and}\quad\la \xi(t)\,\xi(s)\ra= 2 D \,\delta{\left(t-s\right)}\,,
\end{split}
\end{equation}
where $D$ is the diffusion coefficient. Despite its apparently simple form the Langevin equation \eqref{eq:Langevin} exhibits several mathematical peculiarities. How to properly handle the stochastic force, $\xi(t)$, led to the study of stochastic differential equations, for which we refer to the  literature \cite{Risken1989}. 

It is interesting to note that the Langevin equation \eqref{eq:Langevin} is equivalent to Einstein's treatment of Brownian motion \cite{Einstein1905}. This can be seen by explicitly deriving the \emph{Fluctuation-Dissipation theorem} from Eq.~\eqref{eq:Langevin}.

\paragraph{Fluctuation-Dissipation Theorem.}

The Langevin equation \eqref{eq:Langevin} for the case of a free particle, $V(x)=0$, can be expressed in terms of the velocity $v = \dot x$ as,
\begin{equation}
\label{eq:Langevin_v}
m\, \dot v + m\gamma\, v = \xi(t) \,.
\end{equation}
The solution of the latter first-order differential equation \eqref{eq:Langevin_v} reads,
\begin{equation}
\label{eq:solution_v}
v_t=v_0\,\e{-\gamma t} + \frac{1}{m} \int\limits_{0}^{t} d s\, \xi(s)\, \e{-\gamma\left(t-s\right)}\,,
\end{equation}
where $v_0$ is the initial velocity. Since the Langevin force is of vanishing mean \eqref{eq:correlation_noise}, the averaged solution $\la v_t\ra$ becomes,
\begin{equation}
\label{eq:average_v}
\la v_t \ra =v_0\,\e{-\gamma t}\,.
\end{equation}
Moreover,  we obtain for the mean-square velocity $\la v^2_t \ra $,
\begin{equation}
\label{eq:var_v}
\la v^2_t \ra=v_0^2\,\e{-2\gamma t} + \frac{1}{m^2} \int\limits_{0}^{t} d s_1 \int\limits_{0}^{t} d s_2 \, \e{-\gamma\left(t-s_1\right)}  \e{-\gamma\left(t-s_2\right)}\la \xi({s_1}) \xi({s_2}) \ra\,.
\end{equation}
With the help of the correlation function \eqref{eq:correlation_noise} the twofold integral can be written in closed form and, thus, Eq.~\eqref{eq:var_v} becomes,
\begin{equation}
\label{eq:var_v_2}
\la v^2_t \ra=v_0^2\,\e{-2\gamma t} + \frac{D}{\gamma m^2}\left(1-\e{-2\gamma t}\right)\,.
\end{equation}
In the stationary state for $\gamma\,t\gg1$, the exponentials become negligible and the mean-square velocity \eqref{eq:var_v_2} further simplifies to,
\begin{equation}
\label{eq:var_v_stat}
\la v^2_t \ra=\frac{D}{\gamma m^2} \,.
\end{equation}
However, we also know from kinetic gas theory \cite{Callen1985} that in equilibrium $\la v^2_t \ra=1/\beta m$ where we introduce the inverse temperature, $\beta=1/k_B T$. Thus, we finally have
\begin{equation}
\label{eq:FDT}
D=\frac{m\gamma}{\beta}\,,
\end{equation}
which is the \emph{Fluctuation-Dissipation Theorem}.

\paragraph{Klein-Kramers equation.}

The Klein-Kramers equation is an equation of motion for distribution functions in position and velocity space, which is equivalent to the Langevin equation \eqref{eq:Langevin}, see also \cite{Risken1989}. For a Brownian particle in one-dimension it takes the form,
\begin{equation}
\label{eq:KK}
\frac{\pd}{\pd t} P(x,v,t)=-\frac{\pd}{\pd x} \left( v \,P(x,v,t)\right)+\frac{\pd}{\pd v}\left(\frac{V'(x)}{m}\,P(x,v,t)+\gamma v\,P(x,v,t)\right)+\frac{\gamma}{m\beta}\,\frac{\pd^2}{\pd v^2} P(x,v,t)\,,
\end{equation}
Note that by construction the stationary solution of the Klein-Kramers equation \eqref{eq:KK} is the Boltzmann-Gibbs distribution, $P_\mathrm{eq}\propto\exp{\left(-\beta/2\,m v^2-\beta V\right)}$. The main advantage of the Klein-Kramers equation \eqref{eq:KK} over the Langevin equation \eqref{eq:Langevin} is that we can compute the entropy production directly, which we will exploit shortly for quantum systems in Sec.~\ref{sec:entropy}.

\subsection{Stochastic energetics}

An important step towards  the discovery of the fluctuation theorems \eqref{eq:FT} was Sekimoto's insight that thermodynamic notions can be generalized to single particle dynamics \cite{Sekimoto1998}. To this end, consider the overdamped Langevin equation
\begin{equation}
\label{eq:Langevin_over}
0=-\left(-m\gamma\, \dot x+\xi_t\right)d x + \pd_x V\left(x,\lambda\right)d x\,,
\end{equation}
where we separated contributions stemming from the interaction with the environment and mechanical forces. Here and in the following, $\lambda$ is an external control parameter, whose variation drives the system. 

Generally, a change in internal energy of a single particle is comprised of changes in kinetic \emph{and} potential energy. In the overdamped limit, however, one assumes that the momentum degrees of freedom equilibrate much faster than any other time-scale of the dynamics. Thus, the kinetic energy is always at its equilibrium value, and thus a change in internal energy, $d e$, for a single trajectory, $x$, is given by
\begin{equation}
\label{eq:stoch_first}
d e(x,\lambda)=d V(x,\lambda)=\pd_x V(x,\lambda)d x+\pd_\lambda V(x,\lambda)d \lambda\,.
\end{equation}
Further, identifying  the heat with the external terms in Eq.~\eqref{eq:Langevin_over}, which are governed by the damping and the noise, we can write
\begin{equation}
\label{eq:heat}
d q(x)=\left(-m\gamma\, \dot x+\xi_t\right)d x\,.
\end{equation}
Thus, we obtain a stochastic, microscopic expression of the first law \eqref{eq:1st}
\begin{equation}
\label{eq:stoch_first_a}
0=-d q(x)+d e(x,\lambda)-\pd_\lambda V(x,\lambda)d \lambda\,,
\end{equation}
which uniquely defines the stochastic work for a single trajectory,
\begin{equation}
\label{eq:work_stoch}
d w(x)=\pd_\lambda V(x,\lambda)d \lambda
\end{equation}
Note that the work increment, $d w$, is given by the partial derivative of the potential with respect to the externally controllable work parameter, $\lambda$. 

\subsection{Jarzynski equality and Crooks theorem}

The stochastic work increment $d w(x)$ uniquely characterizes the thermodynamics of single Brownian particles. However, since $d w(x)$ is subject to thermal fluctuations none of the traditional statements of the second law can be directly applied, and in particular there is no maximum work theorem for $d w(x)$. Therefore, special interest has to be on the distribution of $\mc{P}(W)$, where $W=\int d w(x)$ is the accumulated work performed during a thermodynamic process.

In the following we will briefly discuss representative derivations of the most prominent fluctuation theorems, namely the classical Jarzynski equality and the Crooks theorem, and then the quantum Jarzynski equality in Sec.~\ref{sec:quantum_Jarzynski} and finally a quantum fluctuation theorem for entropy production in Sec.~\ref{sec:entropy}.

\paragraph{Jarzynski equality.}

\begin{figure}
\begin{mdframed}[roundcorner=10pt]
\begin{minipage}[l]{.4\textwidth}
\begin{center}
\includegraphics[height=5cm]{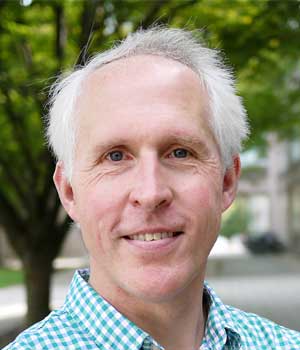}
\end{center}
\end{minipage}
\hfill
\begin{minipage}[r]{.58\textwidth}
Christopher Jarzynski:\\ \emph{If we shift our focus away from equilibrium states, we find a rich universe of non-equilibrium behavior \cite{Jarzynski2015}. }
\end{minipage}
\end{mdframed}
\end{figure}

Thermodynamically, the simplest cases are systems that are isolated from their thermal environment. Realistically imagine, for instance, a small system that is \emph{ultraweakly} coupled to the environment. If left alone, the system equilibrates at inverse temperature $\beta$ for a fixed work parameter, $\lambda$. Then, the time scale of the variation of the work parameter is taken to be much shorter than the relaxation time, $1/\gamma$. Hence, the dynamics of the system during the variation of $\lambda$ can be approximated by Hamilton's equations of motion to high accuracy. 

Now, let $\Gamma=\left(\vec{q},\vec{p}\right)$ denote a \textit{microstate} of the system, which is a point in the many-dimensional phase space including all relevant coordinates to specify the microscopic configurations $\vec{q}$ and momenta $\vec{p}$. Further, $H\left(\Gamma;\lambda\right)$ denotes the Hamiltonian of the system and the Klein-Kramers equation \eqref{eq:KK} reduces for $\gamma\ll1$ to the Liouville equation,
\begin{equation}
\label{eq:Liouville}
\frac{\pd}{\pd t} P(\Gamma,t)=-\{P(\Gamma,t),H\left(\Gamma;\lambda\right)\}\,,
\end{equation}
where $\{\cdot,\cdot\}$ denotes the Poisson bracket.

We now assume that the system was initially prepared in a Boltzmann-Gibbs equilibrium state
\begin{equation}
\label{eq:BG}
p^\mathrm{eq}_\lambda \left(\Gamma\right)=\frac{1}{Z_\lambda}\,\exp{\left(-\beta H\left(\Gamma;\lambda\right)\right)}\,,
\end{equation}
with partition function $Z_\lambda$ and Helmholtz free energy, $F_\lambda$,
\begin{equation}
\label{eq:partition}
Z_\lambda=\int d \Gamma\,\exp{\left(-\beta H\left(\Gamma;\lambda\right)\right)}\quad\mrm{and}\quad \beta F_\lambda=-\ln{Z_\lambda}\,.
\end{equation}
As the system is isolated during the thermodynamic process we can identify the work performed during a single realization with the change in the Hamiltonian,
\begin{equation}
\label{eq:work_hamiltonian}
W=H\left(\Gamma_\tau\left(\Gamma_0\right);\lambda_\tau\right)-H\left(\Gamma_0;\lambda_0\right)\,,
\end{equation}
where $\Gamma_\tau\left(\Gamma_0\right)$ is time-evolved point in phase space given that the system started at $\Gamma_0$.

It is then a simple exercise to derive the Jarzynski equality for Hamiltonian dynamics \cite{Jarzynski1997}. To this end, consider
\begin{equation}
\label{eq:Jarzynski_derive}
\begin{split}
\la \e{-\beta W}\ra&=\int d\Gamma_0\,p^\mathrm{eq}_{\lambda_0} \left(\Gamma_0\right)\,\e{-\beta W\left(\Gamma_0\right)} \\
\quad&=\frac{1}{Z_{\lambda_0}}\,\int d\Gamma_0\,\e{-\beta H\left(\Gamma_\tau\left(\Gamma_0\right);\lambda_\tau\right)}\\
\quad&=\frac{1}{Z_{\lambda_0}}\,\int d\Gamma_\tau\,\left|\frac{\pd \Gamma_\tau}{\pd \Gamma_0}\right|^{-1}\,\e{-\beta H\left(\Gamma_\tau;\lambda_\tau\right)}\,.
\end{split}
\end{equation}
Changing variables and using Liouville's theorem, which ensures conservation of phase space volume, i.e., $\left|\pd \Gamma_\tau/\pd \Gamma_0\right|^{-1}=1$,  we arrive at,
\begin{equation}
\label{eq:Jarzynski_result}
\la \e{-\beta W}\ra=\frac{1}{Z_{\lambda_0}}\,\int d\Gamma_\tau\,\e{-\beta H\left(\Gamma_\tau;\lambda_\tau\right)}=\frac{Z_{\lambda_\tau}}{Z_{\lambda_0}}=\e{-\beta \Delta F}\,.
\end{equation}

The Jarzynski equality \eqref{eq:Jarzynski_result} is one of the most important building blocks of modern thermodynamics \cite{Zarate2011}. It can be rightly understood as a generalization of the second law of thermodynamics to systems far from equilibrium, and it has been shown to hold a in wide range of classical systems, with weak and strong coupling, with slow and fast dynamics, with Markovian and non-Markovian noise etc. \cite{Jarzynski2011}.

\paragraph{Crooks's fluctuation theorem.}

The second most prominent fluctuation theorem is the work relation by Crooks \cite{Crooks1998,Crooks1999}. As before we are interested in the evolution of a thermodynamic system for times $0\leq t\leq \tau$, during which the work parameter, $\lambda_t$, is varied according to some protocol. For the present purposes, we now assume that the thermodynamic process is described as a sequence, $\Gamma_0, \Gamma_1,...,\Gamma_N,$ of microstates visited at times $t_0, t_1,...,t_N$ as the system evolves. For the sake of simplicity we assume the time sequence to be equally distributed, $t_n=n\tau/N$, and, implicitly, $\left(\Gamma_N;t_N\right)=\left(\Gamma_\tau;\tau\right)$. Moreover,  we assume that the evolution is a \textit{Markov process}: given the microstate $\Gamma_n$ at time $t_n$, the subsequent microstate $\Gamma_{n+1}$ is sampled randomly from a transition probability distribution, $P$, that depends merely on $\Gamma_n$, but not on the microstates visited at earlier times than $t_n$ \cite{Kampen1992}. This means that the transition probability to go from $\Gamma_{n}$ to $\Gamma_{n+1}$ depends only on the current microstate, $\Gamma_n$, and the the current value of the work parameter, $\lambda_n$. Finally, we assume that the system fulfills a local detailed balance condition \cite{Kampen1992}, namely
\begin{equation}
\label{eq:detailed_balance}
\frac{P\left(\Gamma\rightarrow \Gamma';\lambda\right)}{P\left(\Gamma\leftarrow \Gamma';\lambda\right)}=\frac{\e{-\beta H\left(\Gamma';\lambda\right)}}{\e{-\beta H\left(\Gamma;\lambda\right)}}\,.
\end{equation}

When the work parameter, $\lambda$, is varied in discrete time steps from $\lambda_0$ to $\lambda_N=\lambda_\tau$, the evolution of the system during one time step can be expressed as a sequence,
\begin{equation}
\label{eq:forward}
\mathrm{forward:}\hspace{2em}\left(\Gamma_n,\lambda_n\right)\rightarrow\left(\Gamma_n,\lambda_{n+1}\right)\rightarrow\left(\Gamma_{n+1},\lambda_{n+1}\right)\,.
\end{equation}
In this sequence first the value of the work parameter is updated and, then, a random step is taken by the system. A trajectory of the whole process between initial, $\Gamma_0$, and final microstate, $\Gamma_\tau$, is generated by first sampling $\Gamma_0$ from the initial, Boltzmann-Gibbs distribution $p^\mathrm{eq}_{\lambda_0}$ and, then, repeating Eq.~\eqref{eq:forward} in time increments, $\delta t=\tau/N$.

Consequently, the net change in internal energy, $\Delta E=H\left(\Gamma_N,\lambda_N\right)-H\left(\Gamma_0,\lambda_0\right)$, can be written as a sum of two contributions. First, the changes in energy due to variations of the work parameter,
\begin{equation}
\label{eq:work_Crooks}
W=\sum\limits_{n=0}^{N-1}\left[H\left(\Gamma_n;\lambda_{n+1}\right)-H\left(\Gamma_n;\lambda_n\right)\right]\,,
\end{equation}
and second, changes due to transitions between microstates in phase space,
\begin{equation}
\label{eq:heat_Crooks}
Q=\sum\limits_{n=0}^{N-1}\left[H\left(\Gamma_{n+1};\lambda_{n+1}\right)-H\left(\Gamma_n;\lambda_{n+1}\right)\right]\,.
\end{equation}
As argued by Crooks \cite{Crooks1998} the first contribution \eqref{eq:work_Crooks} is given by an \textit{internal} change in energy and the second term \eqref{eq:heat_Crooks} stems from the interaction with the environment introducing the random steps in phase space. Thus, Eq.~\eqref{eq:work_Crooks} is a natural definition of stochastic work, and Eq.~\eqref{eq:heat_Crooks} is the stochastic heat for a single trajectory.

The probability to generate a trajectory, $\Xi=\left(\Gamma_0\rightarrow...\Gamma_N \right)$, starting in a particular initial state, $\Gamma_0$, is given by the product of the initial distribution and all subsequent transition probabilities,
\begin{equation}
\label{eq:prob_forward}
P^F[\Xi]=p^\mathrm{eq}_{\lambda_0}\left(\Gamma_0\right)\,\prod\limits_{n=0}^{N-1}\,P\left(\Gamma_n\rightarrow\Gamma_{n+1};\lambda_{n+1}\right)\,,
\end{equation}
where the stochastic independency of the single steps is guaranteed by the Markov assumption.

Analogously to the \emph{forward} process, we can define a \emph{reverse} trajectory with $\left(\lambda_0\leftarrow\lambda_\tau\right)$. However, the starting point is sampled from $p^\mathrm{eq}_{\lambda_\tau}$ and the system first takes a random step and, then, the value of the work parameter is updated,
\begin{equation}
\label{eq:reverse}
\mathrm{reversed:}\hspace{2em}\left(\Gamma_{n+1},\lambda_{n+1}\right)\leftarrow\left(\Gamma_{n+1},\lambda_n\right)\leftarrow\left(\Gamma_n,\lambda_n\right)\,.
\end{equation}
Now, we compare the probability of a trajectory $\Xi$ during a forward process, $P^F[\Xi]$, with the probability of the \textit{conjugated} path, $\Xi^\dagger=\left(\Gamma_0\leftarrow...\Gamma_N \right)$, during the reversed process, $P^R[\Xi^\dagger]$. The ratio of these probabilities reads,
\begin{equation}
\label{eq:forward_vs_reverse}
\frac{P^F[\Xi]}{P^R[\Xi^\dagger]}=\frac{p^\mathrm{eq}_{\lambda_0}\left(\Gamma_0\right)\,\prod\limits_{n=0}^{N-1}\,P\left(\Gamma_n\rightarrow\Gamma_{n+1};\lambda^F_{n+1}\right)}{p^\mathrm{eq}_{\lambda_1}\left(\Gamma_N\right)\,\prod\limits_{n=0}^{N-1}\,P\left(\Gamma_n\leftarrow\Gamma_{n+1};\lambda^R_{N-1-n}\right)}\,.
\end{equation}
Here, $\left\{\lambda_0^F, \lambda_1^F,...,\lambda_N^F \right\}$ is the protocol for varying the external work parameter from $\lambda_0$ to $\lambda_\tau$ during the forward process. Analogously, $\left\{\lambda_0^R, \lambda_1^R,...,\lambda_N^R \right\}$ specifies the reversed process, which is related to the forward process by,
\begin{equation}
\label{eq:alpha}
\lambda_n^R=\lambda_{N-n}^F\,.
\end{equation}
Hence, every factor $P\left(\Gamma\rightarrow\Gamma';\lambda\right)$ in the numerator of the ratio \eqref{eq:forward_vs_reverse} is matched by $P\left(\Gamma\leftarrow\Gamma';\lambda\right)$ in the denominator. 

In conclusion, Eq.~\eqref{eq:forward_vs_reverse} reduces to \cite{Crooks1998},
\begin{equation}
\label{eq:detailed_Crooks}
\frac{P^F[\Xi]}{P^R[\Xi^\dagger]}=\e{\beta\left(W^F[\Xi]-\Delta F\right)}\,,
\end{equation}
where $W^F[\Xi] $ is the work performed on the system during the forward process. 

Forward work, $W^F[\Xi] $, and reverse work, $W^R[\Xi^\dagger]$, are related through
\begin{equation}
\label{eq:work_fr}
W^F[\Xi]=-W^R[\Xi^\dagger]
\end{equation}
for a conjugate pair of trajectories, $\Xi$ and $\Xi^\dagger$. The corresponding work distributions, $\mc{P}_F$ and $\mc{P}_R$, are then given by an average over all possible realizations, i.e. all discrete trajectories of the process,
\begin{equation}
\label{eq:P_work}
\begin{split}
\mc{P}_F\left(+W\right)&=\int d\Xi \,P^F[\Xi]\,\de{W-W^F[\Xi]}\\
\mc{P}_R\left(-W\right)&=\int d\Xi \,P^R[\Xi^\dagger]\,\de{W+W^R[\Xi^\dagger]}\,,
\end{split}
\end{equation}
where $d\Xi=d\Xi^\dagger=\prod_n d \Gamma_n$. Collecting Eqs.~\eqref{eq:detailed_Crooks} and \eqref{eq:P_work} the work distribution for the forward processes can be written as
\begin{equation}
\label{eq:forward_2}
\mc{P}_F\left(+W\right)=\e{\beta\left(W-\Delta F\right)}\,\int d\Xi \,P^R[\Xi^\dagger]\,\de{W+W^R[\Xi^\dagger]}\,,
\end{equation}
from which we obtain the Crooks fluctuation theorem \cite{Crooks1999}
\begin{equation}
\label{eq:Crooks_theorem}
\mc{P}_R\left(-W\right)=\e{-\beta\left(W-\Delta F\right)}\,\mc{P}_F\left(+W\right)\,.
\end{equation}

It is interesting to note that the Crooks theorem  \eqref{eq:Crooks_theorem} is a detailed version of the Jarzynski equality \eqref{eq:Jarzynski_result}, which follows from integrating Eq.~\eqref{eq:Crooks_theorem} over the forward work distribution,
\begin{equation}
\label{eq:Jarzynski_Crooks}
1=\int d W\,\mc{P}_R\left(-W\right)=\int d W\,\e{-\beta\left(W-\Delta F\right)}\,\mc{P}_F\left(+W\right)=\la \e{-\beta\left(W-\Delta F\right)}\ra_F\,.
\end{equation}
Note, however, that the Crooks theorem \eqref{eq:Crooks_theorem} is only valid for Markovian processes \cite{Jarzynski2000}, whereas the Jarzynski equality can also be shown to hold for non-Markovian dynamics \cite{Speck2007}.

\section{\label{sec:envariance} Foundations of statistical physics from quantum entanglement}

In the preceding section we implicitly assumed that there is a well-established theory if and how physical systems are described in a state of thermal equilibrium. For instance, in the treatment of the Jarzynski equality \eqref{eq:Jarzynski_result} and the Crooks fluctuation theorem \eqref{eq:Crooks_theorem} we assumed that the system is initially prepared in a Boltzmann-Gibbs distribution. In standard textbooks of statistical physics this description of canonical thermal equilibria is usually derived from the fundamental postulate, Boltzmann's H-theorem, the ergodic hypothesis, or the maximization of the statistical entropy in equilibrium \cite{Toda1983,Callen1985}. However, none of these concepts are particularly well-phrased for quantum systems.

It is important to realize that statistical physics was developed in the XIX century, when the fundamental physical theory was classical mechanics. Statistical physics was then developed to translate between microstates (points in phase space) and thermodynamic macrostates (given by temperature, entropy, pressure, etc). Since microstates and macrostates are very different notions a new theory became necessary that allows to ``translate'' with the help of fictitious, but useful, concepts such as ensembles. However, ensembles consisting of infinitely many copies of the same system seem rather ill-defined from the point of view of a fully quantum theory.

Only relatively recently this conceptual problem was repaired by showing that the famous representations of microcanonical and canonical equilibria can be obtained from a fully quantum treatment -- from symmetry considerations of entanglement \cite{Deffner_Zurek2016}. 
This novel approach to the foundations of statistical mechanics relies on \emph{entanglement assisted invariance} or in short on \emph{envariance} \cite{Zurek2003a,Zurek2005,Zurek2011}.

In the following we summarize the main conceptual steps that were originally published in Ref.~\cite{Deffner_Zurek2016}.

\subsection{Entanglement assisted invariance}

Consider a quantum system, $\mc{S}$, which is maximally entangled with an environment, $\mc{E}$, and let $\ket{\psi_\mc{SE}}$ denote the composite state in $\mc{S}\otimes\mc{E}$. Then $\ket{\psi_\mc{SE}}$ is called \emph{envariant} under a unitary map $U_\mc{S}=u_\mc{S}\otimes\id_\mc{E}$ if and only if there exists another unitary $U_\mc{E}=\id_\mc{S}\otimes u_\mc{E}$ such that,
\begin{equation}
\begin{split}
\label{eq:U}
U_\mc{S} \ket{\psi_\mc{SE}}& =\left(u_\mc{S}\otimes\id_\mc{E}\right)\ket{\psi_\mc{SE}}=\ket{\eta_\mc{SE}}\\
U_\mc{E}\ket{\eta_\mc{SE}}&=\left(\id_\mc{S}\otimes u_\mc{E}\right)\ket{\eta_\mc{SE}}=\ket{\psi_\mc{SE}}\,.
\end{split}
\end{equation}
Thus, $U_\mc{E}$ that does not act on $\mc{S}$ ``does the job'' of the inverse map of $U_\mc{S}$ on $\mc{S}$ -- assisted by the environment $\mc{E}$.

The principle is most easily illustrated with a simple example. Suppose $\mc{S}$ and $\mc{E}$ are each given by two-level systems, where $\{\ket{\uparrow}_\mc{S}, \ket{\downarrow}_\mc{S}\}$ are the eigenstates of $\mc{S}$ and $\{\ket{\uparrow}_\mc{E}, \ket{\downarrow}_\mc{E}\}$ span $\mc{E}$. Now, further assume $\ket{\psi_\mc{SE}}\propto\ket{\uparrow}_\mc{S}\otimes\ket{\uparrow}_\mc{E}+\ket{\downarrow}_\mc{S}\otimes\ket{\downarrow}_\mc{E}$ and $U_\mc{S}$ is a \textit{swap} in $\mc{S}$ -- it ``flips'' its spin. Then, we have
\begin{equation}
\label{eq02a}
 \begin{tikzpicture}[>=stealth,baseline,anchor=base,inner sep=0pt]
      \matrix (foil) [matrix of math nodes,nodes={minimum height=0.5em}] {
         & {\color{blue}\ket{\uparrow}_\mc{S}} & \otimes & \ket{\uparrow}_\mc{E} &  & + &  & {\color{blue}\ket{\downarrow}_\mc{S}} & \otimes & \ket{\downarrow}_\mc{E} &  \xrightarrow{\quad {\color{blue}U_\mc{S}}\quad} 
{\color{blue}\ket{\downarrow}_\mc{S}}\otimes\ket{\uparrow}_\mc{E}+{\color{blue}\ket{\uparrow}_\mc{S}}\otimes\ket{\downarrow}_\mc{E}\,.\\
      };
      \path[->] ($(foil-1-2.north)+(0,1ex)$)   edge[blue,bend left=45]    ($(foil-1-8.north)+(0,1ex)$);
      \path[<-] ($(foil-1-2.south)-(0,1ex)$)   edge[blue,bend left=-45]    ($(foil-1-8.south)-(0,1ex)$);
    \end{tikzpicture}
\end{equation}
The action of $U_\mc{S}$ on $\ket{\psi}_\mc{SE}$ can be restored by a swap, $U_\mc{E}$, on $\mc{E}$,
  \begin{equation}
    \begin{tikzpicture}[>=stealth,baseline,anchor=base,inner sep=0pt]
      \matrix (foil) [matrix of math nodes,nodes={minimum height=0.5em}] {
         & \ket{\downarrow}_\mc{S} & \otimes & {\color{red}\ket{\uparrow}_\mc{E}} &  & + &  & \ket{\uparrow}_\mc{S} & \otimes & {\color{red}\ket{\downarrow}_\mc{E}} & \xrightarrow{\quad {\color{red}U_\mc{E}}\quad} 
\ket{\downarrow}_\mc{S}\otimes{\color{red}\ket{\downarrow}_\mc{E}}+\ket{\uparrow}_\mc{S}\otimes{\color{red}\ket{\uparrow}_\mc{E}}\,. \\
      };
      \path[->]  ($(foil-1-4.north)+(0,1ex)$) edge[red,bend left=45] ($(foil-1-10.north)+(0,1ex)$);
       \path[<-]  ($(foil-1-4.south)-(0,1ex)$) edge[red,bend left=-45] ($(foil-1-10.south)-(0,1ex)$);
    \end{tikzpicture}
  \end{equation}
Thus, the swap $U_\mc{E}$ on $\mc{E}$ restores the pre-swap $\ket{\psi}_\mc{SE}$ without ``touching'' $\mc{S}$, i.e., the global state is restored by solely acting on $\mc{E}$. Consequently, local probabilities of the two swapped spin states are both exchanged and unchanged. Hence, they have to be equal. This provides the fundamental connection of quantum states and probabilities \cite{Zurek2003a}, and leads to Born's rule \cite{Zurek2005}.

Recent experiments in quantum optics \cite{Vermeyden2015,Karimi2016} and on IBM's Q Experience \cite{Deffner_heliyon2017} have shown that envariance is not only a theoretical concept, but a physical reality. Thus, envariance is a valid and purely quantum mechanical concept that we can use as a stepping stone to \emph{motivate} and \emph{derive} quantum representations of thermodynamic equilibrium states.

\subsection{Microcanonical state from envariance}

We begin by considering the microcanonical equilibrium. Generally, thermodynamic equilibrium states are characterized by extrema of physical properties, such as maximal phase space volume, maximal thermodynamic entropy, or maximal randomness \cite{Uffink2007}.  We will define the microcanonical equilibrium as the quantum state that is ``maximally envariant'', i.e., envariant under all unitary operations on $\mc{S}$. To this end, we write the composite state $\ket{\psi_\mc{SE}}$ in Schmidt decomposition \cite{Nielsen2010},
\begin{equation}
\label{eq:Schmidt}
\ket{\psi_\mc{SE}}=\sum_k a_k \ket{s_k}\otimes\ket{\varepsilon_k}\,,
\end{equation}
where by definition $\{\ket{s_k}\}$ and $\{\ket{\varepsilon_k}\}$ are orthocomplete in $\mc{S}$ and $\mc{E}$, respectively. The task is now to identify the ``special'' state that is maximally envariant.

It has been shown \cite{Zurek2005} that $\ket{\psi_\mc{SE}}$ is envariant under all unitary operations if and only if the Schmidt decomposition is even, i.e., all coefficients have the same absolute value, $\left|a_k\right|=\left|a_l\right|$ for all $l$ and $k$. We then can write,
\begin{equation}
\label{eq:Schmidt_even}
\ket{\psi_\mc{SE}}\propto \sum_k \e{i\phi_k} \ket{s_k}\otimes\ket{\varepsilon_k}\,,
\end{equation}
where $\phi_k$ are phases. Recall that in classical statistical mechanics equilibrium ensembles are identified as the states with the largest corresponding volume in phase space \cite{Uffink2007}. In the present context this ``identification'' readily translates into an equilibrium state that is envariant under the maximal number of, i.e.,  \emph{all}  unitary operations.

To conclude the derivation we note that the microcanonical state is commonly identified as the state that is also fully energetically degenerate \cite{Callen1985}. To this end,  denote the Hamiltonian of the composite system by
\begin{equation}
\label{eq:energy}
H_\mc{SE}=H\otimes \id_\mc{E}+\id_\mc{S}\otimes H_\mc{E}\,.
\end{equation}
Then, the internal energy of $\mc{S}$ is given by the quantum mechanical average
\begin{equation}
\label{eq:energy_1}
E=\bra{\psi_\mc{SE}} \left(H\otimes \id_\mc{E}\right)\ket{\psi_\mc{SE}}=\sum_k \bra{s_k} H\ket{s_k}/Z_\mrm{mic}\,,
\end{equation} 
where $Z_\mrm{mic}$ is the energy-dependent dimension of the Hilbert space of $\mc{S}$, which is commonly also called the microcanonical partition function \cite{Callen1985}. Since $\ket{\psi_\mc{SE}}$ \eqref{eq:energy_1} is envariant under all unitary maps we can assume without loss of generality that $\{s_k\}_{k=1}^{Z_\mrm{mic}}$ is a representation of the energy eigenbasis corresponding to $H$, and we have $  \bra{s_k} H\ket{s_k}=e_k$ with $E=e_k=e_{k'}$ for all $k,k'\in\{1,\dots,Z_\mrm{mic}\}$.

Therefore, we have identified the fully quantum mechanical representation of the microcanonical state by two conditions. Note that in our framework the microcanonical equilibrium is not represented by a unique state, but rather by an equivalence class of all maximally envariant states with the same energy: the state representing the microcanonical equilibrium of a system $\mc{S} $ with Hamiltonian $H$ is the state that is (i) envariant under all unitary operations on $\mc{S}$ and (ii) fully energetically degenerate with respect to $H$.

\paragraph{Reformulation of the fundamental statement.}

Before we continue to rebuild the foundations of statistical mechanics using envariance, let us briefly summarize and highlight what we have achieved so far. All standard treatments of the microcanonical state relied on notions such as probability, ergodicity, ensemble, randomness, indifference, etc. However, in the context of (quantum) statistical physics none of these expressions are fully well-defined. Indeed,  in the early days of statistical physics seminal researchers such as Maxwell and Boltzmann struggled with the conceptual difficulties \cite{Uffink2007}. Modern interpretation and understanding of statistical mechanics, however, was invented by Gibbs, who simply ignored such foundational issues and made full use of the concept of probability.

In contrast, in this approach we only need a quantum symmetry induced by entanglement -- envariance -- instead of relying on mathematically ambiguous concepts. Thus, we can reformulate the fundamental statement of statistical  mechanics in quantum physics:
\begin{quote}
\textit{The microcanonical equilibrium of a system $\mc{S}$ with Hamiltonian $H$ is a fully energetically degenerate quantum state envariant under all unitaries.}
\end{quote}
We will further illustrate this fully quantum mechanical approach to the foundations of statistical mechanics by also treating the canonical equilibrium. 

\subsection{Canonical state from quantum envariance}

Let us now imagine that we can separate the total system $\mc{S}$ into a smaller subsystem of interest $\mf{S}$ and its complement, which we call heat bath $\mf{B}$. The Hamiltonian of $\mc{S}$ can then be written as 
\begin{equation}
\label{eq:energy_2}
H=H_\mf{S}\otimes\id_\mf{B}+\id_\mf{S}\otimes H_\mf{B}+h_\mf{S, B}\,,
\end{equation}
where $h_\mf{S, B}$ denotes an interaction term. Physically this term is necessary to facilitate exchange of energy between the $\mf{S}$ and the heat bath $\mf{B}$. In the following, however, we will assume that $h_\mf{S, B}$ is sufficiently small so that we can neglect its contribution to the total energy, $E=E_\mf{S}+E_\mf{B}$, and its effect on the composite equilibrium state  $\ket{\psi_\mc{SE}}$. These assumptions are in complete analogy to the ones of classical statistical mechanics \cite{Toda1983,Callen1985} and it is commonly referred to as \emph{ultraweak coupling} \cite{Spohn1978}. 

Under these assumptions every composite energy eigenstate $\ket{s_k}$ can be written as a product,
\begin{equation}
\label{eq:states}
\ket{s_k}= \ket{\mf{s}_k}\otimes\ket{\mf{b}_k}\,,
\end{equation}
where the states $ \ket{\mf{s}_k}$ and $\ket{\mf{b}_k}$ are energy eigenstates in $\mf{S}$ and $\mf{B}$, respectively. At this point envariance is crucial in our treatment: All orthonormal bases are equivalent under envariance. Therefore, we can choose $\ket{s_k}$ as energy eigenstates of $H$.

For the canonical formalism we are now interested in the number of states accessible to the total system $\mc{S}$ under the condition that the total internal energy $E$ \eqref{eq:energy_1} is given and constant.  When the subsystem of interest, $\mf{S}$, happens to be in a particular energy eigenstate $\ket{\mf{s}_k}$ then the internal energy of the subsystem is given by the corresponding energy eigenvalue $\mf{e}_k$. Therefore, for the total energy $E$ to be constant, the energy of the heat bath, $E_\mf{B}$, has to obey,
\begin{equation}
\label{eq:energy_bath}
E_\mf{B}\left(\mf{e}_k\right)=E-\mf{e}_k\,.
\end{equation}
This condition can only be met if the energy spectrum of the heat reservoir is at least as dense as the one of the subsystem. 

The number of states, $\mf{N}\left(\mf{e}_k\right)$, accessible to $\mc{S}$ is then given by the fraction
\begin{equation}
\label{eq:number_of_states}
\mf{N}\left(\mf{e}_k\right)=\frac{\mf{N}_\mf{B}\left(E-\mf{e}_k\right)}{\mf{N}_\mc{S}(E)}\,,
\end{equation} 
where $\mf{N}_\mc{S}(E) $ is the total number of states in $\mc{S}$  consistent with Eq.~\eqref{eq:energy_1}, and $\mf{N}_\mf{B}\left(E-\mf{e}_k\right)$ is the number of states available to the heat bath, $\mf{B}$, determined by condition \eqref{eq:energy_bath}. In other words, we are asking for nothing else but the degeneracy in $\mf{B}$ corresponding to a particular energy state of the system of interest $\ket{\mf{s}_k}$.

\paragraph{\label{sec:example}Example: Composition of multiple qubits.}

The idea is most easily illustrated with a simple example, before we will derive the general formula in the following paragraph. Imagine a system of interest, $\mf{S}$, that interacts with $N$ non-interacting qubits with energy eigenstates $\ket{0}$ and $\ket{1}$ and corresponding eigenenergies $e^\mf{B}_0$ and $e^\mf{B}_1$.  Note once again that the composite states $\ket{s_k}$ can  always be chosen to be energy eigenstates, since the even composite state $\ket{\psi_\mc{SE}}$ \eqref{eq:Schmidt_even} is envariant under all unitary operations on $\mc{S}$.  

We further assume the qubits to be non-interacting. Therefore, all energy eigenstates can be written in the form
\begin{equation}
\ket{s_k}=\ket{\mf{s}_k}\otimes\underbrace{\ket{\delta^1_k \delta^2_k \cdots \delta^{N}_k}}_{N-qubits}\,.
\end{equation}
Here $\delta^i_k \in \{0,1\}$ for all $i\in {1,\dots,N}$  describing the states of the bath qubits. Let us   denote the number of qubits of $\mf{B}$ in $\ket{0}$ by $n$. Then the total internal energy $E$ becomes a simple function of $n$ and is given by,
\begin{equation}
E=\mf{e}_k+n\, e^\mf{B}_0 + (N-n)\, e^\mf{B}_1\,.
\end{equation}
Now it is easy to see that the total number of states corresponding to a particular value of $E$, i.e., the degeneracy in $\mf{B}$ corresponding to $\mf{e}_k$, \eqref{eq:number_of_states} is given by,
\begin{equation}
\label{eq:number_of_states_simple}
\mf{N}\left(\mf{e}_k\right)=\frac{N!}{n!\,(N-n)!}\,.
\end{equation} 
Equation~\eqref{eq:number_of_states_simple} describes nothing else but the number of possibilities to distribute $n\, e^\mf{B}_0$ and $ (N-n)\, e^\mf{B}_1$ over $N$ qubits.

It is worth emphasizing that in the arguments leading to Eq.~\eqref{eq:number_of_states_simple} we explicitly used that the $\ket{s_k}$ are energy eigenstates in $\mc{S}$ and the subsystem $\mf{S}$ and heat reservoir $\mf{B}$ are non-interacting. The first condition is not an assumption, since the composite $\ket{\psi_{\mc{SE}}}$ is envariant under all unitary maps on $\mc{S}$, and the second condition  is in full agreement with conventional assumptions of thermodynamics \cite{Toda1983,Callen1985}.

\paragraph{Boltzmann's formula for the canonical state.}

The example treated in the preceding section can be easily generalized. We again assume that the heat reservoir $\mf{B}$ consists of $N$ non-interacting subsystems with identical eigenvalue spectra $\{e_j^\mf{B}\}_{j=1}^{m}$.  In this case the internal energy \eqref{eq:energy_bath} takes the form
\begin{equation}
E=\mf{e}_k+ n_1\, e_1^\mf{B} +n_2\, e_2^\mf{B}+\cdots +n_{m} e_{m}^\mf{B}\,,
\end{equation}
with $\sum_{j=1}^{m} n_j =N$. Therefore, the degeneracy \eqref{eq:number_of_states} becomes
\begin{equation}
\label{eq:nos}
\mf{N}\left(\mf{e}_k \right)=\frac{N!}{n_1! n_2!\cdots n_{m}!}\,.
\end{equation}
This expression is readily recognized as a quantum envariant formulation of Boltzmann's counting formula for the number of classical microstates \cite{Uffink2007}, which quantifies the volume of phase space occupied by the thermodynamic state. However, instead of having to equip  phase space with an (artificial) equispaced grid, we simply count degenerate states.

We are now ready to derive the Boltzmann-Gibbs formula. To this end consider that in the limit of very large, $N\gg 1$, $\mf{N}\left(\mf{e}_k \right)$ \eqref{eq:nos} can be approximated with Stirling's formula. We have
\begin{equation}
\loge{\mf{N}\left(\mf{e}_k \right)}\simeq N\loge{N}-\sum_{j=1}^m n_j \loge{n_j}\,.
\end{equation}
As  pointed out earlier, thermodynamic equilibrium states are characterized by a maximum of symmetry or maximal number of ``involved energy states'', which corresponds classically to a maximal volume in phase space. In the case of the microcanonical equilibrium this condition was met by the state that is maximally envariant, namely envariant under all unitary maps. Now, following Boltzmann's line of thought we identify the canonical equilibrium by the configuration of the heat reservoir $\mf{B}$ for which the maximal number of energy eigenvalues are occupied. Under the constraints,
\begin{equation}
\sum_{j=1}^{m} n_j =N\quad\text{and}\quad E-\mf{e}_k=\sum_{j=1}^{m} n_j\,e_j^\mf{B}
\end{equation}
this problem can be solved by variational  calculus. One obtains
\begin{equation}
\label{eq:Boltzmann}
n_j=\mu\,\e{\beta\,e_j^\mf{B}}\,,
\end{equation}
which is the celebrated Boltzmann-Gibbs formula. Notice that Eq.~\eqref{eq:Boltzmann} is the number of states in the heat reservoir $\mf{B}$ with energy $e_j^\mf{B} $ for $\mf{S}$ and $\mf{B}$ being in thermodynamic, canonical equilibrium. In this treatment temperature merely enters through the Lagrangian multiplier $\beta$.

What remains to be shown is that $\beta$, indeed, characterizes the unique temperature of the system of interest, $\mf{S}$. To this end, imagine that the total system $\mc{S}$ can be separated into two small systems $\mf{S}_1$ and $\mf{S}_2$ of comparable size, and the thermal reservoir, $\mf{B}$. It is then easy to see that the total number of accessible states $\mf{N}\left(\mf{e}_k \right)$ does not significantly change in comparison to the previous case. In particular, in the limit of an infinitely large heat bath $\mf{B}$ the total number of accessible states for $\mf{B}$ is still given by Eq.~\eqref{eq:nos}. In addition, it can be shown that the resulting value of the Lagarange multiplier, $\beta$, is unique \cite{Wachsmuth2013a}. Hence, we can formulate a statement of the zeroth law of thermodynamics from envariance -- namely, two systems $\mf{S}_1$ and $\mf{S}_2$, that are in equilibrium with a large heat bath $\mf{B}$, are also in equilibrium with each other, and they have the same temperature corresponding to the unique value of $\beta$.

The present discussion is exact, up to the approximation with the Stirling's formula, and only relies  on the fact that the total system $\mc{S}$ is in a microcanonical equilibrium as defined in terms of envariance \eqref{eq:states}. The final derivation of the Boltzmann-Gibbs formula \eqref{eq:Boltzmann}, however, requires additional thermodynamic conditions. In the case of the microcanonical equilibrium we replaced conventional arguments by maximal envariance, whereas for the canonical state  we required the maximal number of energy levels of the heat reservoir to be ``occupied''. 

\section{\label{sec:quwork} Work, heat, and entropy production} 

Equipped with a classical understanding of thermodynamic phenomenology, the fluctuation theorems \eqref{eq:FT} and understanding of equilibrium states from a fully quantum theory, we can now move on to define work, heat, and entropy production for quantum systems. The following treatment was first published in Ref.~\cite{Gardas2015}.

\subsection{Quantum work and quantum heat}

\paragraph{Quasistatic processes.}

In complete analogy to the standard framework of thermodynamics as discussed in Sec.~\ref{sec:thermo}, we begin the discussion by considering quasistatic processes during which the quantum system, $\mc{S}$, is always in equilibrium with a thermal environment. However, we now further assume that the Hamiltonian of the system, $H(\lambda)$, is parameterized by a control parameter $\lambda$. The parameter can be, e.g., the volume of a piston, the angular frequency of an oscillator, the strength of a magnetic field, etc.

Generally, the dynamics of $\mc{S}$ is then described by the Liouville type equation $\dot{\rho}=L_{\lambda}(\rho)$, where the superoperator $L_{\lambda}$ reflects both the unitary dynamics generated by $H$ and the non-unitary contribution induced by the interaction with the  environment. We further have to assume that the equation for the steady state, $L_{\lambda}(\rho^\mrm{ss})=0$, has a unique solution~\cite{Spohn1978} to avoid any ambiguities.  As before, we will now be interested in thermodynamic state transformations, for which $\mc{S}$ remains in equilibrium corresponding to the value of $\lambda$. 

\paragraph{Thermodynamics of Gibbs equilibrium states.}

As we have seen above, in the the limit of ultraweak coupling the equilibrium state is given by the Gibbs state,
\begin{equation}
\label{eq:Gibbs_state}
\rho^\mrm{eq}=\e{-\beta H}/Z, \quad\text{where}\quad Z=\tr{\e{-\beta H}}\,,
\end{equation}
and where $\beta$ is the inverse temperature of the environment. In this case, the thermodynamic entropy is given by the Gibbs entropy \cite{Callen1985}, $ S=-\tr{\rho^\mrm{eq}\lo{\rho^\mrm{eq}}} = \beta \left(E - F \right)$, where as before $E=\tr{\rho^\mrm{eq}\, H}$ is the internal energy of the system, and $F=-1/\beta\,\lo{Z}$ denotes the Helmholtz free energy. 

For isothermal, quasistatic processes the change of thermodynamic entropy $d S$ can be written as
\begin{equation}
d S  = \beta\, \left( \tr{d\rho^\mrm{eq}\, H} + ( \tr{\rho^\mrm{eq}\, d H } - d F ) \right) = \beta\, \tr{d\rho^\mrm{eq}\, H}\,,
\end{equation}
where the second equality follows from simply evaluating terms. Therefore, two forms of energy can be identified \cite{Gemmer2009a}: heat is the change of internal energy associated with a change of entropy; work is the change of internal energy due to the change of an extensive parameter, i.e., change of the Hamiltonian of the system. We have,
\begin{equation}
\label{eq:quantum_energy}
d E = \dbar Q + \dbar W \equiv \tr{d\rho^\mrm{eq}\, H} + \tr{\rho^\mrm{eq}\, d H}\,. 
\end{equation} 
The identification of heat $\dbar Q$, and work $\dbar W$ \eqref{eq:quantum_energy} is consistent with the second law of thermodynamics for 
quasistatic processes \eqref{eq:ent_0} \emph{if}, and as will shortly see, \emph{only if} $\rho^\mrm{eq}$ is a Gibbs state \eqref{eq:Gibbs_state}.

It is worth emphasizing that for isothermal, quasistatic processes we further have,
\begin{equation}
d S=\beta\, \dbar Q \quad \text{and}\quad d F= \dbar W\,,
\end{equation}
for which the first law of thermodynamics takes the form
\begin{equation}
d E=T\,d S +d F\,.
\end{equation}
In this particular formulation it becomes apparent that changes of the internal energy $d E$ can be separated into ``useful" work $d F$ and an  additional contribution, $T\,d S$, reflecting the \emph{entropic cost} of the process.

\paragraph{Thermodynamics of non-Gibbsian equilibrium states.}

As we have seen above in Sec.~\ref{sec:envariance}, however, quantum systems in equilibrium are only described by Gibbs states \eqref{eq:Gibbs_state} if they are ultraweakly coupled to the environment. Typically, quantum systems are correlated with their surroundings and interaction energies are \emph{not} negligible \cite{Hu1992,Hanggi2008,Gelin2009}. For instance, it has been seen explicitly in the context of quantum Brownian motion \cite{Horhammer2008} that system and environment are generically entangled. 

In such situations  the identification of heat only with changes of the state of the system \eqref{eq:quantum_energy} is no longer correct \cite{Hanggi2008}. Rather, to formulate thermodynamics consistently the energetic back action due to the correlation of system and environment has to be taken into account \cite{Hanggi2008,Horhammer2008}. This means that during quasistatic processes parts of the energy exchanged with the environment are not related to a change of the thermodynamic entropy of the system, but rather constitute the energetic price to maintain the non-Gibbsian state, i.e., coherence and correlations between system and environment. 

Denoting the non-Gibbsian equilibrium state by $\rho^\mrm{ss}$ we can write
\begin{equation}
\begin{split}
 \mc{H} & = -\tr{\rho^\mrm{ss}\lo{\rho^\mrm{ss}}} + \left( \tr{\rho^\mrm{ss}\lo{\rho^\mrm{eq}}} - \tr{\rho^\mrm{ss}\lo{\rho^\mrm{eq}}} \right) \\
			     & = \beta \left[E -(F + T\, S(\rho^\mrm{ss} || \rho^\mrm{eq} )) \right]  = \beta\, ( \mc{E} -\mc{F} ),
	\end{split}
\end{equation}
where as before $E=\tr{\rho^\mrm{ss} H}$ is the internal energy of the system, and $\mc{F} \equiv F + T\, S(\rho^\mrm{ss} || \rho^\mrm{eq} )$ is the so-called the information
free energy \cite{Sagawa2015}. Further, $S(\rho^\mrm{ss}|| \rho^\mrm{eq} )\equiv \tr{\rho^\mrm{ss}\left(\lo{\rho^\mrm{ss}} - \lo{\rho^\mrm{eq}}\right)}$ is the quantum relative entropy \cite{Vedral2002}.

In complete analogy to the standard, Gibbsian case \eqref{eq:quantum_energy} we now consider isothermal, quasistatic processes, for which the infinitesimal change of the entropy reads
\begin{equation}
\begin{split}
		d\mc{H} & = \beta\, \left[ \tr{d\rho^\mrm{ss} H} + ( \tr{\rho^\mrm{ss} d H } - d\mc{F} ) \right] \\
				  & \equiv \beta\, (\dbar Q_\mrm{tot} - \dbar Q_{\rm c} ) \,.
\end{split}
\end{equation}
where we identified the total heat as $\dbar Q_\mrm{tot}\equiv \tr{d\rho^\mrm{ss} H} $ and energetic price to maintain coherence and quantum correlations as $\dbar Q_\mrm{c}\equiv d\mc{F}-\tr{\rho^\mrm{ss} d H}$. 

The \emph{excess heat} $\dbar Q_\mrm{ex}$ is the only contribution that is associated with the entropic cost,
\begin{equation}
	d \mc{H} = \beta\, \dbar Q_{\rm ex}, \quad\text{and}\quad \dbar Q_{\rm ex} = \dbar Q_\mrm{tot} - \dbar Q_{\rm c}\,.
\end{equation}
Accordingly, the first law of thermodynamics takes the form
\begin{equation}
d E=\dbar W_\mrm{ex}+\dbar Q_\mrm{ex}
\end{equation}
where $\dbar W_\mrm{ex}\equiv\dbar W+\dbar Q_\mrm{c}$ is the excess work. Finally, Eq.~\eqref{eq:quantum_energy} generalizes for isothermal, quasistatic processes in generic quantum systems to
\begin{equation}
\label{eq:quantum_energy_ness}
d E= T\,d\mc{H}+d\mc{F}\,.
\end{equation}
It is worth emphasizing at this point once again that thermodynamics is a phenomenological theory, and as one expects, the fundamental relations hold for any physical system. Equation~\eqref{eq:quantum_energy_ness} has exactly the same form as Eq.~\eqref{eq:quantum_energy}, however, the ``symbols'' have to be interpreted differently when translating between the thermodynamic relations and the underlying statistical mechanics.

As an immediate consequence of this analysis, we can now derive the efficiency of any quantum system undergoing a Carnot cycle.

\paragraph{Universal efficiency of quantum Carnot engines.}

To this end, imagine a generic quantum system that operates between two heat reservoirs with hot, $T_{\rm hot}$, and cold, $T_{\rm cold}$, temperatures, respectively. Then, the Carnot cycle consists of two isothermal processes during which the system absorbs/exhausts heat and two thermodynamically adiabatic, i.e., isentropic strokes while the extensive control parameter $\lambda$ is varied.

During the first isothermal stroke, the system is put into contact with the hot reservoir. As a result,  the excess heat $Q_{\rm{ex},1}$ is absorbed
at temperature $T_{\rm hot}$ and excess work $ W_{\rm{ex},1}$ is performed,
\begin{equation}
	\begin{split}
		W_{\rm{ex},1}  &= \mc{F}(\lambda_2,T_{\rm hot}) - \mc{F}(\lambda_1,T_{\rm hot})     \\
		Q_{\rm{ex},1} &= T_{\rm hot}\,(\mc{H}(\lambda_2,T_{\rm hot}) - \mc{H}(\lambda_1,T_{\rm hot})).
	\end{split}
\end{equation}
Next, during the isentropic stroke, the system performs work $W_{\rm{ex},2}$ and no excess heat is exchanged with the reservoir, $\Delta\mc{H}=0$. Therefore, the temperature of the engine drops from $T_{\rm hot}$ to $T_{\rm cold}$, 
\begin{equation}
	\begin{split}
		W_{ex,2}  & = \Delta E = E(\lambda_3,T_{\rm cold}) - E(\lambda_2,T_{\rm hot}) \\
  		 &=\Delta\mc{F} - \left(T_{\rm hot}-T_{\rm cold}\right)\,\mc{H}(\lambda_3,T_{\rm cold}).
	\end{split}
\end{equation}
In the second line, we employed the thermodynamic identity $E=\mc{F} + T\,\mc{H}$, which follows from the definition of $\mc{F}$. During the second isothermal stroke, the excess work 
$  W_{\rm{ex},3} $ is performed on the system by the cold reservoir. This allows for the system to exhaust the excess heat $ Q_{\rm{ex},3} $ at temperature $T_{\rm cold}$. Hence we have
\begin{equation}
	\begin{split}
		W_{ex,3} &= \mc{F}(\lambda_4,T_{\rm cold}) - \mc{F}(\lambda_3,T_{\rm cold}) 		\\
		Q_{ex,3}  &= T_{\rm cold}(\mc{H}(\lambda_c,T_{\rm cold}) - \mc{H}(\lambda_3,T_{\rm cold})).
	\end{split}
\end{equation}
Finally, during the second isentropic stroke, the cold reservoir performs the excess work $ W_{\rm{ex},4} $ on the system. No excess heat is exchanged and the temperature of the engine increases from $T_{\rm cold}$ to $T_{\rm hot}$,
\begin{equation}
	\begin{split}
		 W_{ex,4}  & = \Delta E = E(\lambda_1,T_{\rm hot}) - E(\lambda_4,T_{\rm cold}) \\
         & = \Delta\mc{F} + \left(T_{\rm hot}-T_{\rm cold}\right)\,\mc{H}(\lambda_1,T_{\rm hot}).
	\end{split}
\end{equation}

The efficiency of a thermodynamic device is defined as the ratio of  ``output" to ``input". In the present case the ``output'' is the total work performed
during each cycle, i.e., the total excess work, $W_\mrm{ex}= W +  Q_{\rm c}$. There are two physically distinct contributions:  work in the usual sense, $W$, that can be utilized, e.g., to power external devices, and $Q_{\rm c}$, which cannot serve such purposes as it is the thermodynamic cost to maintain the non-Gibbsian equilibrium state. Therefore, the only thermodynamically consistent definition of the efficiency is
\begin{equation}
\eta = \frac{\sum_{i}  W_{\rm ex, i} }{ Q_{{\rm ex},1} }  = 1 - \frac{T_{\rm cold}}{T_{\rm hot}}\equiv \eta_\mrm{C},
\end{equation}
which is identical to the classical Carnot efficiency. 

\subsection{Quantum entropy production}
\label{sec:quantum_entropy}
Having established a conceptual framework for quantum work and heat, the next natural step is to determine the \emph{quantum entropy production}. To this end, we now imagine that $\mc{S}$ is initially prepared in an equilibrium state, which however is not necessarily a Gibbs state \eqref{eq:Gibbs_state} with respect to the temperature of the environment. For a variation of the external control parameter $\lambda$ we can write the change of internal energy $\Delta E$ and a change of the von-Neumann entropy as
\begin{equation}
 \Delta E=  W +  Q \quad\mrm{and}\quad\Delta \mc{H}= \beta  Q +  \Sigma \ ,
\end{equation}
where here $Q$ is the total heat exchanged during the process with the environment at inverse temperature $\beta$. Thus, we can write for the mean nonequilibrium entropy production,
\begin{equation}
\label{eq:qu_entropy}
 \Sigma= \Delta \mc{H} -\beta \Delta E +\beta W  \ .
\end{equation}
Now, expressing the internal energy with the help of the Gibbs state $\rho^\mrm{eq}$ \eqref{eq:Gibbs_state} we have
\begin{equation}
\beta E= \beta\,\tr{\rho^\mrm{ss} H}=-\tr{\rho^\mrm{ss}\, \lo{\rho^\mrm{eq}}}+\lo{Z}\ ,
\end{equation}
Thus, we can write for a process that varies $\lambda$ from $\lambda_0$ to $\lambda_1$ [compare with the classical expression \eqref{eq:work_stoch} and the quantum case \eqref{eq:quantum_energy}]
\begin{equation}
\label{eq:qu_work}
\begin{split}
 \beta W &=\beta \int_{\lambda_0}^{\lambda_1} d\lambda\,\tr{\rho^\mrm{ss}(\lambda)\,\pd_\lambda H(\lambda)} \\
 &=-\int_{\lambda_0}^{\lambda_1} d\lambda\, \tr{\rho^\mrm{ss}(\lambda) \,\pd_\lambda \lo{\rho^\mrm{eq}(\lambda)}}-\lo{Z_1}+\lo{Z_0}\ .
 \end{split}
\end{equation}
Combining Eqs.~\eqref{eq:qu_entropy}-\eqref{eq:qu_work}, we obtain the general expression for the entropy production along a nonequilibrium path [compare Fig.~\ref{fig:path}], 
\begin{equation}
\label{eq:quantum_entropy}
\Sigma=S\left(\rho^\mrm{ss}(\lambda_0)||\rho^\mrm{eq}(\lambda_0)\right)-S\left(\rho^\mrm{ss}(\lambda_1)||\rho^\mrm{eq}(\lambda_1)\right)-\int_{\lambda_0}^{\lambda_1} d\lambda \, \tr{\rho^\mrm{ss}(\lambda)\, \pd_\lambda \lo{\rho^\mrm{eq}(\lambda)}}
\end{equation}
Equation \eqref{eq:quantum_entropy} is the exact microscopic expression for the mean  nonequilibrium entropy production for a driven  open quantum system weakly coupled to a single heat reservoir. It is valid for intermediate states that can be arbitrarily far from equilibrium.

\subsection{Two-time energy measurement approach \label{sec:quantum_Jarzynski}}

Having identified expressions for the average, work, heat, and entropy production, we can now continue building \emph{\textbf{Quantum Stochastic Thermodynamics}}. In complete analogy to the classical case, Quantum Stochastic Thermodynamics is built upon fluctuation theorems. Conceptually, the most involved problem is how to identify heat and work for single realizations -- and even what a ``single realization'' means for a quantum system. 

The most successful approach has become known as two-time energy measurement approach \cite{Campisi2011}. In this paradigm, one considers an isolated quantum system that evolves under the time-dependent Schr\"odinger equation
\begin{equation}
i\hbar\,\pd_t \ket{\psi_t}=H_t\,\ket{\psi_t}\,.
\end{equation}
As before, we are interested in describing thermodynamic processes that are induced by varying an external control parameter $\lambda_t$ during time $\tau$, so that $H_t=H(\lambda_t)$. Within the two-time energy measurement approach quantum work is determined by the following protocol: At initial time $t=0$ a projective energy measurement is performed on the system; then the system is let to evolve under the time-dependent Schr\"odinger equation, before a second projective energy measurement is performed at $t=\tau$. 

Therefore, work $W$ becomes a stochastic variable, and for a single realization of this protocol we have
\begin{equation}
W\left[\ket{m}; \ket{n}\right]=E_m(\lambda_{0})-E_n(\lambda_{\tau})\,,
\end{equation}
where $\ket{m}$ is the initial eigenstate with eigenenergy $E_{m}(\lambda_0)$ and $\ket{n} $ with $E_{n}(\lambda_{\tau}) $ denotes the final state. 

The distribution of work values is then given by averaging over an ensemble of realizations of the same process,
\begin{equation}
\mc{P}(W)=\la \de{W-W\left[\ket{m}; \ket{n}\right]} \ra \,,
\end{equation}
which can be rewritten as
\begin{equation}
\label{eq:P_quwork}
\mathcal{P}(W)=\sumint_{m,n} \de{W-W\left[\ket{m};\ket{n}\right]}\,p\left(\ket{m}\rightarrow\ket{n}\right).
\end{equation}
In the latter equation the symbol $\sumint$ denotes that we have to sum over the discrete part of the eigenvalue spectrum and integrate over the continuous part. Therefore, for systems with spectra that have both contributions the work distribution will have a continuous part and delta-peaks, see for instance for the Morse oscillator in Ref.~\cite{Leonard2015}.

Further, $p\left(\ket{m}\rightarrow\ket{n}\right)$ denotes the probability to observe a specific transition $\ket{m}\rightarrow\ket{n} $. This probability is given by \cite{Leonard2015},
\begin{equation}
\label{eq:P_transition}
p\left(\ket{m}\rightarrow\ket{n}\right)=\tr{\Pi_{n}\, U_{\tau}\, \Pi_{m}\,\rho^\mrm{eq}\,\Pi_{m}\, U_{\tau}^\dagger}\,,
\end{equation}
where $\rho^\mrm{eq}$ is the initial, Gibbsian density operator \eqref{eq:Gibbs_state} of the system\footnote{Generally, the initial state can be chosen according to the physical situation. However, in Quantum Stochastic Thermodynamics it is often convenient to assume an initially thermal state.}, and $U_{\tau}$ is the unitary time evolution operator, $U_{\tau}=\mc{T}_> \e{-i/\hbar\,\int_0^{\tau}d t\,H_t}$. Finally, $\Pi_{\nu}$ denotes the projector into the space spanned by the $\nu$th eigenstate. For Hamiltonians with non-degenerate spectra we simply have $\Pi_\nu=\ket{\nu}\bra{\nu}$.

\paragraph{The quantum Jarzynski equality.}

It is then a relatively simple exercise to show that such a definition of quantum work fulfills a quantum version of the Jarzynski equality. To this end, we compute the average of the exponentiated work
\begin{equation}
\la \e{-\beta W}\ra=\int dW\,\mc{P}(W)\, \e{-\beta W}= \sumint_{m,n}  \e{\beta\,W\left[\ket{m};\ket{n}\right]}\,p\left(\ket{m}\rightarrow\ket{n}\right)
\end{equation}
Using the explicit expression for the transition probabilities \eqref{eq:P_transition} and for the Gibbs state \eqref{eq:Gibbs_state}, we immediately have
\begin{equation}
\label{eq:qu_Jarzynski}
\la \e{-\beta W}\ra=\e{-\beta \Delta F}\,.
\end{equation}
The latter theorem looks analogous to the classical Jarzynski equality \eqref{eq:Jarzynski_result}. However, quantum work is a markedly different quantity than work in classical mechanics. It has been pointed out that work as defined from the two-time measurement is not a quantum observable in the usual sense, namely that there is no Hermitian operator whose eigenvalues are the classical work values \cite{Talkner2007,Talkner2016}. The simple reason is that the final Hamiltonian does not necessarily commute with the initial Hamiltonian, $\com{H_\tau}{H_0}\neq 0$. Rather, quantum work is given by a time-ordered correlation function, which reflects that thermodynamically work is a non-exact, i.e., path dependent quantity.

\paragraph{Neglected informational cost.}

Another issue arises from the fact that generally the final state $\rho_\tau$ is a complicated nonequilibrium state. This means, in particular, that also $\rho_\tau=\rho(\lambda_\tau)$ does not commute with the final Hamiltonian $H_\tau$, and one has to consider the back-action on the system due to the projective measurement of the energy \cite{Nielsen2010}. For a single measurement, $\Pi_{n}$, the post-measurement state is given by $\Pi_{n} \rho_\tau\Pi_{n}/p_n$, where $p_n=\tr{\Pi_{n}\,\rho_\tau}$. Thus, the system can be found on average in
\begin{equation}
\rho^M_\tau=\sum_{n}\Pi_{n}\,\rho_\tau\,\Pi_{n}\,.
\end{equation}
Accordingly, the final measurement of the energy is accompanied by a change of information, i.e., by a change  of the von Neumann entropy of the system
\begin{equation}
\Delta \mc{H}^M=-\tr{\rho^M_\tau\lo{\rho^M_\tau}}+\tr{\rho_\tau\lo{\rho_\tau}}\geq 0\,.
\end{equation}
Information, however, is physical \cite{Landauer1991} and its acquisition ``costs'' work.  This additional work has to be paid by the external observer -- the measurement  device. In a fully consistent thermodynamic framework this cost should be taken into consideration.

\paragraph{Quantum work without measurements.}

To remedy this conceptual inconsistency arising from neglecting the informational contribution of the projective measurements, an alternative paradigm has been proposed \cite{Deffner2016_work}. For isolated systems quantum work is clearly given by the change of internal energy. As a statement of the first law of thermodynamics this holds true no matter whether the system is measured or not.

Actually, for thermal, Gibbs states \eqref{eq:Gibbs_state} measuring the energy is superfluous as state and energy commute. Hence, an alternative notion of quantum work can be formulated that is fully based on the time-evolution of energy eigenstates. Quantum work for a single realization is then determined by considering 
how much the expectation value for a single energy eigenstate changes 
under the unitary evolution. Hence, we define
\begin{equation}
\label{eq:work_one}
\widetilde{W}_{m}\equiv \bra{m} U^\dagger_\tau\,H_\tau\,U_\tau\ket{m}-E_m(\lambda_0)\,.
\end{equation}
We can easily verify that the so defined quantum work \eqref{eq:work_one}, indeed, fulfills the first law. To this end, we compute the average work $\la W\ra_{\widetilde{\mc{P}}}$,
\begin{equation}
\begin{split}
\la W\ra_{\widetilde{\mc{P}}}&=\sumint_{m} \bra{m} U^\dagger_\tau\,H_\tau\,U_\tau\ket{m}\,p_m-\tr{\rho^\mrm{eq}_0\, H_0}\\
&=\tr{\rho^\mrm{ss}_\tau\,H_\tau}-\tr{\rho^\mrm{eq}_0\,H_0}=\la W\ra \,,
\end{split}
\end{equation}
where $p_m=\e{-\beta\, E_m(\lambda_0)}/Z_0$ is the probability to find the system in the $m$th eigenstate at time $t=0$. It is important to note that the average quantum work determined from two-time energy measurements is identical to the (expected) value given only knowledge from a single measurement at $t=0$.  Most importantly, however, in this paradigm the external observer does not have to pay a thermodynamic cost associated with the change of information due to measurements.

\paragraph{Modified quantum Jarzynski equality.}

We have now seen that the first law of thermodynamics is immune to whether the energy of the system is measured or not, since projective measurements of the energy do not affect the internal energy. However, the informational content of the system of interest, i.e., the entropy, crucially depends on whether the system is measured. Therefore, we expect that the statements of the second law have to be modified to reflect the informational contribution \cite{Deffner2013PRX}. In this paradigm the modified quantum work distribution becomes
\begin{equation}
\widetilde{\mc{P}}(W)=\sumint_{m}~\delta(W-\widetilde{W}_{m})\, p_m\,,
\end{equation}
where as before $p_m=\e{-\beta\, E_m(\lambda_0)}/Z_0$. Now, we can compute the average exponentiated work,
\begin{equation}
\label{eq:Jarzynski_modified}
\la \e{-\beta W}\ra_{\widetilde{\mc{P}}}=\frac{1}{Z_0}\,\sumint_{m}\,\e{-\beta \,\bra{m} U^\dagger_\tau\,H_\tau\,U_\tau\ket{m}}\,.
\end{equation}
The right side of Eq.~\eqref{eq:Jarzynski_modified} can be interpreted as the ratio  of two partition functions, where $Z_0$ describes the initial thermal 
state. The second partition function,
\begin{equation}
\widetilde{Z}_\tau\equiv \sumint_{m}\,\e{-\beta \,\bra{m} U^\dagger_\tau\,H_\tau\,,
U_\tau\ket{m}}
\end{equation}
corresponds to the best possible guess for a thermal  state of the final system given only the time-evolved energy eigenbasis. 
This state can be written as
\begin{equation}
\widetilde{\rho}_\tau\equiv \frac{1}{\widetilde{Z}_\tau}\,\sumint_{m}\e{-\beta \,\bra{m} U^\dagger_\tau\,H_\tau\,U_\tau\ket{m}}\,U_\tau\ket{m}\bra{m}U^\dagger_\tau,
\end{equation}
which differs from the true thermal state, $\rho_\tau^\mrm{eq}=\e{-\beta H_\tau}/Z_\tau$.

As noted above, in information theory the ``quality'' of such a best possible guess is quantified  by the relative entropy \cite{Vedral2002}, which measures the 
distinguishability of two (quantum) states. Hence, let us consider
\begin{equation}
S(\widetilde{\rho}_\tau||\rho_\tau^\mrm{eq})=\tr{\widetilde{\rho}_\tau\lo{\widetilde{\rho}_\tau}}-\tr{\widetilde{\rho}_\tau\lo{\rho_\tau^\mrm{eq}}}\,,
\end{equation}
for which we compute both terms separately. For the first term, the negentropy of $\widetilde{\rho}_\tau$ we obtain,
\begin{equation}
\begin{split}
\tr{\widetilde{\rho}_\tau\lo{\widetilde{\rho}_\tau}}&=-\ln{(\widetilde{Z})}-\beta\, \tr{\widetilde{\rho}_\tau \sumint_{m} \bra{m} U^\dagger_\tau\,H_\tau\,U_\tau\ket{m}\,U_\tau\ket{m}\bra{m}U^\dagger_\tau }\\
&=-\ln{(\widetilde{Z})}-\beta \widetilde{E}\,,
\end{split}
\end{equation}
where we introduced the expected value of the energy, $\widetilde{E}$, under the time-evolved eigenstates,
\begin{equation}
\widetilde{E}=\frac{1}{\widetilde{Z}} \sumint_{m}\e{-\beta \,\bra{m} U^\dagger_\tau\,H_\tau\,U_\tau\ket{m}}\bra{m} U^\dagger_\tau\,H_\tau\,U_\tau\ket{m}\,.
\end{equation}
The second term, the cross entropy of $\widetilde{\rho}_\tau $ and $\rho_\tau^\mrm{eq} $, simplifies to
\begin{equation}
\begin{split}
\tr{\widetilde{\rho}_\tau\lo{\rho_\tau^\mrm{eq}}}&=-\lo{Z_\tau}-\beta \,\tr{\sumint_{m}\frac{1}{\widetilde Z} \e{-\beta \,\bra{m} U^\dagger_\tau\,H_\tau\,U_\tau\ket{m} U_\tau\ket{m}\bra{m}U^\dagger_\tau\,H_\tau}}\\
&=-\lo{Z_\tau}-\beta \widetilde{E}\,.
\end{split}
\end{equation}
Hence, the modified quantum Jarzynski equality \eqref{eq:Jarzynski_modified} becomes
\begin{equation}
\la \e{-\beta W}\ra_{\widetilde{\mc{P}}}=\e{-\beta \Delta F}\,\e{-S(\widetilde{\rho}_\tau||\rho_\tau^\mrm{eq})}\,,
\end{equation}
where as before $\Delta F=-1/\beta\, \lo{Z_\tau/Z_0}$. Jensen's inequality further implies,
\begin{equation}
\label{eq:max_work}
\beta\,\la W\ra \geq \beta\,\Delta F + S(\widetilde{\rho}_\tau||\rho_\tau^\mrm{eq})
\end{equation}
where we used $\la W\ra_{\widetilde{\mc{P}}}=\la W\ra$. 

By defining quantum work as an average over time-evolved eigenstates we obtain a modified quantum Jarzynski equality \eqref{eq:Jarzynski_modified} and a generalized maximum work theorem \eqref{eq:max_work}, in which the thermodynamic cost of projective measurements becomes apparent. These results become even more transparent by noting that similar versions of the maximum work theorem have been derived in the thermodynamics of information \cite{Sagawa2015}. As mentioned above, it has proven useful to introduce the notion of an information free energy, 
\begin{equation}
\label{eq:infoF_var}
\widetilde{F}_\tau=F_\tau+S(\widetilde{\rho}_\tau||\rho_\tau^\mrm{eq})/\beta\,.
\end{equation}
Here, $\widetilde{F}_\tau$ accounts for the additional capacity of a thermodynamic system to perform work due to information \cite{Deffner2013PRX}. Note that in Eq.~\eqref{eq:infoF_var} $\widetilde{F}_\tau$ is computed for the fictitious equilibrium state $\widetilde{\rho}_\tau$.

We can rewrite Eq.~\eqref{eq:max_work} as
\begin{equation}
\beta\,\la W\ra \geq \beta\,\Delta \widetilde{F}\,.
\end{equation}
The latter inequality constitutes a sharper bound than the usual maximum work theorem, and it accounts for the extra free energy available to the system. Free energy, however, describes the usable, extractable work. In real-life applications one is more interested in the maximal free energy the system has available, than in the work that could be extracted by intermediate, disruptive measurements of the energy. Therefore, this treatment could be considered thermodynamically more relevant than the two-time measurement approach. 

\subsection{Quantum fluctuation theorem for arbitrary observables \label{sec:gft}}

Another issue with the two-time energy measurement approach is that in many experimental situations projective measurements of the energy are neither feasible nor practical.  Rather, only other observables such as the spatial density or the magnetization are accessible. Then, the natural question is whether there is a fluctuation theorem for the observable that can actually be measured. 

To answer this question, let us consider a more general paradigm, which was first published in Ref.~\cite{Kafri2012}: Information about a quantum system and its dynamics is obtained by performing measurements on $\mc{S}$ at the beginning and end of a specific process. Initially a quantum measurement is made of observable $\Omega^i$, with eigenvalues $\omega_m^i$. As before, $\Pi_{m}^i$ denote the orthogonal projectors into the eigenspaces of $\Omega^i$, and we have $\Omega^i=\sum_m \omega_m^i\Pi_{m}^i$. Note that the eigenvalues $\{\omega_m^i\}$ can be degenerate, so the projectors $\Pi_{m}^i$ may have rank greater than one. Unlike the classical case, as long as $\rho_0$ and $\Omega^i$ do not have a common set of eigenvectors - i.e. they do not commute - performing a measurement on $\mc{S}$ alters its statistics. Measuring $\omega_m^i$ maps $\rho_0$ to the state $\Pi_{m}^i \rho_0\Pi_{m}^i/p_m$, where $p_m = \tr{ \Pi_{m}^i \rho_0\Pi_{m}^i}$ is the probability of the measurement outcome $\omega_m^i$. Generally accounting for all possible measurement outcomes, the statistics of $\mc{S}$ after the measurement are given by the weighted average of all projections, 
\begin{equation}
M^i(\rho_0) = \sum_m \Pi_{m}^i\, \rho_0\, \Pi_{m}^i\,.
\end{equation}
 If $\rho_0$ commutes with $\Omega^i$, it commutes with each $\Pi_{m}^i$, so $M^i(\rho_0) =\sum_m \Pi_{m}^i \Pi_{m}^i \rho_0  =\rho_0$ and the statistics of the system are unaltered by the measurement. After measuring $\omega_m^i$, $\mc{S}$ undergoes a generic time evolution, after which it is given by $\mbb{E}( \Pi_{m}^i \rho_0 \Pi_{m}^i)/p_m$. Here $\mbb{E}$ represents any linear (unitary or non-unitary) quantum transformation, which is trace-preserving and maps non-negative operators  to non-negative operators. Moreover, we require that this holds whenever $\mbb{E}$ is extended to an operation $\mbb{E} \otimes \id_\mc{E}$ on any  enlarged Hilbert space $\mc{H}_S\otimes\mc{H}_\mc{E}$ ($\id_\mc{E}$ being the identity map on $\mc{H}_\mc{E}$). Such a transformation is called a completely positive, trace preserving (CPTP) map \cite{Nielsen2010}. 
 
After this evolution, a measurement of a second (not necessarily the same) observable, $\Omega^f = \sum_n \omega_n^f \Pi_{n}^f$, is performed on $\mc{S}$. The probability of measuring $\omega_n^f$, conditioned on having first measured $\omega_m^i$, is $p_{n|m} =\tr{\Pi_{n}^f\, \mbb{E} \left( \Pi_{m}^i \rho_0\Pi_{m}^i\right)}/p_m$. Accordingly, the joint probability distribution $p_{m\rightarrow n}$ reads
\begin{equation}
p(\ket{m}\rightarrow \ket{n})= p_m \cdot p_{n|m} = \tr{\Pi_{n}^f\, \mbb{E} \left( \Pi_{m}^i \rho_0\Pi_{m}^i\right)}\,.
\end{equation}
We are interested in the probability distribution of possible measurement outcomes, $\mathcal{P}\left( \Delta \omega\right)$, where $\Delta \omega_{n,m} = \omega_n^f - \omega_m^i$ is a random variable determined in a single measurement run. Its probability distribution is given by averaging over all possible realizations,
\begin{equation}
\mathcal{P}\left( \Delta \omega\right)=\la \de{\Delta \omega-\Delta \omega_{n,m}}\ra=\sum\limits_{m,n} \de{\Delta \omega-\Delta \omega_{n,m}}\, p(\ket{m}\rightarrow \ket{n})\,.
\end{equation}
To derive the integral fluctuation theorem we follow another standard approach and compute its characteristic function, $\mc{G}(s)$, which is the Fourier transform of $\mc{P}(\Delta \omega)$ \cite{Campisi2011}
\begin{equation}
\begin{split}
\mc{G}(s)& = \int d(\Delta \omega)\,\mc{P}(\Delta \omega)\,\e{i s\,\Delta \omega} \\
& = \tr{ \e{i s \Omega^f}\,  \mbb{E} \left( M^i(\rho_0) \e{-i s \Omega^i}\right)  }\,.
\end{split}
\end{equation}
Choosing $s = i$, we obtain the identity 
\begin{equation}
\label{eq:GFT}
\la \e{- \Delta \omega} \ra = \varepsilon\,.
\end{equation}
Since it is explicitly dependent on the map $\mbb{E}$, the quantity $\varepsilon$ accounts for the  information \textit{lost} by not measuring the environment,
\begin{equation}
\label{eq:qu_efficacy}
\varepsilon = \tr{\e{-\Omega^f}\, \mbb{E} \left(M^i(\rho_0) \e{ \Omega^i} \right) }\,.
\end{equation}
which has been called the \emph{quantum efficacy}.

We emphasize that Eq.~\eqref{eq:GFT} is \emph{not} an integral fluctuation theorem in the strict sense. Generally, the quantum efficacy \eqref{eq:qu_efficacy} explicitly depends on the choice of the observables, $\Omega^i$ and $\Omega^f$, the initial state $\rho_0$, and the CPTP map $ \mbb{E}$. In a fluctuation theorem the right hand side, i.e, $\varepsilon$ should be a c-number independent of the details of the measurement protocol.

However, it is also easy too see when  Eq.~\eqref{eq:GFT} becomes a fluctuation theorem. This is the case, if the initial state $\rho_0$ is proportional to $\e{-\Omega^i}$, and if the CPTP map is unital, which means $\mbb{E}(\id)=\id$. These conditions are naturally fulfilled for initial Gibbs states \eqref{eq:Gibbs_state}, energy measurements, and unitary Schr\"odinger dynamics. However, we also immediately observe that the quantum Jarzynski equality \eqref{eq:qu_Jarzynski} remains valid in purely decohering or purely dephasing models \cite{Albash2013,Rastegin2013,Rastegin2014,Gardas2016,Smith2018}.

\subsection{Quantum entropy production in phase space \label{sec:entropy}}

We conclude this section with an alternative approach to stochastic thermodynamics of quantum systems, which was first published in Ref.~\cite{Deffner2013EPL}. We have seen above that for classical systems the irreversible entropy production is defined along a path in phase space \eqref{eq:work_hamiltonian}. If we want to define an analogous entropy production for a quantum process, we have to choose a representation of quantum phase space.

A particularly convenient representation of quantum states is given by the Wigner function \cite{Wigner1932},
\begin{equation}
\mc{W}_t(x,p)=\frac{1}{2\pi\hbar}\int d y\,\e{-\frac{i}{\hbar}\,p y}\,\bra{x+\frac{y}{2}}\rho_t\ket{x-\frac{y}{2}}\,.
\end{equation}
The Wigner function contains the full classical information, and its marginals are the probability distributions for the position $x$ and the momentum $p$, respectively. In addition,  $\mc{W}_t(x,p)$ contains the full quantum information about a state, as, e.g., areas in phase space where $\mc{W}_t(x,p)$ takes negative values are indicative for quantum interference.

In complete analogy to the classical case, the quantum Liouville equation can  be written as
\begin{equation}
\label{eq:wigner_dyn}
\pd_t\,\mc{W}(\Gamma,t)=\mc{L}_{\lambda}\, \mc{W}(\Gamma,t)\,,
\end{equation}
where $\Gamma=(x,p)$ denotes again a point in phase space. It is worth emphasizing that a Liouvillian, $\mc{L}_\lambda$, does not generally exist for all quantum systems. In particular, for a thermally open harmonic oscillator it was shown in \cite{Karrlein1997} that the existence and explicit form of $\mc{L}_\lambda$ are determined by the initial preparation of the environment.

The stationary solution of Eq.~\eqref{eq:wigner_dyn} is determined by
\begin{equation}
\mc{L}_{\lambda}\,\mc{W}_\mrm{stat}(\Gamma,\lambda)=0\,.
\end{equation}
Generally the stationary Wigner function $\mc{W}_\mrm{stat}(\Gamma,\lambda)$ for an open quantum system in equilibrium is not given by the Wigner representation of the Gibbs state \eqref{eq:Gibbs_state}. For instance, the exact master equation for a harmonic oscillator coupled to an environment consisting of an ensemble of harmonic oscillators is known \cite{Hu1992} and can be solved analytically \cite{Ford2001}. In a high temperature approximation the quantum Liouville equation \eqref{eq:wigner_dyn} becomes, in leading order of $\hbar$, \cite{Dillenschneider2009}
\begin{equation}
\label{eq:qu_KK}
\mc{L}_t=-\frac{p}{m}\, \pd_x+V'(x,t)\,\pd_p\,+\pd_p\left(\gamma p+ D_{pp}\,\pd_p\right)+D_{xp}\,\pd_{xp}^2
\end{equation}
where $\gamma$ is again the coupling coefficient to the environment,  $D_{pp}=m\gamma/\beta +m \beta\gamma\hbar^2(\omega^2-\gamma^2)/12$, and $D_{xp}=\beta\gamma\hbar^2/12$. Note that in the high-temperature limit, $\beta\hbar\omega\ll 1$, Eq.~\eqref{eq:qu_KK} reduces to the classical Klein-Kramers equation \eqref{eq:KK}. The stationary solution can be written as
\begin{equation}
\begin{split}
\mc{W}_\mrm{stat}(x,p)&=\frac{m \gamma\omega}{2\pi}\frac{1}{\sqrt{D_{pp}\left(D_{pp}+m\gamma\,D_{xp}\right)}}\,\e{-\frac{\gamma}{2}\left(\frac{p^2}{D_{pp}}+\frac{m^2\omega^2\,x^2}{D_{pp}+m\gamma\,D_{xp}}\right)}\,.
\end{split}
\end{equation}

We will now prove that the quantum entropy production $\Sigma$ for \emph{any} quantum dynamics described by Eq.~\eqref{eq:wigner_dyn} and defined as
\begin{equation}
\label{eq:wigner_ent}
\Sigma[\Gamma_\tau;\,\lambda_\tau]\equiv-\int_0^\tau d t\,\dot\lambda_t\,\frac{\pd_\lambda\mc{W}_\mrm{stat}(\Gamma_t,\lambda_t)}{\mc{W}_\mrm{stat}(\Gamma_t,\lambda_t)}\,,
\end{equation}
fulfills an integral fluctuation theorem. Note that writing $\Sigma$ as a functional of a trajectory in (quantum) phase space is a mathematical construct, which is convenient for the following proof. More formally, we understand the entropy produced along a quantum trajectory in analogy to Feynman path integrals. Here a quantum trajectory is a mathematical tool defined as a generalization of the classical trajectory. Physical quantities are given by averages over an ensemble of such trajectories.

Consider the accumulated entropy $\sigma$ produced up to time $t$, $\sigma(t)=-\int_0^t d s\,\dot\lambda_s\,\pd_\lambda\mc{W}_\mrm{stat}/\mc{W}_\mrm{stat}$, and thus $\sigma(\tau)=\Sigma$. Then the joint (quasi) probability distribution for the point in phase space and the accumulated entropy production, $P(\Gamma,\sigma,t)$, evolves according to,
\begin{equation}
\label{eq:joint}
\pd_t\,P(\Gamma,\sigma,t)=\left[\mc{L}_\lambda-j_\mrm{stat}(\Gamma,\lambda_t)\,\pd_\sigma\right]\,P(\Gamma,\sigma,t)\,,
\end{equation}
where $j_\mrm{stat}(\Gamma,\lambda_t)$ is the (quasi) probability flux associated with the accumulated entropy production $\sigma$,
\begin{equation}
j_\mrm{stat}(\Gamma,\lambda_t)=\dot\lambda_t\,\frac{\pd_\lambda\mc{W}_\mrm{stat}(\Gamma,\lambda_t)}{\mc{W}_\mrm{stat}(\Gamma,\lambda_t)}\,.
\end{equation}
Now we define the auxiliary density $\Psi(\Gamma,t)$ which is the exponentially weighted marginal of $P(\Gamma,\sigma,t)$. We have
\begin{equation}
\Psi(\Gamma,t)=\int d\sigma\,P(\Gamma,\sigma,t)\,\e{-\sigma}\,,
\end{equation}
for which the evolution equation \eqref{eq:joint} becomes
\begin{equation}
\label{eq:joint_simple}
\pd_t\,\Psi(\Gamma,t)=\left[\mc{L}_\lambda-j_\mrm{stat}(\Gamma,\lambda_t)\right]\,\Psi(\Gamma,t)\,.
\end{equation}
It is easy to see that a solution of Eq.~\eqref{eq:joint_simple} is given by the stationary solution of the original master equation \eqref{eq:wigner_dyn} and we obtain
\begin{equation}
\Psi(\Gamma,t)=\mc{W}_\mrm{stat}(\Gamma,\lambda_t)\,.
\end{equation}
Using the normalization of the stationary Wigner function  we calculate with the latter solution for $\Psi(\Gamma,t)$,
\begin{equation}
\label{eq:qu_FT}
1=\int d \Gamma\,\mc{W}_\mrm{stat}(\Gamma,\lambda_\tau)=\int d \Gamma\,\Psi(\Gamma,\tau)=\la \e{-\Sigma}\ra\,,
\end{equation}
which concludes the proof. For any quantum system, open or closed, the entropy production fulfilling an integral fluctuation theorem is given by Eq.~\eqref{eq:wigner_ent}. 

\clearpage

\section{Checklist for ``The principles of modern thermodynamics"}

\begin{enumerate}
\item Thermodynamics is a phenomenological theory to describe the average behavior of heat and work.
\item Reversible processes can be understood as paths on the thermodynamic manifold described by the equation of state.
\item Systems can be locally in equilibrium and at the same time not in equilibrium with the rest of the Universe.
\item Heat, work, and entropy can be defined along single trajectories of classical systems.
\item Fluctuation theorems are symmetry relations for the distribution of work values expressing that ``violations'' of the second law are exponentially unlikely.
\item Statistical mechanics can be built from a purely quantum framework using symmetries of entanglement.
\item Quantum work is not an observable in the usual sense.
\item There are many different and equally justifiable notions of quantum work and entropy production.
\end{enumerate}

\section{Problems}

\subsection*{A phenomenological theory of heat and work \ref{sec:thermo}}

\begin{itemize}
\item[\textbf{[1]}] Consider a single quantum particle in an infinite square well, and whose density operator is a Gibbs state. Compute the equation of state for the length of the box $L$, the temperature $T$, and the mass $m$, and plot the thermodynamic manifold. How does the manifold change if an additional (identical) particle is added? Does it matter whether the particles are fermions of bosons?
\item[\textbf{[2]}] Quantum heat engines are thermodynamic devices with small quantum systems as working medium. A stereotypical example is a single quantum particle trapped in a harmonic potential. The natural external control parameter is the angular frequency. Determine the equation of state assuming that the quantum particle is ultra-weakly coupled to a thermal environment, which means the density operator is a Gibbs state. Compute the efficiency of such a device as it undergoes an Otto cycle.
\end{itemize}

\subsection*{The advent of stochastic thermodynamics \ref{sec:fluc}}

\begin{itemize}
\item[\textbf{[3]}] Consider a 1-dimensional, classical harmonic oscillator, whose dynamics is described by the classical Liouville equation \eqref{eq:Liouville}. Compute the probability density function of the work done during a variation of the angular frequency, if the oscillator was initially prepared in a Maxwell-Boltzmann distribution. Verify the Jarzynski equality \eqref{eq:Jarzynski_result}.
\item[\textbf{[4]}] Consider a 1-dimensional classical harmonic oscillator in contact with a thermal bath, whose dynamics is described by the classical Klein-Kramers equation \eqref{eq:KK}. Compute the probability density function  of the work while dragging the oscillator along the $x$-axis,  if the oscillator was initially prepared in a Maxwell-Boltzmann distribution. Verify the Crooks fluctuation theorem \eqref{eq:Crooks_theorem}.
\end{itemize}

\subsection*{Foundations of statistical physics from quantum entanglement \ref{sec:envariance}}

\begin{itemize}
\item[\textbf{[5]}] Illustrate the concept of envariance for a Universe consisting of two harmonic oscillators and parity preserving unitary maps.
\item[\textbf{[6]}] Repeat the arguments leading to Eq.~\eqref{eq:Boltzmann}, but by including the next two terms of the Stirling approximation $$\loge{n!}\simeq n \loge{n}-n+\frac{1}{2}\loge{2 \pi n} +\frac{1}{12\,n}\,.$$
How would one identify the temperature in this case?
\end{itemize}

\subsection*{Work, quantum heat, and quantum entropy production \ref{sec:quwork}}

\begin{itemize}
\item[\textbf{[7]}] Consider a thermally isolated, quantum harmonic oscillator in one dimension. Compute the probability density function for the work done during an infinitely slow variation of the angular frequency, if the oscillator was initially prepared in a Gibbs state.
\item[\textbf{[8]}]  Consider a 1-dimensional quantum harmonic oscillator in contact with a thermal bath, whose dynamics is described by the quantum Klein-Kramers equation \eqref{eq:qu_KK}. Compute the probability density function  of the entropy production while dragging the oscillator along the $x$-axis. Verify the quantum fluctuation theorem \eqref{eq:qu_FT}.
\end{itemize}

\addcontentsline{toc}{section}{References}
\bibliographystyle{plain}
\bibliography{book}

\clearpage


\chapter{\label{chap:devices}Thermodynamics of Quantum Systems}

We have seen from the previous chapter that when dealing with quantum systems their thermodynamic description requires careful consideration. Beyond the obvious fundamental interest in developing the theory to faithfully describe a quantum system's thermodynamic properties, a natural question arises: Do we gain any advantage using quantum systems over their classical counterparts? This question cuts right to the heart of the practical applicability of quantum devices. Indeed, already it is well accepted that exploiting quantum features, such as entanglement, allows for superior information processing and cryptographic devices \cite{Sanders2017}. It is therefore not far fetched to imagine that in thermodynamic processes quantum systems may also offer some remarkable advantages. Rather than attempting to fully address this question here, in this chapter we instead focus on several paradigmatic settings aiming to provide the basic theoretical framework on which  investigations along this line or reasoning have, and continue to, develop.

We start with a few remarks on how temperature is actually measured in quantum systems in Sec.~\ref{sec:qu_thermoetry}. Then, in Sec.~\ref{sec:HeatEngines} we discuss the quantum version of heat engines, where the working substance is a genuinely quantum material, and examine both, the ideal, reversible as well as the endoreversible quantum Otto cycle. Following this in Sec.~\ref{sec:WorkExtraction}, we then explore the notion of quantum batteries, where there is evidence of a clear quantum advantage emerging. We close the chapter in Sec.~\ref{sec:Decoherence} and examine open quantum systems and briefly introduce the notion of quantum Darwinism.

\section{Quantum thermometry}
\label{sec:qu_thermoetry}

\begin{figure}
\centering
\includegraphics[width=.8\textwidth]{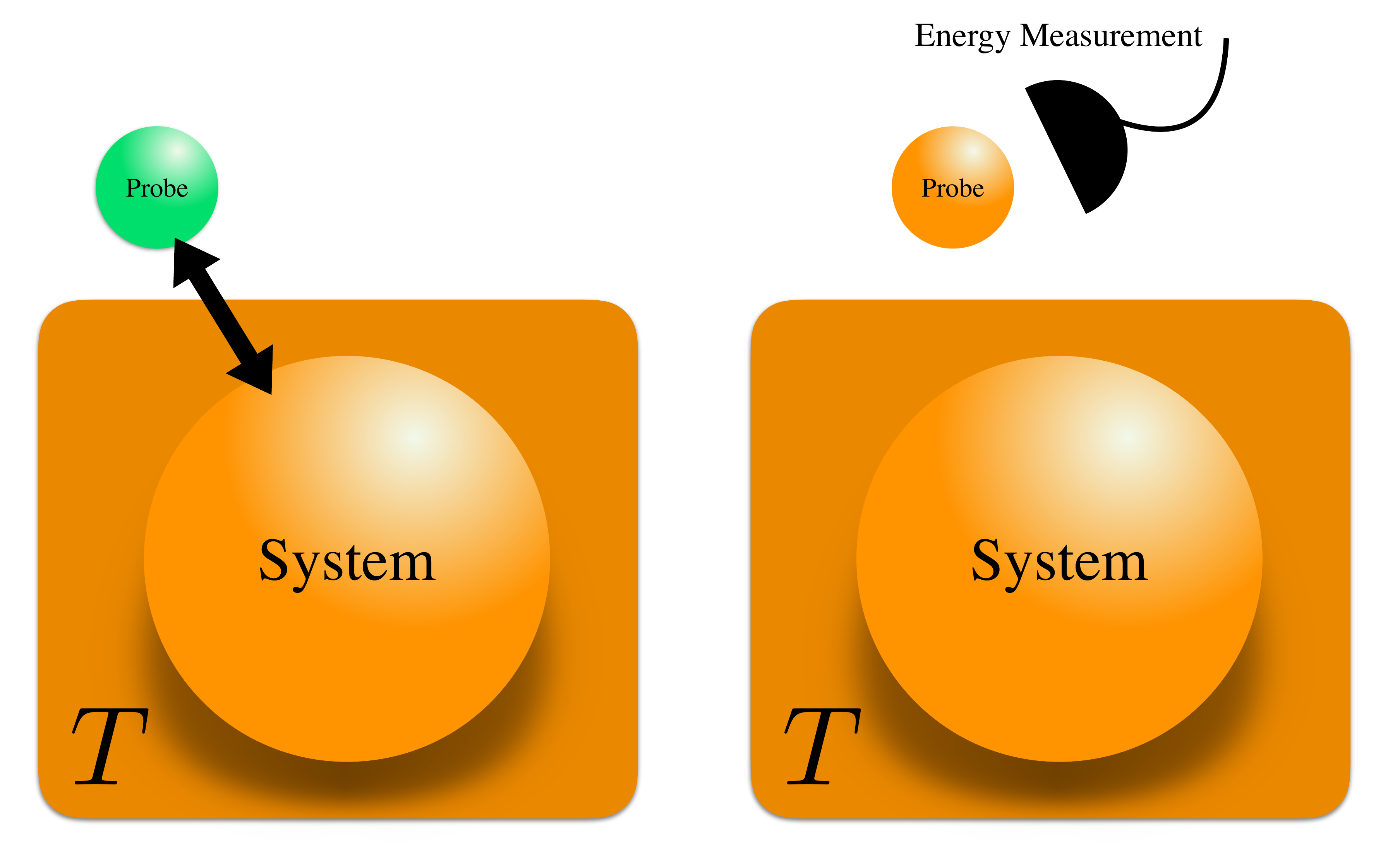}
\caption{\label{fig:thermo} Illustration of the set-up in quantum thermometry. A quantum probe is allowed to equilibrate with a system of interest, and then decoupled. The temperature of the quantum system can then be determined by measuring the energy of the probe.}
\end{figure}

In the following we will be interested in better understanding quantum heat engines. However, we have already seen above that quantifying heat is intimately connected with being able to distinguish hot and cold --  the ability to measure temperature. In this section, as originally presented in Ref.~\cite{CampbellQST2018}, we assess how precisely temperature can be measured using a single quantum probe. The general set-up is illustrated in Fig.~\ref{fig:thermo}

Whereas the temperature of a classical system is one of the best understood and most commonly used physical quantities, assigning a meaningful and unique temperature to quantum systems is \textit{a priori} a significantly harder task \cite{Gemmer2009}. Indeed, generally the temperature of quantum systems is neither a classical nor a quantum observable. Thus, one has to resort to quantum estimation techniques \cite{HelstromBook,ParisIJQI2009} to derive the ultimate limits on its determination. It is not surprising then that recent years have witnessed intense efforts in the design of `optimal quantum thermometers' and in accurately determining the temperature of a variety of quantum systems \cite{LuisPRL, ParisJPA2016, DePasqualeNatComm2016}. 

To assess the ability of a quantum system to act as a thermometer, we must first introduce the main tools in (quantum) estimation theory~\cite{ParisIJQI2009}. Information about an unknown parameter, $\mu$, which is imprinted in a quantum system $\rho(\mu)$, can be revealed by measuring any arbitrary observable over the system. By repeating such a measurement a large number of times, a dataset of outcomes is collected, upon which one might build up an \textit{estimator} $\hat{\mu}$ in order to estimate the parameter. Since statistical error---arising from the uncertainty in the outcomes of the measurement---is inescapable, a crucial task in metrology is its identification and optimization. For any unbiased estimator, i.e. $\big<\hat{\mu}\big>=\mu$, the statistical error is quantified by the (square root of the) variance of the estimator, which according to the Cram\'er-Rao inequality is lower bounded by
\begin{equation}
\label{bound_2}
\mbox{Var}(\hat{\mu})\ge\frac{1}{MF(\mu)}.
\end{equation}
Here $M$ denotes the number of measurements employed and $F(\mu)$ the so called Fisher information (FI) associated with the parameter $\mu$. For measurements having a discrete set of outcomes, the FI is given by
\begin{equation}
\label{classical}
F(\mu)=\sum_{j}p_{j}(\partial_\mu\ln p_{j})^2=
\sum_{j}\frac{|\partial_\mu p_j|^2}{p_j},
\end{equation}
where $p_{j}$ represents the probability to get outcome $j$ from the performed measurement. As Eq.~\eqref{classical} suggests, the FI can be taken as a measure of \textit{sensitivity} to the parameter: The larger the FI the more sensitive this measurement is to the unknown parameter, hence the smaller is the statistical error. The dependence of the FI on $p_j$ makes it clear that the quality of the estimation depends on the measurement protocol. However, one may be interested in the ultimate achievable sensitivity, optimized over all possible measurements. This maximum value is called the quantum Fisher Information (QFI)~\cite{HelstromBook,ParisIJQI2009}. The QFI only depends on $\rho(\mu)$, the density matrix of the system and is given by
\begin{equation}
\label{explicit}
\mathfrak{F}(\mu)=\sum_p\frac{[\partial_\mu\rho_p(\mu)]^2}{\rho_p(\mu)}+2\sum_{m\neq n}2\left[\frac{\rho_n(\mu)-\rho_m(\mu)}{\rho_n(\mu)+\rho_m(\mu)}\right]^2\,|\langle\psi_m|\partial_\mu\psi_n\rangle|^2,
\end{equation}
with $\rho_p$ the eigenvalues of the density matrix of the system, and $\ket{\psi_i}$ are the eigenstates of the system. Replacing $F(\mu)$ with $\mathfrak{F}(\mu)$ in Eq.~\eqref{bound_2} gives us the \emph{quantum} Cram\'er-Rao bound.

If we assume the system is already at thermal equilibrium, and therefore in a canonical Gibbs state, $\rho^\mrm{eq} = \e{-\beta H}/Z$, the QFI associated to temperature can be simplified by noticing that the eigenstates entering the second term of the RHS of Eq.~\eqref{explicit} do not change with temperature, and therefore this term is identically zero. Thus for a thermal state, $\mathfrak{F}(\mu)$ is fully determined by the change of the density matrix eigenvalues with temperature. Hence, for a $d$-dimensional Gibbs state $\rho^\mrm{eq}$, $\mathfrak{F}(\mu)$ can be easily evaluated and is equal to the classical Fisher information corresponding to a measurement described by the eigenstates of the Hamiltonian $H$. In formula one obtains \cite{ZanardiPRA2008,LuisPRL,ParisJPA2016,DePasqualeNatComm2016}
\begin{equation}
\label{eq:QFI}
\mathfrak{F}(\beta)= \sum_{n=1}^d \frac{|\partial_T p_n |^2}{p_n} =\beta^4\, [\text{Var}(H)]^2\,,
\end{equation}
where as before $\beta=1/k_B T$.

\subsection{Thermometry for Harmonic Spectra} \label{s:harmonic}
Let us consider using a harmonically spaced quantum system, with energy gap $\Delta$, as a thermometer. For a thermal two level system with free Hamiltonian $H=\Delta/2\, \sigma_z$, the corresponding Gibbs state is 
\begin{equation}
\label{eq:rho_qubit}
\rho^\mrm{eq} = \frac{1}{2} \left(
\begin{array}{cc}
1-\tanh \left(\beta\Delta/2\right)\ & 0 \\
 0 & 1+\tanh \left(\beta\Delta/2 \right) \\
\end{array}
\right).
\end{equation}
We can determine the QFI using Eq.~\eqref{eq:QFI} and find
\begin{equation}
\label{QFIqubit}
\mathfrak{F}(\beta) = \frac{\beta^4\Delta ^2 }{2}\,\text{sech}^2\left(\beta\Delta/2\right)\,.
\end{equation}
It is also straight forward to consider the infinite-dimensional quantum harmonic oscillator, $H\!=\!\Delta(a^\dagger a + 1/2)$ ($a$ representing the bosonic annihilation operator satisfying $[a,a^\dagger] = \id$), with the same spectral gap $\Delta$. The QFI for a thermal state is given by
\begin{equation}
\label{QFIHO}
\mathfrak{F}(\beta)= \frac{\beta^4\Delta ^2}{4}\, \text{csch}^2\left(\beta\Delta/2\right)\,.
\end{equation}

\begin{figure}[t]
\begin{centering}
{\bf (a)} \hskip0.5\columnwidth {\bf (b)} \\
\includegraphics[width=0.45\columnwidth]{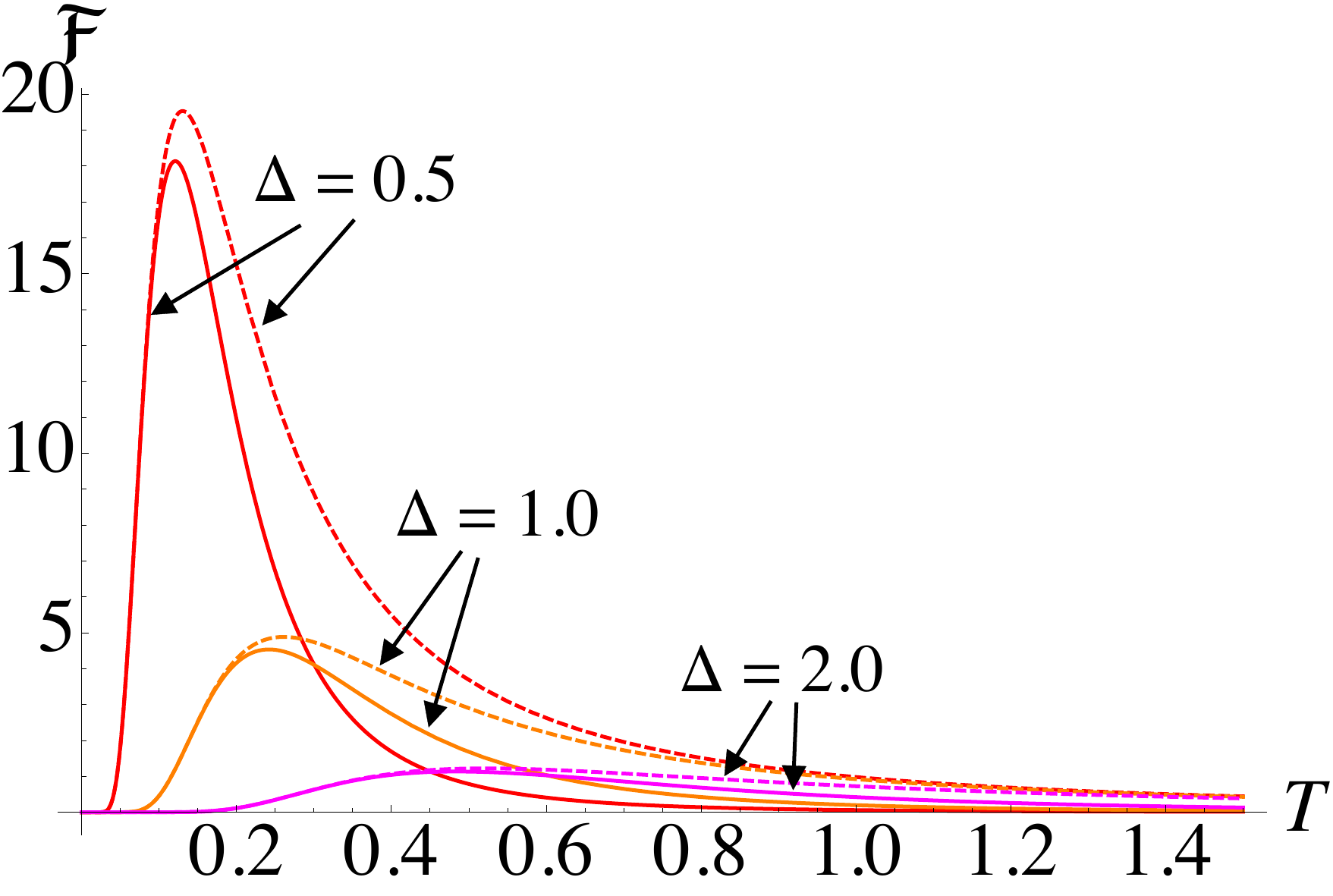}
\includegraphics[width=0.45\columnwidth]{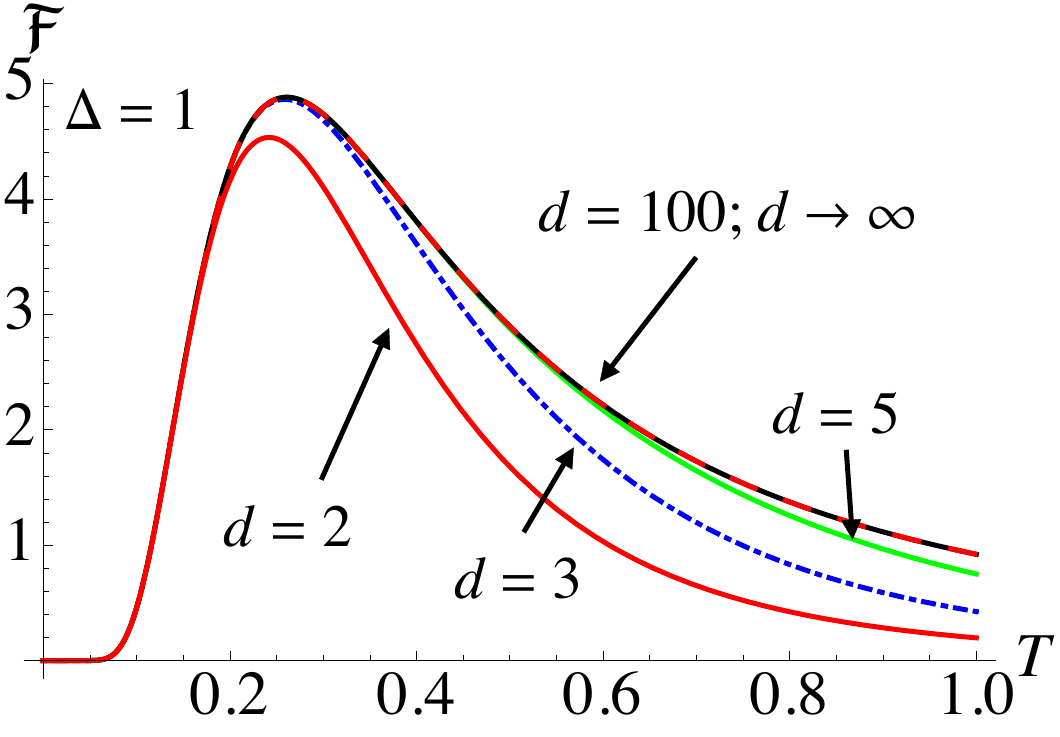}
\caption{{\bf (a)} The QFI for three values of energy spacing, $\Delta$, against temperature, $k_B T$. The solid (dashed) curves correspond to a two-level (harmonic oscillator) system. {\bf (b)} QFI for several different dimensional systems with harmonic spectra. We have fixed $\Delta=1.0$.}
\label{fig:Therm1}
\end{centering}
\end{figure}

In Fig.~\ref{fig:Therm1} {\bf (a)} the solid (dashed) curves show the QFI for several values of the energy level splitting, $\Delta$, for the two-level system (harmonic oscillator). Clearly, smaller energy gaps can lead to significantly better precision, however an important point to note is that the QFI peaks at a single value of $T$. This means there is a single temperature that a given system with a specified energy gap is optimized to probe. This temperature corresponds to the value of $T$ maximizing the QFI, $\mathfrak{F}_\text{max}$, and as we change $\Delta$ the position of this peak shifts. An intuitive understanding of this can be drawn by closer consideration of Eq.~\eqref{eq:QFI}: for sufficiently low temperatures, regardless of dimensionality, the quantum thermometer will be in its ground state. Since the sensitivity is related to how the thermal populations change, smaller gaps between the eigenstates result in rapid rates of change for low temperatures since more energy levels become populated at these temperatures. Conversely, when the spectral gap is larger, the system requires larger temperatures before any of the excited states become populated. 

From Fig.~\ref{fig:Therm1} we clearly see that the two disparate dimensional systems exhibit qualitatively identical behaviors, thus implying that the achievable precision for thermometry with harmonic systems is solely dependent on the single characteristic spectral gap, $\Delta$, while dimensionality plays only a minor role. We can show this more explicitly by considering arbitrary $d$-dimensional harmonic systems, described by the Hamiltonian $H\!\!=\!\! \sum_{n=1}^d n\Delta \ket{E_n}\bra{E_n}$, and calculating the corresponding QFI. For a thermal state, the probe is in Gibbs form and the energy level occupations (eigenvalues) are simply given by a Boltzmann distribution. Thus, for a $d$-dimensional system with energy spacing $\Delta$, the $n$th eigenvalue of the thermal state is
\begin{equation}
p_n = \frac{\left(e^{\beta\Delta}-1\right) e^{\beta\Delta(d-n)}}{e^{\beta\Delta d}-1}.
\end{equation}
From Eq.~\eqref{eq:QFI} we know that the QFI is based solely on the rate of change of these occupations with respect to temperature, and we obtain
\begin{equation}
\begin{split}
&\mathfrak{F}_d(\beta) = \frac{\beta^4\Delta ^2\left[-d^2 \left(1+ e^{2\beta\Delta}\right) e^{\beta\Delta  d}+2 \left(d^2-1\right) e^{\beta\Delta (d+1)}+e^{\beta\Delta(1 +2 d)}+e^{\beta\Delta}\right]}{\left(e^{\beta\Delta}-1\right)^2 \left(e^{\beta\Delta d}-1\right)^2} .
\end{split}
\end{equation}
It is easy to check that we recover Eq.~\eqref{QFIqubit} (Eq.~\eqref{QFIHO}) by setting $d=2$ ($d\to\infty$).

We depict $\mathfrak{F}_d(\mu)$ for various values of $d$ in Fig.~\ref{fig:Therm1} \textbf{(b)} where we have (arbitrarily) fixed $\Delta=1$. It is immediately evident that for low temperatures, $k_B T\lesssim 0.2$, all systems perform identically, while differences arise only at comparatively large temperatures. This again can be intuitively understood in light of the fact that at low temperatures all systems are constrained to the low energy portion of the spectrum and therefore in this region only the ground and first excited state will play a significant role. We can conclude (i) the constant energy level spacing in harmonically gapped systems plays the most crucial role in thermometry. Therefore, to probe low temperatures one should seek to use a system with a small energy spacing, while for larger temperatures larger gapped systems are significantly more useful. (ii) We can gain some enhancement by going from two- to a three-level system, however higher dimensional systems offer no advantage regarding the optimal achievable precision. (iii) Regardless, such systems are only designed to estimate a single temperature with the optimal precision.

\subsection{Optimal Thermometers}
While the study of quantum thermometers with harmonically gapped spectra is a natural starting point, it leaves us questioning what makes an optimal thermometer? In Ref.~\cite{LuisPRL} Correa \etal addressed this question by explicitly considering under what conditions the quantum Fisher information in Eq.~\eqref{eq:QFI} is maximized. It is immediately evident that a probe with the maximum possible energy variance at equilibrium fulfills the task. Considering an arbitrary $N$-dimensional system, this is equivalent to insisting that the heat capacity of the probe is maximal. Such an optimal thermometer is an effective two-level system, with a unique ground state and an $(N-1)$-fold degenerate excited state. The corresponding precision scales with the dimensionality of the probe, however this comes at a cost of having a reduced specified temperature range for which it operates efficiently~\cite{LuisPRL}. Clearly these optimal thermometers again have a single characteristic energy splitting and therefore are designed to estimate a single temperature precisely. However, just as exploiting degeneracy is shown to enhance precision, including anharmonicity can also increase the range of temperatures that can be precisely probed~\cite{CampbellQST2018}.\footnote{For a more expansive review of thermometry we refer to Ref.~\cite{ThermometryReview}.}

\section{Quantum heat engines -- engines with atomic working fluids}
\label{sec:HeatEngines}

\begin{figure}
\centering
\includegraphics[width=.75\textwidth]{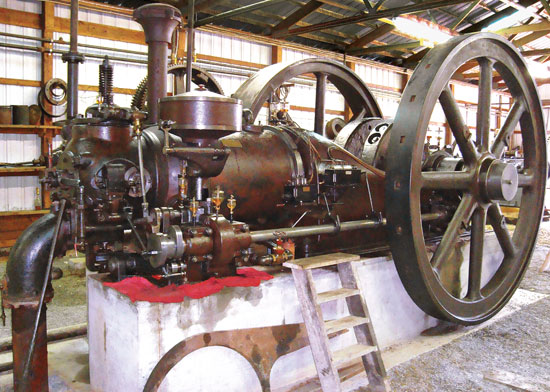}
\caption{\label{huge_Otto}The 1925 175 HP Otto is likely the largest single-cylinder, conventional two-flywheel gas engine in the world.}
\end{figure}

Now that we have understood the limitations in measuring temperature in quantum systems, we can continue to apply the conceptual framework to the most important application of quantum thermodynamics -- quantum heat engines. We have already seen an example of a quantum thermodynamic cycle in Sec.~\ref{sec:quwork} -- the Carnot cycle. In the following, we will instead focus on arguably the most widely used cyclic process: the Otto cycle~\cite{OttoPRE, OttoPRE2, Abah2012, Rossnagel2016, KosloffEntropy, STAHeatEngine}. The Otto cycle underlies the working principles of all internal combustion heat engines and is therefore of significant practical and fundamental relevance [see Fig.~\ref{huge_Otto} for a real example].

In what follows we will begin outlining the basic strokes involved in the classical formulation of the Otto cycle before looking into its quantum mechanical description. Some time will be dedicated to highlighting the subtle differences between the Carnot and Otto cycles and we will explore two paradigmatic instances of the quantum Otto cycle where the working substance is a two-level system and a quantum harmonic oscillator, the latter which was recently used to realize the first experimental demonstration of a genuinely quantum heat engine.

\subsection{The Otto Cycle: Classical to quantum formulation}
\label{OttoCycleSection}
The Otto cycle, named after the German engineer Nikolaus Otto who is accredited with building the first working four-stroke engine based on the design by Alphonse Beau de Rochas, consists of the following thermodynamic processes between local equilibrium states $A,B,C,D$:
\begin{enumerate}
\item {\it Adiabatic (isentropic) compression} ($A\to B$): The working medium is compressed. This stroke involves both volume and temperature changes, while the entropy remains constant.
\item  {\it Isochoric heating} ($B\to C$): The volume of the working medium is fixed, while the temperature is increased.
\item {\it Adiabatic (isentropic) expansion} ($C\to D$): The power stroke, when useful work is extracted from the engine. Again this stroke involves both volume and temperature changes, at fixed entropy.
\item  {\it Isochoric cooling} ($D\to A$): The working medium is cooled at a fixed volume and returned to its initial state, ready to begin the cycle again.
\end{enumerate}
In order to clearly define the quantum Otto cycle it is important to determine how these four strokes are realized when the working medium is a quantum system. To this end, it is important to examine adiabatic and isochoric processes for quantum systems~\cite{OttoPRE, OttoPRE2}.

\paragraph{Quantum adiabatic processes.}
The quantum adiabatic condition posits that if a system is perturbed at a slow enough rate, so as to avoid any excitations being generated, the energy level populations will remain constant. This then implies that that there is no heat exchanged with the external heat bath during the process, and all changes in internal energy of the working substance is therefore in the form of work. In textbook expositions of the classical Otto cycle the working substance is assumed to be perfectly isolated from the thermal baths during the quasistatic expansion/compression strokes. As such the quantum and classical adiabatic processes appear quite similar. However, in practice one can envisage achieving the classical adiabatic processes by rapid compression/expansion strokes. Indeed, for an ideal classical gas, such fast processes ensure no heat exchange with the bath. This is in stark contrast to the quantum Otto cycle, where to ensure the entropy is constant, and therefore zero heat exchange, these strokes must be performed quasistatically. Thus, quantum adiabatic processes form only a subset of classical adiabatic processes.

\paragraph{Quantum isochoric processes.}
These processes involve changing the energy level occupations, and therefore also the entropy of the working substance, until it is in thermal equilibrium with the heat bath. In this case the energy eigenvalues remain unchanged, and therefore these processes involve only heat exchanges with no work being done to/by the working substance. Quantum and classical isochoric processes are quite similar insofar as they both involve a change in temperature of the system through heat exchange with zero work performed.

\paragraph{Quantum work and heat.} Consider the following Hamiltonian of the working substance
\begin{equation}
H(\lambda)=\sum_{n} E_n(\lambda) \ketbrad{n(\lambda)}
\end{equation}
with $\ket{n(\lambda)}$ the energy eigenstates, $E_n$ the associated energy eigenvalues, and $\lambda$ a variable work parameter. It is convenient to rescale the energy with respect to the ground state energy
\begin{equation}
\label{chap2:ham}
H(\lambda)=\sum_{n} (E_n-E_0)(\lambda) \ketbrad{n(\lambda)}.
\end{equation}
The internal energy is then simply given by the expectation value of this Hamiltonian over the state of the working substance
\begin{equation}
E=\langle H \rangle = \sum_{n} p_n E_n
\end{equation}
where as before, $p_n=\e{-\beta E_n}/Z$. From the first law (see Sec.~\ref{sec:thermo}), we can then split the change in internal energy into a work and heat contribution
\begin{equation}
\label{InternalEn}
dE=\sum_{n} (p_n dE_n + E_n dp_n) = \dbar W + \dbar Q.
\end{equation}
From Eq.~\eqref{InternalEn} it is easy to identify when a process is quantum adiabatic and/or isochoric. In an adiabatic process, the eigenenergies vary while their populations, and therefore the entropy of the working substance, are kept constant. Thus, $d p_n = 0$ and the change in internal energy is attributed solely to work. Conversely, isochoric processes change the populations without affecting the energy eigenvalues, hence $d E_n = 0$ and as such the change in internal energy is due purely to heat exchanges. At this point it is useful to highlight one significant difference between idealized Carnot and Otto cycles. While both involve two adiabatic transformations, in a Carnot cycle heat is absorbed/emitted through isothermal processes. The system must remain in thermal equilibrium, at temperature $T$ of the bath, while the heating/cooling process occurs. Therefore, an ideal isothermal process will require changes in {\it both} $E_n$ and $p_n$, implying work done and heat exchange. Under the strict condition that the temperature of the working substance is equal to that of the heat bath at the end of the adiabatic stroke in the Carnot cycle, the ensuing isothermal process is thermodynamically reversible, i.e. results in zero irreversible entropy production. The Otto cycle, on the other hand, only requires the working substance to be in \emph{local thermal equilibrium} at the end of each stroke. Therefore, at the end of the adiabatic power stroke $C\to D$ the working substance will be in a thermal state with temperature given by the hot heat bath, $T_h$. It is then connected to the cool heat bath in order to thermalize to the initial state through the isochoric cooling. This cooling is accompanied by an irreversible entropy production. Therefore, it is important to keep in mind that while the expressions for the maximal efficiency of the ideal Carnot and Otto cycles share several similar features, the underlying physical processes are fundamentally different.

\subsection{A two-level Otto cycle}
To solidify the ideas let us consider an Otto cycle when the working substance is a two-level system (TLS) with energy eigenvalues $\{E_g,~E_e \}$ and corresponding eigenstates $\{\ket{\psi_g},~\ket{\psi_e} \}$~\cite{OttoPRE}. Note that we have dropped the explicit dependence on the work parameter $\lambda$ for brevity, since in what follows we will only be concerned with the state of the TLS at the beginning and end of each stroke. Following from Eq.~\eqref{chap2:ham} we can rescale the Hamiltonian such that $E_g=0$ and therefore 
\begin{equation}
H = (E_e-E_g)\ket{\psi_e}\!\!\bra{\psi_e} =  \Delta \ket{\psi_e}\!\!\bra{\psi_e}
\end{equation}
where $\Delta$ is the energy difference between ground and excited states. The corresponding thermal state at temperature $T$ is given by Eq.~\eqref{eq:rho_qubit}, which we can also write as
\begin{equation}
\begin{split}
\rho^\mrm{eq}\equiv p_g\ketbrad{\psi_g} + p_e\ketbrad{\psi_e}.
\end{split}
\end{equation}
Now let us examine how the state changes during the various strokes of the Otto-cycle. In what follows we will denote the temperature of the TLS at the end of each stroke as $T_i$ for $i=A,B,C,D$ and allow the energy splitting $\Delta$ to vary between $\Delta_i$ and $\Delta_f$ with $\Delta_f > \Delta_i$.

\paragraph{Isentropic compression.}
We assume the working substance is initially in thermal equilibrium with a cold bath at temperature $T_c$, and therefore $T_c = T_A$. During the adiabatic stroke work is performed on the TLS to increase energy splitting from $\Delta_i \to \Delta_f$ without any exchange of heat. This requires that the populations at the start and end of the process are equal, i.e. 
\begin{equation}
\label{chap2:conditions}
p_g^A = p_g^B~\qquad~\text{and}~\qquad~p_e^A = p_e^B.
\end{equation}
In order for these conditions to hold, the state of the working substance at $B$ cannot be in thermal equilibrium with the cold bath anymore. However, as there is only a single energy splitting, we can define an effective local temperature of the TLS at $B$ from Eq.~\eqref{chap2:conditions}
\begin{equation}
\label{chap2:TBeff}
T_B = \frac{\Delta_f}{\Delta_i} T_A.
\end{equation}
Since no heat is exchanged, the change in internal energy of the TLS is work. From Eq.~\eqref{InternalEn} 
\begin{equation}
\begin{split}
\label{workAB}
W_{A \to B} &= \sum_n \int_A^B p_n d E_n = \left(E_e^B-E_g^B\right) p_e^B - \left(E_e^A-E_g^A\right) p_e^A \\
		&= \frac{(\Delta_f - \Delta_i)}{2} \left[1 - \tanh\left(\beta_A\Delta_i/2 \right)  \right]. 
\end{split}
\end{equation}

\paragraph{Isochoric Heating.}
The working substance is then connected to the hot bath at temperature $T_h$. During the heating processes, the energy eigenvalues of the TLS remain constant while the system equilibrates with the bath and therefore its temperature goes from $T_B \to T_C = T_h$. As no work is performed during an isochoric process, the heat exchanged from the bath to the TLS is readily evaluated from Eq.~\eqref{InternalEn}
\begin{equation}
\begin{split}
\label{heatBC}
Q_{B\to C} &= \sum_n \int_B^C E_n dp_n = \left(E_e^C - E_g^C\right) p_e^C - \left(E_e^B - E_g^B\right) p_e^B \\
		&= \frac{\Delta_f}{2} \left[\tanh\left(\beta_B\Delta_f/2\right) - \tanh\left(\beta_C\Delta_f/2 \right)  \right]
\end{split}
\end{equation}

\paragraph{(Power) Stroke: isentropic expansion}
Work is now extracted during the third stroke by adiabatically reducing the energy of the working substance from $\Delta_f \to \Delta_i$. Analogously to the isentropic compression, the energy change during expansion is purely due to the work extraction and no heat is exchanged meaning again the energy occupations remain invariant, i.e.
\begin{equation}
\label{chap2:conditions2}
p_g^C = p_g^D~\quad~\text{and}~\quad~p_e^C = p_e^D,
\end{equation}
thus, we can also determine the local effective temperature of the working substance at $D$
\begin{equation}
\label{chap2:TDeff}
T_D = \frac{\Delta_i}{\Delta_f} T_C.
\end{equation}
Clearly at $D$ (and in fact at any point during the work extraction process) the working substance is not in thermal equilibrium with the hot bath. From Eq.~\eqref{InternalEn} the work extracted is
\begin{equation}
\begin{split}
\label{workCD}
W_{C \to D} &= \sum_n \int_C^D p_n d E_n= \left(E_e^D-E_g^D\right) p_e^D - \left(E_e^C-E_g^C\right) p_e^C \\
		&= -\frac{(\Delta_f - \Delta_i)}{2} \left[1 - \tanh\left(\beta_C\Delta_f/2 \right)  \right]. 
\end{split}
\end{equation}

\paragraph{Isochoric Cooling.}
The final isochoric cooling stroke is required to return the TLS to its initial state. The working substance is connected to the cold bath at temperature $T_c = T_A$. From Eq.~\eqref{InternalEn} the heat removed from the TLS is
\begin{equation}
\begin{split}
\label{heatDA}
Q_{D\to A} &= - \sum_n \int_D^A E_n dp_n = - \left[ \left(E_e^D - E_g^D\right) p_e^D - \left(E_e^A - E_g^A\right) p_e^A\right]\\
		&= - \frac{\Delta_i}{2} \left[\tanh\left(\beta_D\Delta_i/2 \right) + \tanh\left(\beta_A\Delta_i/2 \right)  \right] 
\end{split}
\end{equation}
with the global minus sign indicating the removal of heat.

\begin{figure}
\centering
\includegraphics[height=.3\textheight]{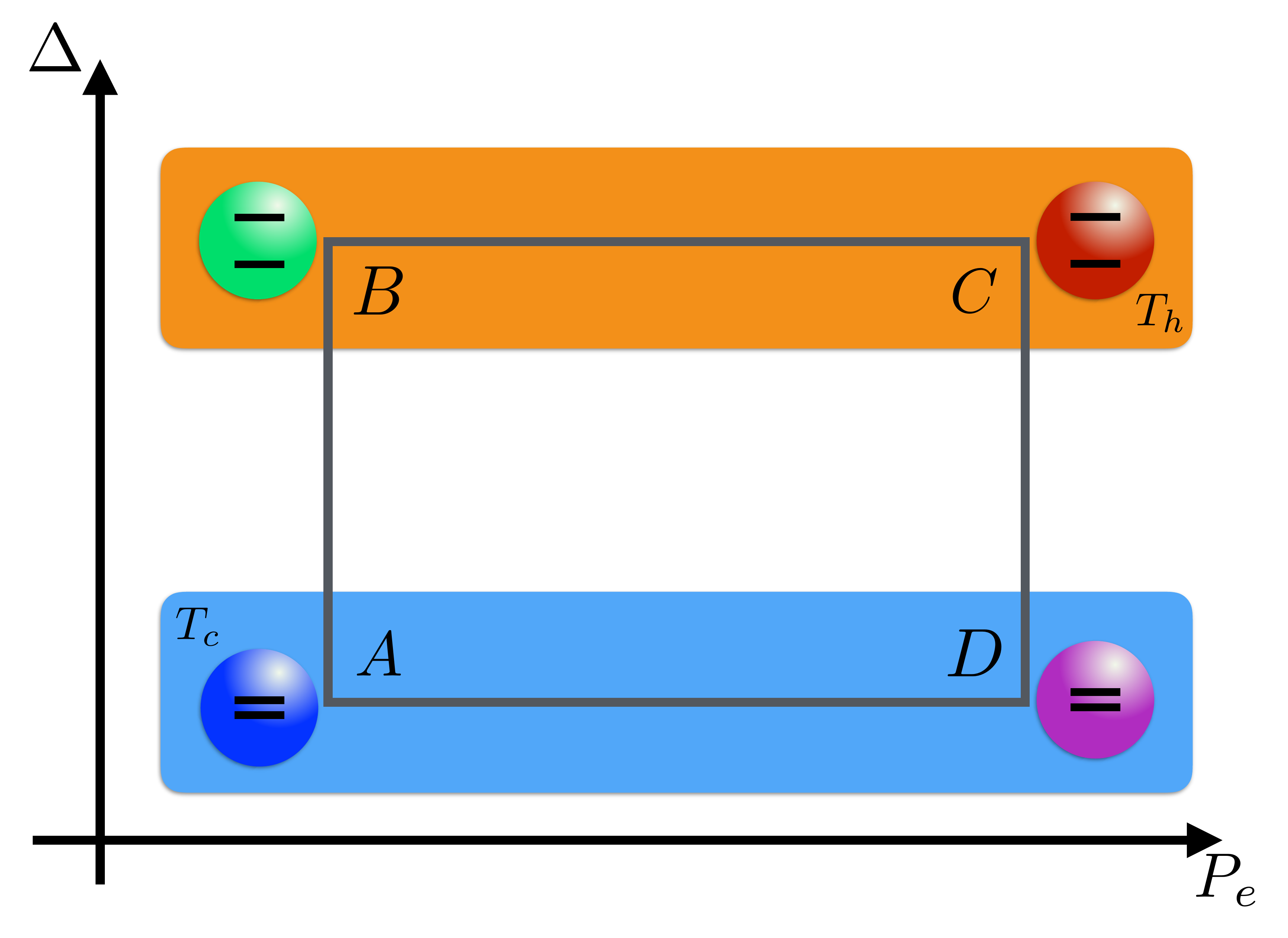}
\caption{Otto cycle for a quantum two-level system (TLS). The TLS is initially in thermal equilibrium with the cold bath. The change in internal energy during the strokes is in the form of work $A\to B$ and $C \to D$, and heat $B \to C$ and $D \to A$. The TLS is only ever in thermal equilibrium with the baths at $A$ and $C$, nevertheless the TLS can always be described as some locally thermal state.}
\label{figOtto}
\end{figure}

From the above analysis it should be clear that the \emph{quantum} Otto-cycle is an irreversible process. Indeed, only at $A$ and $C$ is the TLS at thermal equilibrium with the cold and hot baths, respectively. For most of the cycle the working substance is in fact out-of-equilibrium with respect to the baths and the requirement to thermalize at $A$ and $C$ leads to the irreversibility. The Otto cycle for the TLS can be captured diagrammatically in Fig.~\ref{figOtto}.

\paragraph{Positive work condition.}
In order for the Otto-cycle to be useful we need it to produce a net-output of work and this leads to what is known as the {\it positive work condition}. This constraint can be intuitively seen by considering what should happen during the heating stroke $B \to C$, where the temperature of the working substance is raised until it is in thermal equilibrium with the hot bath at temperature $T_h = T_C$. Clearly, this implies that 
\begin{equation}
\label{PWC}
T_B<T_C\quad\mrm{and}\quad\frac{\Delta_f}{\Delta_i}< \frac{T_h}{T_c}
\end{equation}
which is fully analogous to its classical, macroscopic counterpart.

\paragraph{Efficiency.}
Using the effective temperatures in Eqs.~\eqref{chap2:TBeff} and \eqref{chap2:TDeff} we can express the work and heat exchanges in Eqs.~\eqref{workAB} and \eqref{heatBC}, \eqref{workCD}, and \eqref{heatDA} purely in terms of the bath temperatures, $T_h$ and $T_c$, and the energy splittings $\Delta_i$ and $\Delta_f$ 
\begin{eqnarray}
W_{A \to B} & = &    \frac{\Delta_f - \Delta_i}{2} \left[ 1- \tanh\left(\beta_c\Delta_i/2  \right) \right], \\
Q_{B \to C} & = &    \frac{\Delta_f}{2} \left[\tanh\left(\beta_c\Delta_i/2 \right) - \tanh\left(\beta_h\Delta_f/2 \right)  \right], \\
W_{C \to D} & = &   \frac{\Delta_i - \Delta_f}{2} \left[ 1-\tanh\left(\beta_h\Delta_f /2 \right) \right] \\
Q_{D \to A} & = &    \frac{\Delta_i}{2} \left[\tanh\left(\beta_c\Delta_i /2\right) -\tanh\left(\beta_h\Delta_f/2 \right) \right]. 
\end{eqnarray}
The net work performed during the cycle is then
\begin{equation}
\begin{split}
W 	&= - (W_{A\to B} + W_{C \to D}) =  Q_{B \to C} - Q_{D \to A} \\
	&= \frac{\Delta_f - \Delta_i}{2} \left[\tanh\left(\beta_c\Delta_i/2 \right)  - \tanh\left(\beta_h\Delta_f/2 \right)   \right]
\end{split}
\end{equation}
from which we can readily determine the Otto efficiency
\begin{equation}
\label{eq:eff_otto}
\eta_O = \frac{W}{Q_{B\to C}} = 1- \frac{\Delta_i}{\Delta_f}\equiv 1-\kappa.
\end{equation}
It is worth noting that in complete analogy to classical engines the quantum Otto efficiency for a TLS is governed by the ``compression'' ratio $\kappa\equiv\Delta_i/\Delta_f$ and, due to the constraints set by the positive work condition, Eq.~\eqref{PWC}, the Otto efficiency will always be smaller than the Carnot efficiency.

\subsection{Endoreversible Otto cycle}

We have seen in the previous section that the Otto cycles for quantum and classical systems are remarkably similar. However, for quantum engines, even more so than  for classical engines, one is more interested in the efficiency at maximal power -- rather then simply determining the maximal efficiency. In Sec.~\ref{sec:finite_time_thermo} we have already seen that the efficiency for endoreversible Carnot engines at maximum power output is given by the Curzon-Ahlborn efficiency \eqref{eq:efficency_CA}.

Quite remarkably, the same efficiency has been found for many different systems, such as an endoreversible Otto engine with an ideal gas as working medium \cite{Leff1987}, the endoreversible Stirling cycle \cite{Erbay1997}, Otto engines in open quantum systems in the quasistatic limit \cite{Rezek2006}, or a single ion in a harmonic trap undergoing a quantum Otto cycle \cite{Abah2012,Rossnagel2014}. Recent experimental breakthroughs in the implementation of nanosized heat engines \cite{Rossnagel2016,Klaers2017} that could principally exploit quantum resources \cite{Scovil1959,Scully2002,Scully2003,Scully2011,Zhang2014PRL,Gardas2015,Hardal2015,Roulet2018,Niedenzu2018,Ronzani2018} pose the question whether or not $\eta_\mrm{CA}$ is more universal that one could expect. In Sec.~\ref{sec:finite_time_thermo} we derived the Curzon-Ahlborn efficiency  \eqref{eq:efficency_CA}  for endoreversible \emph{Carnot cycles}, which are clearly independent on the nature of the working medium. However, a standard textbook exercise shows that the efficiency of \emph{Otto cycles} should be dependent on the equation of state, i.e., on the specific working medium \cite{Callen1985}. 

As we have seen above, the standard Otto cycle is a four-stroke cycle consisting of isentropic compression, isochoric heating, isentropic expansion, and ischoric cooling \cite{Callen1985}. Thus, we have in the endoreversible regime:

\paragraph{Isentropic compression.}
During the isentropic strokes the working substance does not exchange heat with the environment. Therefore, the thermodynamic state of the working substance can be considered independent of the environment, and the endoreversible description is identical to the equilibrium cycle. From the first law of thermodynamics, $\Delta E=Q+W$, we have,
\begin{equation}
Q_\mrm{comp}=0\quad\mrm{and}\quad W_\mrm{comp}=E(T_B,\omega_f)-E(T_A,\omega_i)
\end{equation}
where $Q_\mrm{comp}$ is the heat exchanged, and $W_\mrm{comp}$ is the work performed during the compression. Moreover, we denote the work parameter by $\omega$, which will be interpreted as the  frequency of a harmonic oscillator \eqref{eq:harm}, shortly.

\paragraph{Isochoric heating.}

During the isochoric strokes the work parameter is held constant, and the system exchanges heat with the environment. Thus, we have for isochoric heating 
\begin{equation}
Q_h=E(T_C,\omega_f)-E(T_B,\omega_f)\quad\mrm{and}\quad W_h=0\,.
\end{equation}

In complete analogy to Curzon and Ahlborn's original analysis \cite{Curzon1975} we now assume that the working substance is in a state of local equilibrium, but also that the working substance never fully equilibrates with the hot reservoir. Therefore, we can write
\begin{equation}
T(0)=T_B\quad\mrm{and}\quad T(\tau_h)=T_C\quad\mrm{with}\quad T_B<T_C \leq T_h\,,
\end{equation}
where as before $\tau_h$ is the duration of the stroke.

Note that in contrast to the Carnot cycle the Otto cycle does not involve isothermal strokes, and, hence, the rate of heat flux is not constant. Rather, we have to explicitly account for the change in temperature from $T_B$ to $T_C$. To this end, Eq.~\eqref{eq:heat_hot} is replaced by Fourier's law \cite{Callen1985},
\begin{equation}
\label{eq:fourier_hot}
\frac{d T}{dt}=-\alpha_h \left(T(t)-T_h\right)
\end{equation}
where $\alpha_h$ is a constant depending on the heat conductivity and heat capacity of the working substance. 

Equation~\eqref{eq:fourier_hot} can be solved exactly, and we obtain the relation
\begin{equation}
\label{eq:rel_hot}
T_C-T_h=\left(T_B-T_h\right)\,\e{-\alpha_h \tau_h}\,.
\end{equation}
In the following, we will see that Eq.~\eqref{eq:rel_hot} is instrumental in reducing the number of free parameters.

\paragraph{Isentropic expansion.}
 In complete analogy to the compression, we have for the isentropic  expansion,
\begin{equation}
Q_\mrm{exp}=0\quad\mrm{and}\quad W_\mrm{exp}=E(T_D,\omega_i)-E(T_C,\omega_f)\,.
\end{equation}

\paragraph{Isochoric cooling.} 

Heat and work during the isochoric cooling read,
\begin{equation}
Q_c=E(T_A,\omega_i)-E(T_D,\omega_i)\quad\mrm{and}\quad W_c=0\,,
\end{equation}
where we now have
\begin{equation}
T(0)=T_D\quad\mrm{and}\quad T(\tau_c)=T_A\quad\mrm{with}\quad T_D>T_A \geq T_c\,.
\end{equation}

Similarly to above \eqref{eq:fourier_hot} the heat transfer is described by Fourier's law
\begin{equation}
\label{eq:fourier_cold} 
\frac{d T}{dt}=-\alpha_c \left(T(t)-T_c\right)\,,
\end{equation}
where $\alpha_c$ is a constant characteristic for the cold stroke. From the solution of Eq.~\eqref{eq:fourier_cold} we now obtain
\begin{equation}
\label{eq:rel_cold}
T_A-T_c=\left(T_D-T_c\right)\,\e{-\alpha_c \tau_c}\,,
\end{equation}
which properly describes the decrease in temperature from $T_D$ back to $T_A$.

\paragraph{Classical harmonic engine.}

To continue the analysis we now need to specify the internal energy $E$. As a first example, we consider the working medium to be described as a classical harmonic oscillator. The bare Hamiltonian reads,
\begin{equation}
\label{eq:harm}
H(p,x)=\frac{p^2}{2m}+\frac{1}{2}m \omega^2 x^2\,,
\end{equation}
where $m$ is the mass of the particle. 

For a particle in thermal equilibrium the Gibbs entropy, $S$, and the internal energy, $E$, are
\begin{equation}
\label{eq:harm_ent}
\frac{S}{k_B}=1+\lo{\frac{k_B T}{\hbar\omega}}\quad\mrm{and}\quad E=k_B T\,,
\end{equation}
where we introduced Boltzmann's constant, $k_B$. 

Note, that from Eq.~\eqref{eq:harm_ent} we obtain a relation between the frequencies, $\omega_i$ and $\omega_f$ and the four temperatures, $T_A$, $T_B$, $T_C$, and $T_D$. To this end, consider the isentropic strokes, for which we have
\begin{equation}
S(T_B,\omega_f)=S(T_A,\omega_i)\quad\mrm{and}\quad S(T_D,\omega_i)=S(T_C,\omega_f)\,,
\end{equation}
which is fulfilled by
\begin{equation}
\label{eq:temp}
T_A\, \omega_f=T_B\,\omega_i\quad\mrm{and}\quad T_C\, \omega_i=T_D\,\omega_f\,.
\end{equation}

We are now equipped with all the ingredients necessary to compute the endoreversible efficiency, 
\begin{equation}
\label{eq:eff}
\eta=-\frac{W_h+W_c}{Q_h}\,.
\end{equation}
In complete analogy to fully reversible cycles \cite{Callen1985}, Eq.~\eqref{eq:eff} can be written as
\begin{equation}
\label{eq:eff_temp}
\eta=1-\frac{T_D-T_A}{T_C-T_B}\,,
\end{equation}
where we used the explicit from of the internal energy $E$ \eqref{eq:harm_ent}. Further, using Eqs.~\eqref{eq:temp} the endoreversible Otto efficiency becomes
\begin{equation}
\eta=1-\frac{\omega_i}{\omega_f}\equiv1-\kappa\,,
\end{equation}
where we again introduced the compression ratio, $\kappa$. Observe that the endoreversible efficiency takes the same form as its reversible counter part \cite{Callen1985}, and also Eq.~\eqref{eq:eff_otto}. However, in Eq.~\eqref{eq:eff_temp} the temperatures correspond the local equilibrium state of the working substance, and not to a \emph{global} equilibrium with the environment. 

Similarly to Curzon and Ahlborn's treatment of the endoreversible Carnot cycle \cite{Curzon1975} we now compute the efficiency for a value of $\kappa$, at which the power \eqref{eq:power} is maximal. We begin by re-writing the total work with the help of the compression ratio $\kappa$ and Eqs.~\eqref{eq:temp} as,
\begin{equation}
W_\mrm{tot}=k_B\,\left(\kappa -1\right)\left(T_B-T_C\right)\,.
\end{equation}
Further using Eq.~\eqref{eq:rel_hot} we obtain
\begin{equation}
W_\mrm{tot}=k_B\,\left(\kappa -1\right)\left[1-\e{-\alpha_h\tau_h}\right]\left(T_B-T_h\right)\,,
\end{equation}
which only depends on the free parameters $T_B$, $\kappa$, and $\tau_h$. Of these three, we can eliminate one more, by combining Eqs.~\eqref{eq:rel_hot} and \eqref{eq:rel_cold}, and we have
\begin{equation}
T_B=\frac{T_c\left[\e{\alpha_c\tau_c}-1\right]+\kappa\,T_h\left[1-\e{-\alpha_h\tau_h}\right]}{\kappa\left[\e{\alpha_c\tau_c}-\e{-\alpha_h\tau_h}\right]}\,.
\end{equation}
Finally, the power output \eqref{eq:power} takes the form, 
\begin{equation}
\label{eq:power_clas}
P=\frac{2k_B\,(\kappa-1)(T_c-\kappa\,T_h)}{\zeta\kappa (\tau_c+\tau_h)}\,\frac{\sinh{\left(\alpha_c\tau_c/2\right)}\sinh{\left(\alpha_h\tau_h/2\right)}}{\sinh{\left[(\alpha_c\tau_c+\alpha_h\tau_h)/2\right]}}\,.
\end{equation}

Remarkably the power output factorizes into a contribution that only depends on the compression ratio, $\kappa$, and another term that is governed by the stroke times, $\tau_c$ and $\tau_h$,
\begin{equation}
P(\kappa,\tau_h,\tau_c)=f_1(\kappa) f_2(\tau_h,\tau_c)\,.
\end{equation}
It is then a simple exercise to show that $P(\kappa,\tau_h,\tau_c)$ is maximized for any value of $\tau_h$ and $\tau_c$ if we have,
\begin{equation}
P_\mrm{max}=P(\kappa_\mrm{max})\quad\mrm{with}\quad \kappa_\mrm{max}=\sqrt{\frac{T_c}{T_h}}\,.
\end{equation}
Therefore, the efficiency at maximal power reads,
\begin{equation}
\label{eq:eff_result}
\eta=1-\sqrt{\frac{T_c}{T_h}}\,.
\end{equation}
In conclusion, we have shown analytically that for the classical harmonic oscillator the efficiency at maximal power of an endoreversible Otto cycle \eqref{eq:eff} is indeed given by the Curzon-Ahlborn efficiency \eqref{eq:efficency_CA}.

It is worth emphasizing that for the endoreversible Otto cycle we started with six free parameters, the four temperatures $T_A$, $T_B$, $T_C$, and $T_D$, and the two stroke times, $\tau_h$ and $\tau_c$. Of these, we succeeded in eliminating three, by explicitly using Fourier's law for the heat transfer, Eqs.~\eqref{eq:fourier_hot} and \eqref{eq:fourier_cold}, and the explicit expressions for the entropy and the internal energy \eqref{eq:harm_ent}. Therefore, one would not expect to obtain the same result \eqref{eq:eff_result} for other working substances such as the quantum harmonic oscillator.

\paragraph{Quantum Brownian engine.}

Next, we will analyze a quantum harmonic engine in the ultra-weak coupling limit \cite{Spohn1978}. This situation is similar to the model studied in Ref.~\cite{Rezek2006}, however in the present case we will \emph{not} have to solve the full quantum dynamics.

Accordingly, the internal energy reads
\begin{equation}
\label{eq:qu_energy}
E=\frac{\hbar\omega}{2}\coth{\left(\frac{\hbar\omega }{2 k_B T}\right)}
\end{equation}
and the entropy becomes
\begin{equation}
\label{eq:qu_entropy_CA}
\frac{S}{k_B}=\frac{\hbar\omega }{2 k_B T}\coth{\left(\frac{\hbar\omega }{2 k_B T} \right)}-\ln{\left[\frac{1}{2}\sinh{\left(\frac{\hbar\omega }{2 k_B T}\right)}\right]}\,.
\end{equation}

\begin{figure}
\centering
{\bf (a)} \hskip0.5\textwidth {\bf (b)}\\
\includegraphics[width=.5\textwidth]{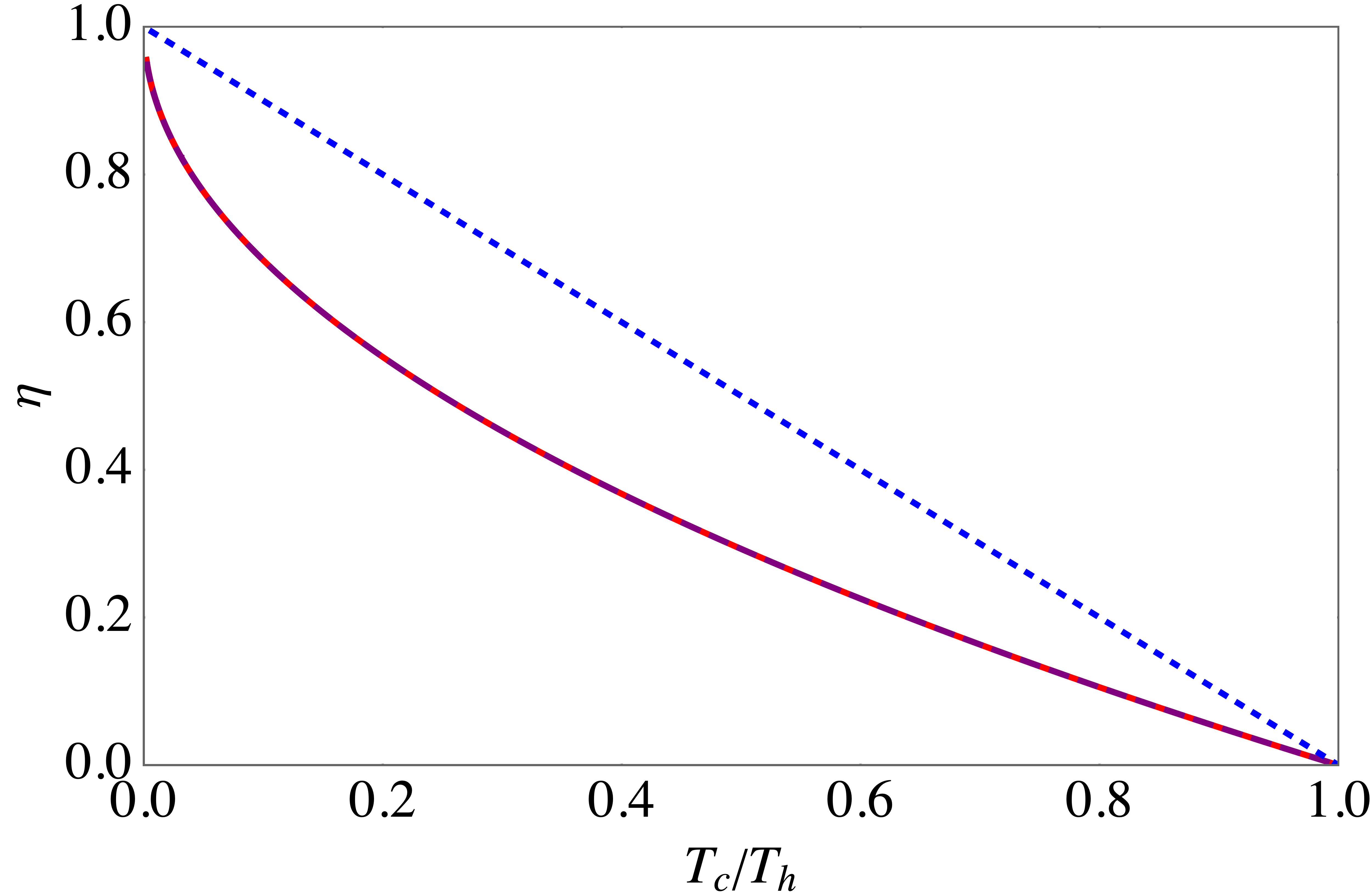}~\includegraphics[width=.5\textwidth]{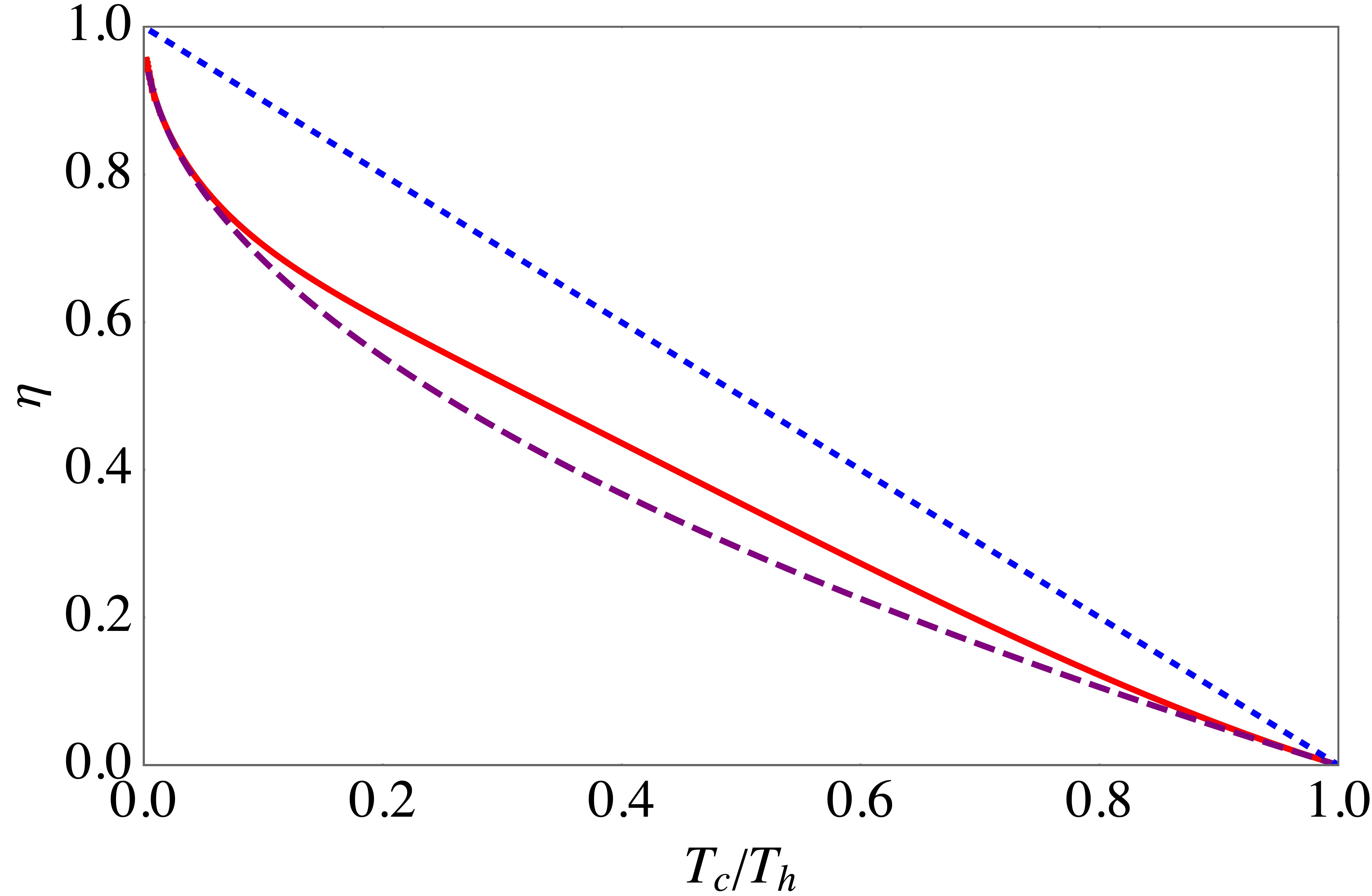}
\caption{\label{fig:efficiency_high} {\bf (a)} Efficiency of the endoreversible Otto cycle at maximal power (red, solid line), together with the Curzon-Ahlborn efficiency (purple, dashed line) and the Carnot efficiency (blue, dotted line) in the high temperature limit, $\hbar\omega_f/(k_B T_c) =0.1$. {\bf (b)} Efficiency of the endoreversible Otto cycle at maximal power (red, solid line), together with the Curzon-Ahlborn efficiency (purple, dashed line) and the Carnot efficiency (blue, dotted line) in the deep quantum regime, $\hbar\omega_f/(k_B T_c) =10$. Other parameters are $\hbar=1$, $\alpha_c=1$, $\alpha_h=1$, and $\zeta=1$.}
\end{figure}

Despite the functional form of $S$ being more involved, we notice that the four temperatures and the two frequencies are still related by the same Eq.~\eqref{eq:temp}. Thus, it can be shown \cite{Rezek2006} that the efficiency of an endoreversible Otto cycle in a quantum harmonic oscillators also reads,
\begin{equation}
\label{eq:eff_qu}
\eta=1-\kappa\,.
\end{equation}

Following the analogous steps that led to Eq.~\eqref{eq:power_clas} we obtain for the power output of an endoreversible quantum Otto engine,
\begin{equation}
\label{eq:power_qu}
\begin{split}
&P=\frac{\hbar\omega_f}{2}\,\frac{1-\kappa}{\zeta\left(\tau_c+\tau_h\right)}\, \mrm{csch}{\left[\frac{\hbar\omega_f\,\kappa}{2}\,\frac {e^{\alpha_c\tau_c+\alpha_h\tau_h}-1}{T_c \left(e^{\alpha_c\tau_c}-1 \right)+\kappa T_h\,e^{\alpha_c\tau_c}\left(e^{\alpha_h\tau_h}-1\right)}\right]}\\
&\times\mrm{csch}{\left[\frac{\hbar\omega_f\,\kappa}{2}\, \frac{e^{\alpha_c\tau_c+\alpha_h\tau_h}-1}{T_c\,e^{\alpha_h\tau_h}\left(e^{\alpha_c\tau_c}-1\right)+\kappa T_h\left(e^{\alpha_h\tau_h}-1\right)}\right]}\\
&\times\,\mrm{sinh}{\left[\frac{\hbar\omega_f\,\kappa}{2}\frac{\left(\kappa T_h-T_c\right)\left(e^{\alpha_c\tau_c+\alpha_h\tau_h}-1\right)\left(e^{\alpha_h\tau_h}-1\right)\left(e^{\alpha_c\tau_c}-1\right)}{\left(T_c \left(e^{\alpha_c\tau_c}-1 \right)+\kappa T_h\,e^{\alpha_c\tau_c}\left(e^{\alpha_h\tau_h}-1\right)\right)\left(T_c\,e^{\alpha_h\tau_h}\left(e^{\alpha_c\tau_c}-1\right)+\kappa T_h\left(e^{\alpha_h\tau_h}-1\right)\right)}\right]}
\end{split}
\end{equation}
where we set $k_B=1$. We immediately observe that in contrast to the classical case \eqref{eq:power_clas} the expression no longer factorizes. Consequently, the value of $\kappa$, for which $P$ is maximal does depend on the stroke times $\tau_h$ and $\tau_c$.

Due to the somewhat cumbersome expression \eqref{eq:power_qu}  the maximum of $P(\kappa,\tau_h,\tau_c)$ has to be found numerically. Fig.~\ref{fig:efficiency_high} {\bf (a)} illustrates the efficiency at maximal power in the high temperature limit, $\hbar\omega_f/(k_B T_c)\ll 1$. Consistently with our classical example, the efficiency is given by Eq.~\eqref{eq:eff_result}, which was also found in Ref.~\cite{Rezek2006} for quasistatic cycles.

Figure~\ref{fig:efficiency_high} {\bf (b)} depicts the efficiency at maximal power \eqref{eq:eff_qu} as a function of $T_c/T_h$ in the deep quantum regime,  $\hbar\omega_f/(k_B T_c)\gg 1$. In this case, we find that the quantum efficiency is larger than the Curzon-Ahlborn efficiency \eqref{eq:eff_result}. From a thermodynamics' point-of-view this finding is not really surprising since already in reversible cycles the efficiency strongly depends on the equation of state. The latter results were first published in Ref.~\cite{Deffner2018Entropy}.

It is then natural to ask whether this additional efficiency can be exploited, or in other words how to extract useful work from purely quantum resources.

\section{Work extraction from quantum systems}
\label{sec:WorkExtraction}
The positive work condition outlined previously establishes a strict requirement for a given thermodynamic cycle to be able to produce useful work. In the above  discussion, our working substance was a single quantum system, either a qubit or a harmonic oscillator, and we obtained a net output by performing work on this substance. At variance with this picture, in this section we depart from examining thermodynamic cycles and rather focus on under what conditions useful work can be extracted through unitary transformations on $d$-dimensional systems acting as {\it quantum batteries}~\cite{Allahverdyan2004, WorkExtractionPRE, WorkExtractionPRL, WorkExtractionJPhysB, BatteriesNJP, BatteriesPRL, CampisiPRL}.  Parts of this section were originally published in Ref.~\cite{WorkExtractionJPhysB}.

\begin{figure}
\begin{mdframed}[roundcorner=10pt]
\begin{minipage}[l]{.4\textwidth}
\begin{center}
\includegraphics[height=5cm]{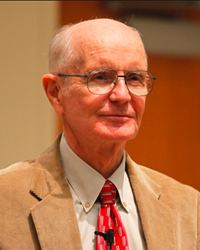}
\end{center}
\end{minipage}
\hfill
\begin{minipage}[r]{.58\textwidth}
Marlan Scully:\\ \emph{The deep physics behind the second law of thermodynamics is not violated; nevertheless, the quantum Carnot engine has certain features that are not possible in a classical engine \cite{Scully2003}. }
\end{minipage}
\end{mdframed}
\end{figure}

A quantum battery is a $d$-dimensional quantum system whose Hamiltonian
\begin{equation}
\label{eq:quantumbattery}
H_B=\sum_{i=0}^{d-1} \epsilon_i \ketbrad{i} \qquad \text{with}~~\epsilon_i < \epsilon_{i+1} 
\end{equation}
where, for simplicity, we have assumed the spectrum to be non-degenerate (however this is not a requirement for what follows). For a generic state of the battery, $\rho$, the internal energy is simply given by $\tr{\rho H_B}$. In order to extract work from a quantum battery, we require a process that transforms the initial state into an energetically lower state, and furthermore, this process must be cyclic (and reversible)~\cite{WorkExtractionPRE}. This last condition is particularly important: one can change the energy of the system through the same adiabatic processes present in the Otto cycle, however this only affects the internal energy and thus requires the full thermodynamic cycle in order to extract anything useful. However, if the process is such that $H_B \to H_B+V_t \to H_B$, where $V_t$ is some external potential which is switched on during the discharging of the battery, it is then clear that the Hamiltonian of the system is invariant before and after the process, but if the final state has a lower energy, the process has extracted work from the battery.

To formalize these ideas we must introduce the notion of {\it passivity}~\cite{PassiveStates}. Let us assume that between $t=0$ and $t=\tau$, the external potential $V_t$ is switched on such that the battery Hamiltonian is given by $H_B+V_t$ for $t\in(0,\tau)$ and $H_B$ otherwise. Passive states, $\rho_P$, are those for which no work can be extracted through this cyclic unitary process. These states can be written in the energy eigenbasis of $H_B$ and are of the form
\begin{equation}
\label{eq:passivestate}
\rho_P=\sum_{i=0}^{d-1} p_i \ketbrad{i} \qquad \text{with}~~p_i > p_{i+1}~\text{and}~\sum_i p_i =1.
\end{equation}
The fact that no work can be extracted from these states through the cyclic unitary process can be intuitively understood: In Eq.~\eqref{eq:passivestate} the populations of the energy levels are in an decreasing order, i.e. the ground state has the largest occupation, followed by the first excited state, followed by the second excited state, and so on. The energy eigenvalues and eigenstates of the battery are the same at the start and the end of the protocol. Therefore, since the entropy is conserved, the final state can, at most, be a reordering of these populations. Therefore any operation that exchanges the populations can only raise the energy of the system as it necessarily will involve increasing the occupation of a higher eigenenergy state, at the cost of correspondingly lowering the occupation of a lower energy eigenstate. For single quantum batteries this is always the case, however the situation becomes more involved for uncoupled arrays as we shall see in the proceeding sections. Thermal states in Gibbs form are a special subset referred to as {\it completely passive states}, they remain passive even if one has multiple copies.

Clearly, in order for work to be extractable we require so-called {\it active states} of the form
\begin{equation}
\label{eq:activestate}
\rho_A=\sum_{i=0}^{d-1} p_i \ketbrad{i} \qquad \text{with}~~p_i < p_{i+1}~~\text{for~at~least~one~value~of~}i.
\end{equation}
The amount of extractable work is called {\it ergotropy} and defined as \cite{Allahverdyan2004}
\begin{equation}
\label{eq:ergotropy}
\mathcal{W} =  \tr{\left(\rho_A - U_\tau \rho_A U_\tau^\dagger \right)H_B },
\end{equation}
where $U$ is the unitary evolution operator. At this point it should be clear that no more work can be extracted from the battery once the final state is passive, i.e. $U_\tau \rho_A U_\tau^\dagger \to \rho_P$, which is achieved for a cyclic unitary process which reorders the energy level occupations in increasing order. Thus, it is clear that the action of the potential will be a collection of swap operations between the energy levels such that the required ordering is achieved.

\subsection{Work extraction from arrays of quantum batteries}
Consider now a register of $n$ identical quantum batteries
\begin{equation}
h_0 =\left(H_B \otimes \id^{\otimes (n-1)}\right) + \left( \id \otimes H_B \otimes \id^{\otimes (n-2)} \right) +\dots+ \left( \id^{\otimes (n-1)} \otimes H_B \right)
\end{equation}
with $H_B$ given by Eq.~\eqref{eq:quantumbattery}. To extract work from these batteries, the register evolves according to the cyclic Hamiltonian
\begin{equation}
h(t) = h_0 + V_t
\end{equation}
where $V_t$ is a potential that acts on the whole register for $t\in(0,\tau)$. It is now a question of what form the potential must take in order to extract the maximum work stored in the batteries, and how this compares to a `classical' work extraction strategy. At this point it is important to clarify that we refer to a `classical' strategy as one which acts locally on each battery, and therefore clearly, if for a single battery starting from the active state, $\rho_A$, the maximal ergotropy is $\mathcal{W}_\text{max}$, then such a classical strategy can extract at most $n \mathcal{W}_\text{max}$ from our register. 

We can now clarify the important difference between {\it passive} and {\it completely passive} states. A single-battery state $\rho$ is called passive with respect to $H_B$ if no energy can be extracted from it during a cycle; it is called completely passive if $\rho^{\otimes n}$ is passive with respect to $h_0$ for all $n$. Passive states are not necessarily completely passive, a remarkable exception being represented by qubits, due to the one-to-one correspondence between the entropy of the state and its ordered eigenvalues. This also implies that, working with qubits, the extractable work achieved from the classical strategy can never be exceeded. As we will demonstrate, the situation becomes more subtle for higher dimensional quantum batteries, where given a register of passive batteries, whether the ensemble is completely passive depends on the particular energy splittings and occupations.

\paragraph{Example: Qubit quantum batteries.}
The above notions can be seen clearly from the case of two qubit batteries initially prepared in
\begin{eqnarray}
\rho & = & \left( p_0 \ket{0} + p_1 \ket{1} \right)^{\otimes 2} \nonumber \\
        & =& p_{0}^2\ketbrad{00}+p_{0}p_{1}(\ketbrad{01}+\ketbrad{10})+p_{1}^2\ketbrad{11},\label{omega}
\end{eqnarray}
with $p_0+p_1=1$, in the presence of the Hamiltonian 
\begin{equation}
h_0=2 \epsilon_0 \ketbrad{00}+(\epsilon_0+\epsilon_1)(\ketbrad{01}+\ketbrad{10})+2 \epsilon_1 \ketbrad{11},
\end{equation}
where $\epsilon_0<\epsilon_1$. As said previously, the case of qubits is somewhat special, as single-battery passive states remain passive in the multi-battery scenario. We can see this explicitly by direct calculation. For a single qubit state to be passive we require $p_0>0.5$. As the single excitation subspace is degenerate for the two-qubits and $p_1=1-p_0$, it follows that in Eq.~\eqref{omega} we have $p_0^2 > p_0-p_0^2 > (1-p_0)^2$ for any $p_0>0.5$, and hence all passive qubit states are also completely passive. 

Conversely, the state $\rho$ in Eq.~(\ref{omega}) is active, that is, it is possible to extract work from it, provided that $p_{0}<p_{1}$. The amount of work extracted is maximum if, at the end of the cycle,
\begin{equation}
\rho(\tau)=p_{1}^2 \ketbrad{00}+p_{0}p_{1}(\ketbrad{01}+\ketbrad{10})+p_{0}^2\ketbrad{11}.
\end{equation}
In this case using Eq.~\eqref{eq:ergotropy} we find the maximum ergotropy is 
\begin{equation}
{\cal W}_{\max}=2(\epsilon_1-\epsilon_0)(1-2 p_0). 
\end{equation}
It is immediate to check that ${\cal W}_{\max}$ is twice the maximal work that could be extracted from a single qubit battery with the same $p_1$ value, thus confirming that for arrays of qubit batteries it is not possible to outperform the classical strategy. As stated previously, in order to achieve this amount of ergotropy the cyclic process must arrange the energy occupations in increasing order through swap operations~\cite{WorkExtractionPRE, WorkExtractionPRL, WorkExtractionJPhysB}. The reordering process can be performed either (i) in three steps or (ii) in one single step. In the case (i) the procedure consists of first swapping, for instance, $|00\rangle$ and $|10\rangle$, then  $|10\rangle$ and $|11\rangle$, and, finally,  $|00\rangle$ and $|10\rangle$ again. In the case (ii) , there is a direct swap between $|00\rangle$ and $|11\rangle$. However, as protocol (i) requires three unitary operations each taking a time $\tau$ to be completed, it follows that the direct swap case is preferable as it can be implemented in a single operation. The swaps are non-local operations acting on the two batteries, and therefore, in principle, are entangling operations. Interestingly, following the multi-step strategy, the state $\rho$ remains separable at all times~\cite{WorkExtractionPRL}, while, as a consequence of the direct swap, $\rho$ may be entangled (depending on the particular value of $p_1)$ for some intermediate times between $t=0$ and $t=\tau$. Thus, the presence of entanglement is related to the speed of the extraction process. By performing the direct swap the maximal ergotropy is achieved three times faster than using a protocol that does not generate any entanglement. Therefore, we can see that by allowing the batteries to establish strong quantum correlations during the extraction process the power can be significantly increased. It should be noted that, although one can extract maximal work without ever entangling the two batteries, the same process nevertheless leads to the establishment of other forms of genuine quantum correlations in the form of quantum discord~\cite{WorkExtractionJPhysB}. \bigskip

The situation becomes significantly richer when we extend the dimensionality of our batteries. This is due to the fact that for quantum batteries with anharmonic spectra and dimension $d>2$ passive states are not necessarily completely passive. As we will show in the following example, under the right conditions it is possible to extract more work from an array of quantum batteries compared to the corresponding classical strategy by exploiting non-local entangling operations. These operations are needed in order for the array of batteries to reach the completely passive state which, when the classical strategy is outperformed, necessarily implies that classical correlations are shared between the batteries~\cite{WorkExtractionJPhysB}.

\paragraph{Example: Qutrit quantum batteries.}
The second case we analyze in detail is for two qutrits. The initial state is  
\begin{equation}
\rho=(p_{0}|0\rangle\langle 0|+p_{1}|1\rangle\langle 1|+p_{2}|2\rangle\langle 2|)^{\otimes 2},
\end{equation}
with $p_0\le p_1\le p_2$, while the Hamiltonian is
\begin{eqnarray}
h_0&=&2 (\epsilon_0 |00\rangle\langle 00|+ \epsilon_1 |11\rangle\langle 11|+ \epsilon_2 |22\rangle\langle 22|)\nonumber\\
&+&(\epsilon_0+\epsilon_1)(|01\rangle\langle 01|+|10\rangle\langle 10|)\nonumber\\&+&(\epsilon_0+\epsilon_2)(|02\rangle\langle 02|+|20\rangle\langle 20|)\nonumber\\&+&(\epsilon_1+\epsilon_2)(|12\rangle\langle 12|+|21\rangle\langle 21|).
\end{eqnarray}
Let us assume also that $p_0<p_1<p_2$, and therefore our state is initially active. Using the classical, i.e. local, protocol where the populations of each battery are reordered independently, the maximum work extracted is achieved by simply swapping populations of $\ket{0}$ and $\ket{2}$, and is therefore independent of $p_1$
\begin{equation}
\label{eq:equtritergo}
{\cal W}_{{\rm cl}}=2(\epsilon_2-\epsilon_0)(p_2-p_0).
\end{equation} 
Taking into account global unitary transformations, i.e. swap operations that act on both qutrits and are therefore entangling operations, the final order of the eigenstates leading to the maximum ergotropy depends on the value of $p_1$. There exists a threshold value $p_1^{\rm th}$, obtained imposing $p_1^2=p_0 p_2$, such that, for $p_1\le p_1^{\rm th}$, the maximum extractable work does not exceed the classical limit. This is no longer true for $p_1 > p_1^{\rm th}$, where, besides the swaps achievable through the classical strategy, we also need a further swap between $|11\rangle$ and either $|02\rangle$ or $|20\rangle$, and in doing so results in a final state with classical correlations shared between the two batteries. 

Let us make these ideas more concrete with a specific example. Let us fix the ground, first, and second excited energies $\epsilon_0=0,\;\epsilon_1=0.579,\;\epsilon_2=1$. Therefore in the Hilbert space spanned by the two qutrit batteries, the respective energies for the various eigenstates appropriately ordered are 
\begin{align}
&\ket{00} \to 2\epsilon_0 = 0 &                         &\ket{01} \to \epsilon_0+\epsilon_1 = 0.579 &          &\ket{10} \to \epsilon_0+\epsilon_1 = 0.579 & \nonumber \\
&\ket{02} \to \epsilon_0+\epsilon_2 = 1 &         &\ket{20} \to \epsilon_0+\epsilon_2 = 1 &                &\ket{11} \to 2\epsilon_1 = 1.158 &  \\
&\ket{12} \to \epsilon_1+\epsilon_2 = 1.579  & &\ket{21} \to \epsilon_1+\epsilon_2 = 1.579 &         &\ket{22} \to 2\epsilon_2 =  2 & \nonumber 
\end{align}
Fixing $p_0=0.224$ and assuming $p_0<p_1<p_2$ the initial state of the two batteries, written in the ordered energy eigenbasis, is
\begin{eqnarray}
\rho &= & 0.224^2 \ketbrad{00}  + 0.224 p_1 ( \ketbrad{01} + \ketbrad{10}) + 0.224 p_2 (\ketbrad{02} + \ketbrad{20}) \nonumber \\ 
          && + p_1^2 \ketbrad{11} + p_1 p_2 (\ketbrad{12} +\ketbrad{21}) + p_2^2 \ketbrad{22}.
\end{eqnarray}
Clearly $p_2^2 > p_1 p_2 > p_1^2$ and $0.224 p_2>0.224 p_1>0.224^2$. Therefore the state is clearly active, however whether it is possible to extract more work than the classical strategy depends on the value of $p_1$. Indeed, notice if the batteries are discharged independently, such that the ergotropy is given by Eq.~\eqref{eq:equtritergo}, the final state is
\begin{eqnarray}
\rho_\text{cl} &=& p_2^2\ketbrad{00} + p_1 p_2 (\ketbrad{01} + \ketbrad{10}) + 0.224 p_2 (\ketbrad{02} + \ketbrad{20}) \nonumber \\
                   &&+ p_1^2 \ketbrad{11} + 0.224 p_1 (\ketbrad{12} +  \ketbrad{21}) + 0.224^2  \ketbrad{22}.
\end{eqnarray}
This state is passive if and only if $p_1^2< 0.224 p_2$, i.e. $p_1\lesssim0.32$. For larger values of $p_1$, in order for the state to be passive a swap between $|11\rangle$ and $|02\rangle$ or $|20\rangle$ is required, which leads to classical correlations in the final state of the two batteries. 

This allows us to establish a simple criterion to determine whether the classical limit is beaten or not based on the use of classical correlations: the work extracted is $n$ times the work that could have been extracted from a single battery if and only if the final state is the tensor product of single-battery states. In fact, this condition is necessary because any product state could be obtained by local manipulation of the initial one; on the other hand, it is sufficient because local unitary operations map product states onto product states. Then, classical correlations can be used to measure the distance from the set of product states.

Therefore by processing multiple quantum batteries at once, under the correct conditions, it is possible to extract more work than by classically (locally) processing each battery. Furthermore, the maximal amount of ergotropy is achievable without generating any quantum entanglement during the process by performing a sequence of ordered swaps between the energy levels. Faster discharge, and therefore more power, is achieved when global operations are applied, as less swap operations are required. In such a case, strong quantum correlations are established during the discharge, thus implying that entanglement and other quantum correlations are useful resources in boosting the power of quantum batteries. 

\subsection{Powerful charging of quantum batteries}
We have seen that entangling operations can boost the power output of arrays of quantum batteries by facilitating a faster extraction of work. While one is generally less concerned with how quickly a battery is discharged, the complementary action of charging these devices is evidently important~\cite{BatteriesNJP, BatteriesPRL}. The charging of quantum batteries can obviously be achieved by applying the same unitary operations needed for work extraction to the initially passive state, and therefore by exploiting the same global entangling operations, a faster, more powerful, charging of quantum batteries is achieved as the two process are essentially equivalent.

Such an insight can be made more rigorous following the approach of Binder \etal~\cite{BatteriesNJP}. Consider a quantum battery with Hamiltonian Eq.~\eqref{eq:quantumbattery}, and let us assume that it is initially in the completely passive pure state $\rho = \ketbrad{0}$, i.e. its ground state. In order to charge the state to its maximally active state we must swap the population of $\ket{0}$ with $\ket{d-1}$. Due to the energy-time uncertainty relation, there exists a minimal time to connect orthogonal states, known as the quantum speed limit\footnote{For a recent and comprehensive review on \emph{quantum speed limits} we refer to our Topical Review.~\cite{QSLReview}},
\begin{equation}
\tau_\text{QSL} = \frac{\pi}{2 \text{min}\{E, \Delta E\}}
\end{equation}
with $E$ and $\Delta E$ being the time averaged energy and variance of the generator of the dynamics, i.e. $H_B+V_t$ evaluated over the ground state\footnote{The factor $\pi/2$ can be intuitively recovered by simply considering the application of the unitary corresponding to a Hamiltonian which swaps states $\ket{0}$ and $\ket{d-1}$.}. At this point it is physically reasonable to assume that the amount of energy available while charging our battery is finite, and therefore we can set a bound on the generator of the charging process
\begin{equation}
\label{eq:batteryconst}
\| H_B + V_t \| \leq E_\text{max}.
\end{equation} 
For a single quantum battery subject to this constraint the minimal charging time is simply $\tau=\pi/(2E_\text{max})$. Now let us consider an array of $n$ batteries. Evidently, charging them in parallel subject to the same constraint, Eq.~\eqref{eq:batteryconst}, implies a minimal charging time
\begin{equation}
\tau_\text{para} = \frac{n \pi}{2 E_\text{max}}.
\end{equation}
Allowing for the global swap operation, that still satisfies constraint \eqref{eq:batteryconst}, given by the Hamiltonian
\begin{equation}
H_\text{opt} =E_\text{max} \left( \ket{\bf 0}\!\bra{\bf{d-1}} + \ket{\bf{d-1}}\!\bra{\bf 0} \right)
\end{equation}
with $\ket{\bf 0}=\ket{0}^{\otimes n}$ and $\ket{\bf d-1}=\ket{d-1}^{\otimes n}$, the minimal time to charge the complete array is then
\begin{equation}
\tau_\text{opt} = \frac{\pi}{2 E_\text{max}},
\end{equation}
i.e. the by allowing for the generation of entanglement among the batteries allows for an $n$ fold increase in the charging power.

\section{Quantum decoherence and the tale of quantum Darwinism}
\label{sec:Decoherence}
As should be evident by now, the presence of an environment is often vital when discussing the thermodynamic aspects of a given system or process. Despite this, we have so far mainly approached the environment in an {\it ad hoc} manner. In the following we look to more carefully assess the role the environment plays in establishing a consistent thermodynamic framework for quantum systems, aspects of which were originally published in Ref.~\cite{CampbellPRA2018}.

\subsection{Work, heat, and entropy production for dynamical semigroups}
\label{chap2:semigroup}
A widely used tool to examine the open dynamics of a quantum system is to describe the evolution through a quantum dynamical semigroup. Without laboring into the particular mathematical details allowing one to arrive at this description, we refer to \cite{BreuerBook} for an expansive introduction, we can nevertheless gain some intuitive motivation for the validity of this approach. A crucial assumption underlying the framework is that the system and its environment are initially factorized
\begin{equation}
\rho_{\mathcal{S}\mathcal{E}}(0) = \rho_\mathcal{S}(0) \otimes \rho_\mathcal{E}(0)
\end{equation} 
and that the total system+environment state evolves unitarily according to some Hamiltonian $H_T=H_\mc{S} + H_\mc{E} + H_\mc{I}$, with $H_\mc{S}$ and $H_\mc{E}$ the free Hamiltonians for the system and environment, and $H_\mc{I}$ describes their mutual interaction. The evolution of the system can be described 
\begin{eqnarray}
\frac{d}{dt} \rho_\mathcal{S}(t) & = & -i/\hbar\, \ptr{\mathcal{E}}{ H_T, \rho_{\mathcal{S}\mathcal{E}} } \nonumber \\
                               & = & \mathcal{L}\rho_\mathcal{S} \label{eq:MEreduced}
\end{eqnarray}
where $\mathcal{L}$ is the generator of the dynamics for the system alone, and accounts for the effect of the environment. Assuming the evolution of the system can be written in terms of a linear map, $\Phi$ acting from $0$ to $t$, applied to the initial state of the system which satisfies a composition law,
\begin{equation}
\rho_\mathcal{S} (t) = \Phi(t,0) \rho_\mathcal{S}(0) = \Phi(t,s)\Phi(s,0)\rho_\mathcal{S}(0),   \hspace{2em} t \geqslant s \geqslant 0,
\end{equation}
then implies a ``memoryless" environment. This means that the characteristic time scales describing any environment correlation function decays faster than the dynamics of the system, and therefore can be viewed as being effectively invariant due to the interaction with the system. For weak coupling between system and environment, it is then possible to express Eq.~\eqref{eq:MEreduced} in the so-called Lindblad form
\begin{equation}
\label{eq:MasterEqGen}
\frac{d}{dt}\rho_\mathcal{S} =\mathcal{L}\rho_\mathcal{S} =  -i/\hbar\,[H, \rho_\mathcal{S}] + \sum_{k} \gamma_k \left( A_k \rho_\mathcal{S} A_k^\dagger -\frac{1}{2}A_k^\dagger A_k \rho_\mathcal{S} - \frac{1}{2} \rho_\mathcal{S} A_k^\dagger A_k \right),
\end{equation}
where $A_k$ are Lindblad operators, $\gamma_k$ are dissipation rates. Note that $H$ is not necessarily the system Hamiltonian, but can account for other interactions with the environment, time dependent driving, etc.~\cite{BreuerBook}.

While a diverse range of physical situations can be described by Eq.~\eqref{eq:MasterEqGen}, we will restrict to the case when the environment is a thermal bath. In the case of a single qubit with frequency $\omega_S$ the corresponding master equation is given by
\begin{equation}
\begin{split}
\label{eq:thermalME}
\frac{d}{dt}\rho_\mathcal{S} &= -i/\hbar\,[H , \rho_\mathcal{S}] + \gamma \left( \sigma_- \rho_\mathcal{S} \sigma_+  -\frac{1}{2} \sigma_+\sigma_- \rho_\mathcal{S} -\frac{1}{2} \rho_\mathcal{S}\sigma_+\sigma_- \right) \\
&\quad+  \Gamma \left( \sigma_+ \rho_\mathcal{S} \sigma_-  -\frac{1}{2} \sigma_-\sigma_+ \rho_\mathcal{S} -\frac{1}{2} \rho_\mathcal{S}\sigma_-\sigma_+ \right).
\end{split}
\end{equation}
where $\gamma$ and $\Gamma$ fix the dissipation rates, while $\sigma_+=\sigma_-^\dagger= \ket{1}\!\bra{0}$ are spin raising and lowering operators (arbitrary dimensional systems can readily be examined by considering the appropriate operators). It is important to note that Eq.~\eqref{eq:thermalME} is constructed such that the stationary state is a Gibbsian equilibrium state \cite{BreuerBook}.

The inverse temperature of the bath is
\begin{equation}
\beta = \frac{1}{2\hbar\omega_\mathcal{S}} \ln \left( \frac{\gamma}{\Gamma} \right).
\end{equation}
where $\omega_\mathcal{S}$ corresponds to the natural frequency of the system. Clearly, for $\gamma>\Gamma$ the bath exhibits a well defined positive temperature, however, we remark that even for $\gamma<\Gamma$ the resulting dynamics is still well defined. From the first law \eqref{eq:quantum_energy} it follows that the work and heat are given by 
\begin{eqnarray}
\la W\ra &= &\int \tr{ \dot{H}_\mathcal{S}  \rho_\mathcal{S} } dt, \label{eq:work_2} \\
\la Q \ra &= &\int \tr{ H_\mathcal{S}  \dot{\rho}_\mathcal{S} } dt. \label{eq:heat_2}
\end{eqnarray}
Clearly, when the system's Hamiltonian is time-independent no work is done and all energy changes are due to heat exchange between the system and the thermal bath. 

Turning our attention to the entropy production, following from Sec.~\ref{sec:quantum_entropy}
\begin{equation}
  \langle \Sigma \rangle = \Delta S- \beta \la Q\ra,  \label{eq:irr}
\end{equation}
where $\Delta S$ is the change in entropy of the system, so that $\langle \Sigma \rangle$ indeed provides the contribution in entropy change which cannot be traced back to a reversible heat flow. Assuming the initial and final times of the transformation to be $0$ and $t$, respectively, Eq.~\eqref{eq:irr} can be equivalently rewritten as
\begin{eqnarray}
  \langle \Sigma \rangle & = & S ( \rho (0)|| \rho_{\beta} ) -S ( \rho (t)|| \rho_{\beta} ) ,  \label{eq:irr1}
\end{eqnarray}
where $\rho_{\beta}$ denotes a Gibbs state for the system at inverse temperature $\beta$, and we have used the quantum relative entropy $ S ( \rho ||w)= \tr{  \rho \ln   \rho } - \tr{   \rho \ln  w}$. As noted above, for dynamics described by Eq.~\eqref{eq:thermalME}, when $H=H_\mathcal{S}$, the dynamics admits $\rho_{\beta}$ as an invariant state, i.e. the system thermalizes with the environment in the long-time limit. The irreversible entropy production as defined by Eq.~(\ref{eq:irr1}) is a positive quantity, in accordance with the second law. One can further consider the quantity
\begin{eqnarray}
  \sigma (t) & = & - \frac{d}{dt} S ( \rho (t)|| \rho_{\beta} ) , 
  \label{eq:rate1}
\end{eqnarray}
which can be naturally interpreted as the (instantaneous) entropy production rate. Due to the fact that the relative entropy is a contraction under the action of a completely positive trace preserving map~\cite{Lindblad1975a}, and as recently shown also for a positive trace preserving map~{\cite{Reeb2017a}}, the entropy production is also a positive quantity. 

It is important to note however, that the form of the steady state of $S$ is dependent on the particular details describing $H$ in Eq.~\eqref{eq:thermalME} and the system may reach a non-equilibrium steady state with respect to the bath. In this case the very existence of an invariant state of the dynamics, say $\bar{\rho}$, not necessarily in Gibbs form, is sufficient to introduce via
\begin{equation}
  \langle \bar{\Sigma} \rangle  = S ( \rho (0)|| \bar{\rho} ) -S ( \rho(t)|| \bar{\rho} ) ,  \label{eq:irr2}
\end{equation}
a quantifier of entropy production which is always positive, and whose associated entropy production rate $\bar{\sigma} (t)$ is also positive provided the dynamics is $P$-divisible~\cite{BreuerBook}. For a quantum dynamical semigroup with generator, $\mathcal{L}$, the entropy production rate is given by the explicit expression
\begin{equation}
  \bar{\sigma} (t) = \tr{ \mathcal{L} [ \rho (t)]( \ln \bar{\rho} - \ln   \rho (t)) } ,  \label{eq:rate2}
\end{equation}
whose positivity, following from the divisibility of the dynamics, is also known as Spohn's inequality~\cite{Spohn, Alicki1979}
\begin{equation}
  \tr{ \mathcal{L} [ \rho (t)]( \ln   \bar{\rho} - \ln   \rho (t))} \geqslant  0. \nonumber
\end{equation}
Both definitions for the entropy production rate provide convex functions of the system state, thus ensuring stability, and they are positive for dynamics arising from a semigroup. However, only $\sigma (t)$ defined in Eq.~(\ref{eq:rate1}) via its relation to Eq.~(\ref{eq:irr1}), and therefore heat transfer, can be directly connected to a thermodynamic interpretation. It is worth noting that for non-Markovian dynamics defining the entropy production becomes significantly trickier~\cite{StrasbergPRE, CampbellPRA2018, MarcantoniSciRep, MarcusArXiv}.

\subsection{Entropy production as correlation}
While Sec.~\ref{sec:quantum_entropy} and the previous section outlined how to determine a meaningful quantifier for the entropy production, nevertheless a clear microscopic understanding of what the entropy production captures has been so far avoided. In this section we follow the results of Esposito, Lindenberg and Van den Broeck~\cite{EspositoNJP} to show the connection between established system-environment correlations and the associated entropy production. We remark, that this insight stems from modelling both the system and the environment as a finite dimensional quantum systems, and as such does not explicitly rely on a weak coupling approximations etc.

Indeed the difficultly in elucidating precisely what the entropy production physically corresponds to lies in the fact that the default quantifier of entropy for a quantum system, the von Neumann entropy, is constant for unitary dynamics. Thus, when we consider the composite system+environment dynamics there is no entropy production. As we shall see, however, it is precisely this invariance that allows us to hone in on the system entropy alone and establish an elegant operational notion of the system entropy production in terms of correlations. 

Following Ref.~\cite{EspositoNJP}, consider the environment to be a collection of finite dimensional quantum systems all taken to be initially canonical thermal states
\begin{equation}
\rho_\mathcal{E}(0) = \bigotimes_i \rho^\text{eq}_{\mathcal{E}_i} = \bigotimes_i e^{-\beta_i H_{\mathcal{E}_i}}/Z_{\mathcal{E}_i}
\end{equation}
with $H_{\mathcal{E}_i}$ the Hamiltonian, $\beta_i$ the inverse temperature, and $Z_{\mathcal{E}_i}$ the partition function for the $i$th environmental sub-unit. We consider the same basic set up as in Sec.~\ref{chap2:semigroup}, namely the initial state of system and environment are factorized, i.e. $\rho(0)=\rho_\mathcal{S}(0) \otimes \rho_\mathcal{E}(0)$, and the total Hamiltonian describing the evolution is $H_T=H_\mathcal{S} + H_\mathcal{E} + H_I$, now with $H_\mathcal{E}=\sum_i H_{\mathcal{E}_i}$. By virtue of the invariance of the entropy of the composite system+environment, we can write
\begin{equation}
\label{eq:St}
\begin{aligned}
S(\rho(t))=S(\rho(0)) = & S(\rho_\mathcal{S}(0)) + \sum_i  S(\rho^\text{eq}_{\mathcal{E}_i})  \\
              = & -\tr{ \rho_\mathcal{S}(0) \ln \rho_\mathcal{S}(0) } -  \sum_i \tr{ \rho_{\mathcal{E}_i}^\text{eq} \ln \rho_{\mathcal{E}_i}^\text{eq}  }. \\
              = & -\tr {\rho_\mathcal{S}(0) \ln \rho_\mathcal{S}(0) } - \tr{ \rho_\mathcal{E}(0) \ln \rho_\mathcal{E}(0) }
\end{aligned}
\end{equation}
We can now examine the change in entropy for the system only and using Eq.~\eqref{eq:St} we have
\begin{equation}
\begin{split}
&\Delta S_\mathcal{S} (t)  =  S_\mathcal{S}(t) - S_\mathcal{S}(0)   \\ 
                      &\quad =  S_\mathcal{S}(t) - S(\rho(t)) + S(\rho_\mathcal{E}(0))  \\
                      &\quad =  S_\mathcal{S}(t) - S(\rho(t)) + S(\rho_\mathcal{E}(0)) + \tr{ \rho_\mathcal{E}(t) \ln \rho_\mathcal{E}(0) } - \tr{ \rho_\mathcal{E}(t) \ln \rho_\mathcal{E}(0) } \\
                      &\quad = - \tr{ \rho_\mathcal{S}(t) \ln \rho_\mathcal{S}(t)  } + \tr{ \rho(t) \ln \rho(t) } -\tr{ \rho_\mathcal{E}(t) \ln \rho_\mathcal{E}(0)  } + \tr{ (\rho_\mathcal{E}(t) - \rho_\mathcal{E}(0)) \ln \rho_\mathcal{E}(0) }   \\ 
                      &\quad =  \tr{ \rho(t) \ln \rho(t) } - \tr{ \rho(t) \ln \left( \rho_\mathcal{S}(t)\otimes\rho_\mathcal{E}(0) \right) } + \tr{ (\rho_\mathcal{E}(t) - \rho_\mathcal{E}(0)) \ln \rho_\mathcal{E}(0) } \\
                      &\quad=  S(\rho(t) \| \rho_\mathcal{S}(t)\otimes\rho_\mathcal{E}(0)) + \sum_i \beta_i Q_i. \label{eq:EntropyChange}
\end{split}
\end{equation}
We see that the second term in Eq.~\eqref{eq:EntropyChange} accounts for the heat flowing from the reservoir, thus corresponding to the reversible change in entropy, and therefore we conclude that the first term accounts for the irreversible entropy change. Hence we can write the system's entropy change as the sum of two well defined contributions stemming from the irreversible and reversible processes
\begin{equation}
\label{eq:EntropyProduction}
\Delta S_\mathcal{S}(t) = \Delta_i S(t) + \Delta_r S(t).
\end{equation}
Notice that by expressing the irreversible entropy production in terms of a relative entropy between the total evolved state and the tensor product of the marginal of the system with the initial environmental state highlights is origin: quantum entropy production is intimately related to correlations established between the system and environment. This connection can be made more rigorous considering when the environmental subunits remain in equilibrium at all times, i.e. $\rho_E(t) = \rho_E(0)$. Such a scenario is precisely inline with the dynamics governed by the master equation Eq.~\eqref{eq:thermalME}. Under such conditions, as the environment is assumed to remain invariant, the last term in Eq.~\eqref{eq:EntropyProduction} is zero and therefore the change in entropy of the system is entirely due to the irreversible contribution in Eq.~\eqref{eq:EntropyProduction}. Therefore the entropy production is exactly (up to a difference in sign) the correlations shared between the system and environment. 

\subsection{Quantum Darwinism: Emergence of classical objectivity}

\begin{figure}
\begin{mdframed}[roundcorner=10pt]
\begin{minipage}[l]{.4\textwidth}
\begin{center}
\includegraphics[height=5cm]{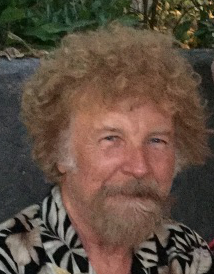}
\end{center}
\end{minipage}
\hfill
\begin{minipage}[r]{.58\textwidth}
Wojciech H. Zurek:\\ \emph{The only ``failure'' of quantum theory is its inability to provide a natural framework for our prejudices about the workings of the Universe  \cite{Zurek1991}.}
\end{minipage}
\end{mdframed}
\end{figure}

A common aspect of the ideas presented thus far has been the focus on the system of interest, while comparatively less attention has been paid to the environment. Indeed an interesting point rarely satisfactorily addressed through any of the techniques for modelling open quantum systems is determining how classicality emerges from the underlying quantum dynamics. More often than not, one implicitly subscribes to the old adage {\it ``shut-up and calculate"}, as in most circumstances we are only interested in the properties of a well defined system and therefore are well justified in effectively ignoring any environmental considerations. Regardless, it is no secret that quantum features have only been witnessed at small scales and when the systems are well isolated.  However, evidently there is nothing {\it a priori} preventing quantum superpositions existing for macroscopic objects. Wojciech H. Zurek then asked the question: Can we explain how classicality emerges while relying only on the known axioms of quantum mechanics? {\it Quantum Darwinism} offers one such explanation~\cite{ZurekRMP, ZurekNatPhys}. In essence, it relies on the notion of classical objectivity, i.e. that several observers will all agree on the outcome of a given measurement. The route to achieving this objectivity then further relies on the notion of redundant information encoding. In what follows we will briefly outline the basic tenets of quantum Darwinism while deliberately avoiding the more philosophical ramifications of the theory.

\paragraph{Pointer states and einselection.}
To begin we must establish the notion of pointer states of a system~\cite{ZurekRMP}. In quantum mechanics, any coherent superposition of legitimate states is also a meaningful quantum state. While this leads to an infinite number of possible configurations for any given system to exist in, not all such superpositions are equal in the eyes of decoherence. As noted in the previous section, when we explicitly consider the system and environment from the outset, such that the overall dynamics is completely unitary, we then see that the initial states of both the system and environment, and the particular details of their interaction, dictate the features exhibited by both during the dynamics. Pointer states are those system states that remain {\it unaffected} by the interaction with the environment. It is important to note however, that while the pointer states themselves are effectively untouched by the environment, the superposition of these pointer states is not, and will decohere by losing phase coherence. Therefore, the pointer states are a special subset of possible states of the system that are singled out by the nature of the system-environment interaction. We refer to this phenomena as environment-induced superselection or {\it einselection}.

\paragraph{Redundant encoding and quantum Darwinism.}
The paradigm for explaining classical emergence comes from treating the environment as a witness of the system's properties. In this regard the environment is elevated to an active participant in how we learn about a quantum system. Such a viewpoint can be intuitively grasped as follows: observations are rarely recorded by directly probing a given system, but rather by collecting information transmitted through some information carrier -- for example photons or phonons. While there will be many such individual information carriers, only a small fraction typically needs to be captured in order for us, the observer, to accurately record the observation. Equally, given two observers, they will both agree on the outcome when they independently intercept different fractions of these information carriers. This simultaneously implies a redundancy in the information carried by each individual photon/phonon, since two independent sets transmitted the same information, and the objectivity of the observation, as the two observers independently reach the same conclusion. 

To illustrate how this can occur, let us assume the system is a qubit in the initial state
\begin{equation}
\ket{\psi_\mathcal{S}} = \alpha \ket{\uparrow} + \beta \ket{\downarrow} \qquad \text{with}~\vert \alpha \vert^2+\vert \beta \vert^2=1,
\end{equation}
i.e. a superposition of the pointer states $\{\ket{\uparrow},~\ket{\downarrow}\}$, and we assume the environment is a collection of $N$ qubits each written in the computational basis $\{\ket{0},~\ket{1}\}$ all in the same initial state $\ket{\psi_{\mathcal{E}_i}}$. Thus, we start with an initially factorized state
\begin{equation}
\ket{\psi_{\mathcal{S}\mathcal{E}}}_i = \left( \alpha \ket{\uparrow} + \beta\ket{\downarrow} \right) \bigotimes \ket{\psi_{\mathcal{E}_i}}. 
\end{equation}
Quantum Darwinism then posits that, if after their mutual interaction, the total state of the system+environment is
\begin{equation}
\label{eq:DarState}
\ket{\psi_{\mathcal{S}\mathcal{E}}} = \alpha \ket{\uparrow}\ket{0^{\otimes N}} + \beta\ket{\downarrow}\ket{1^{\otimes N}}.
\end{equation} 
then classical objectivity emerges. The reason for this seemingly special form becomes apparent when we examine the reduced state of the system or the reduced state of any single environmental unit. Taking the partial trace over the $N$ environmental qubits we are left with the density matrix for $S$
\begin{equation}
\label{eq:DarSysState}
\rho_\mathcal{S} = \vert \alpha \vert^2 \ketbrad{\uparrow} + \vert \beta \vert^2 \ketbrad{\downarrow}.
\end{equation}
While for any single environment qubit we have
\begin{equation}
\label{eq:DarEnvState}
\rho_{\mathcal{E}_i} = \vert \alpha \vert^2 \ketbrad{0} + \vert \beta \vert^2 \ketbrad{1}.
\end{equation}
From Eq.~\eqref{eq:DarSysState} we see that after the interaction with the environment, the system state is left completely decohered, and therefore in a classical state. Crucially however, the populations are unaffected. From Eq.~\eqref{eq:DarEnvState} we see that every environmental qubit has resulted in the populations of the system becoming ``imprinted" onto them. Therefore, by capturing a subset of the $N$ environmental qubits an observer is able to determine the state of the system. Equally a separate observer can capture a different subset of environmental qubits, and when both perform the same type of measurement on their respective sets, will necessarily arrive at the same conclusion. This is due to the entanglement shared between the system and all the environmental degrees of freedom. In fact, notice that after their interaction the system is maximally entangled with the environment. Indeed the states Eq.~\eqref{eq:DarState} are closely related to those discussed in Sec.~\ref{sec:envariance} where the foundations of statistical mechanics where derived from quantum entanglement.

At this point it is important to revisit what we mean by objectivity. In the case of Eq.~\eqref{eq:DarState} it is clear that if each observer independently measures subsets of the environment they will gain the same information about the system. In this respect it should be clear that objectivity will emerge when the {\it amount of information} learned about the system by interrogating portions of the environment exhibits this redundancy. Therefore, a key quantity in assessing quantum Darwinism is the mutual information shared between the system, $\mathcal{S}$, and the fraction of the environment which the observer is measuring, $\mathcal{E}_f$
\begin{equation}
\bar{\mathcal{I}}_{\mathcal{S}\mathcal{E}_f} = S(\rho_\mathcal{S}) + S(\rho_{\mathcal{E}_f}) - S(\rho_{\mathcal{S}\mathcal{E}_f}),
\end{equation}
where $S(\cdot)$ is the von Neumann entropy, and $\rho_j$ are the appropriate reduced density matrices. We now see that the maximum $\bar{\mathcal{I}}_{\mathcal{S}\mathcal{E}_f} = 2 S(\rho_\mathcal{S})$ and occurs when $\mathcal{E}_f=\mathcal{E}$, i.e. the entire environment. Objectivity thus implies that for any other fraction of the environment, the amount of information shared by the system, and therefore accessible to any observer measuring this fraction, will be $\bar{\mathcal{I}}_{\mathcal{S}\mathcal{E}_f}=S(\rho_\mathcal{S})$. Thus, quantum Darwinism is signalled by a characteristic plateau appearing in the behavior of $\bar{\mathcal{I}}_{\mathcal{S}\mathcal{E}_f}$ versus the fraction size. To show this is indeed the case, we consider the following example of a single qubit undergoing decoherence within a spin-bath.

\begin{figure}
\centering
{\bf (a)} \hskip0.5\columnwidth {\bf (b)}
\includegraphics[height=.32\columnwidth]{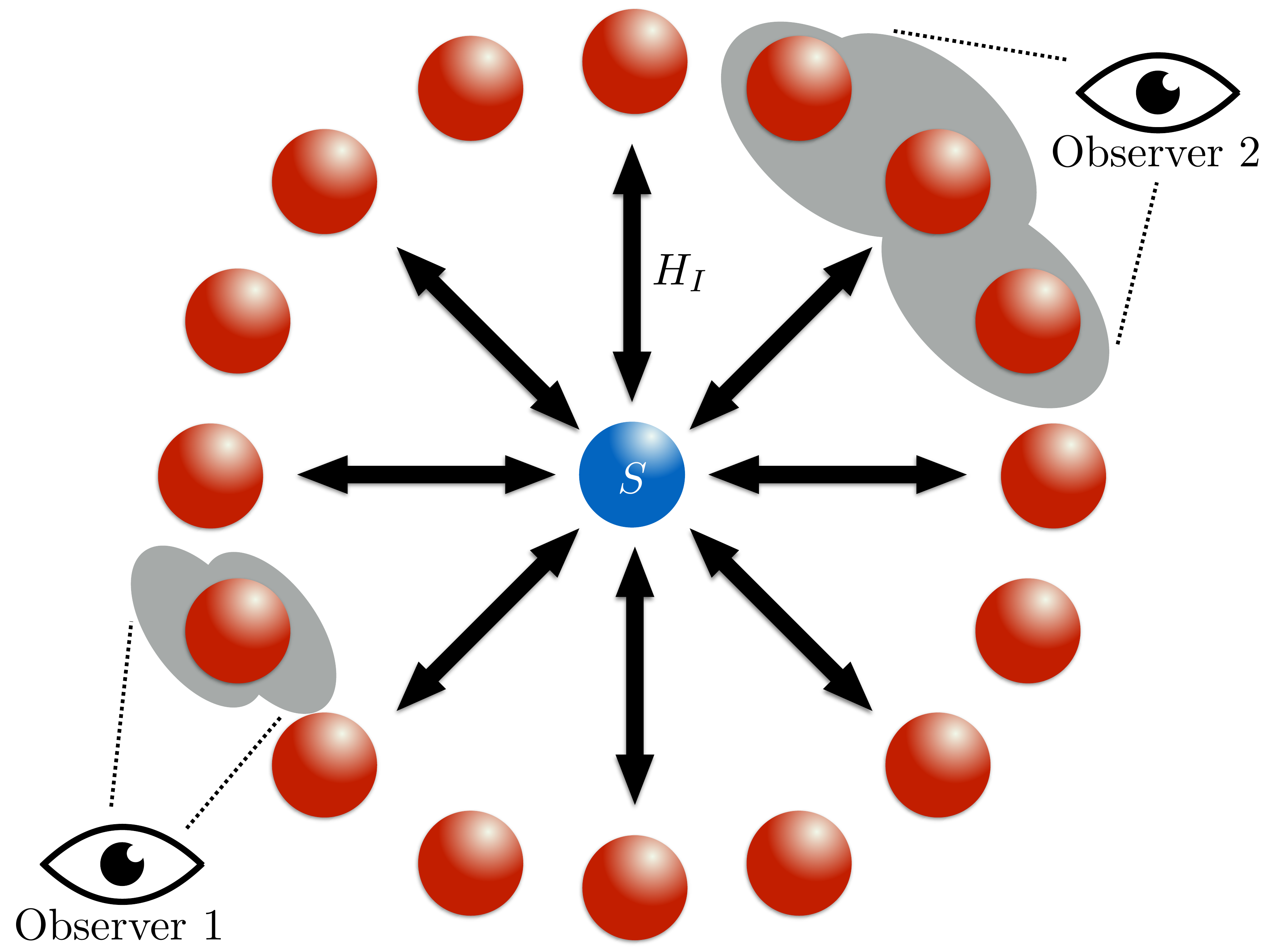}~~~~\includegraphics[height=.32\columnwidth]{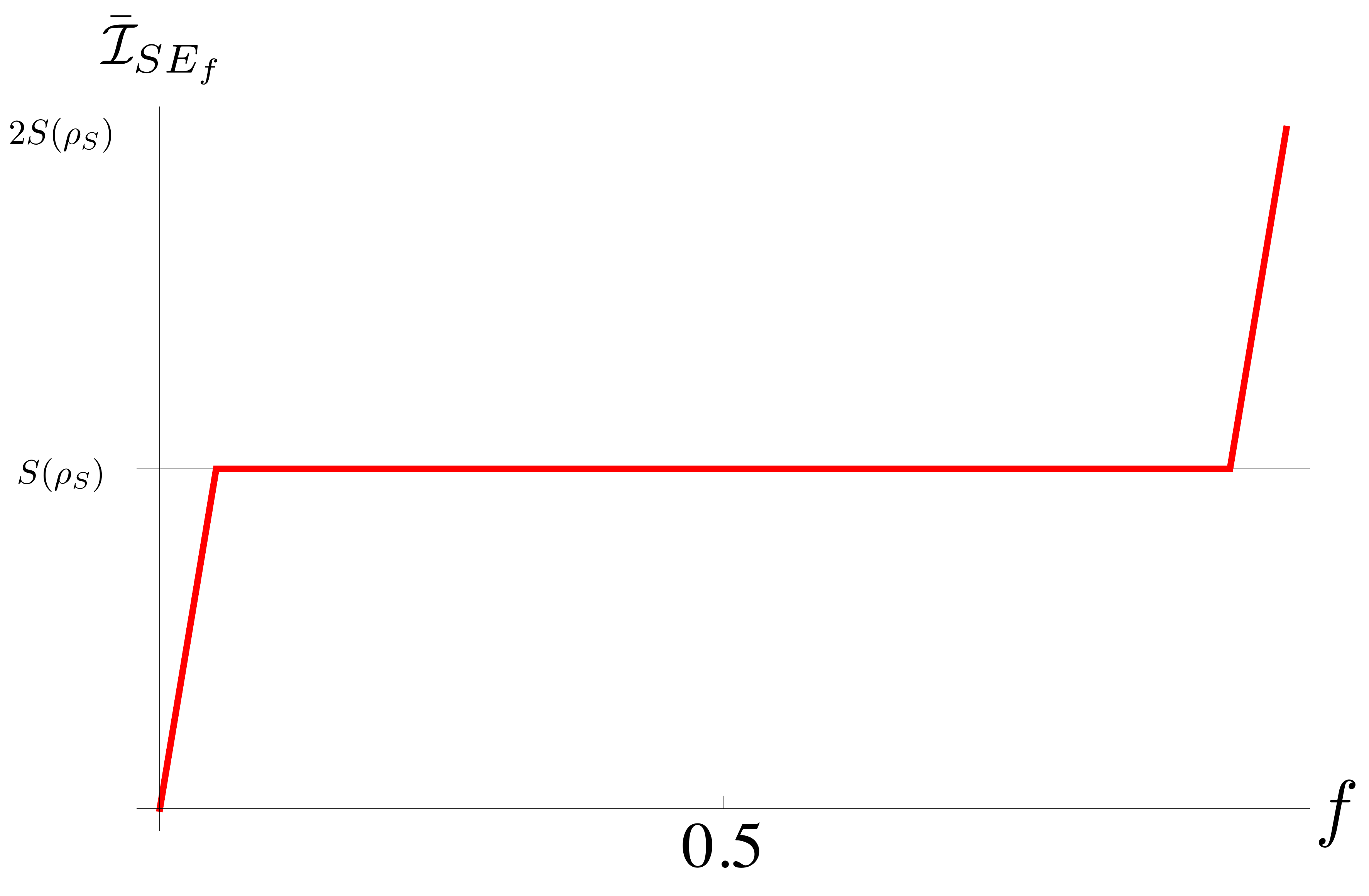}
\caption{{\bf (a)} Spin-star environment -- a central system qubit is coupled to a ``star"-array. For suitably chosen interactions and initial conditions, this model gives rise to a pure dephasing process and exhibits quantum Darwinism, i.e. two observers can intercept different portions of the environment and they will learn the same information about the state of the system. {\bf (b)} Mutual information versus environmental fraction from the spin-star model, for $N=16$. The characteristic plateau is readily visible.}
\label{figDarwin}
\end{figure}

\paragraph{Example: Spin-star environment.}
Consider a single qubit immersed in a spin-bath such that it corresponds to a spin-star configuration, see Fig.~\ref{figDarwin} {\bf (a)}. The system interacts with all of the constituents of the environment equally and independently according to the Hamiltonian
\begin{equation}
H_I = J \sum_{i=1}^N \sigma_z^\mathcal{S} \otimes \sigma_z^{\mathcal{E}_i}
\end{equation}
where $\sigma_z^\mathcal{S} = \ketbrad{\uparrow} - \ketbrad{\downarrow}$, $\sigma_z^{\mathcal{E}_i} = \ketbrad{1} - \ketbrad{0}$, and $J$ is the coupling strength. This system-environment model realizes pure dephasing on the system, i.e. the populations of the system are unaffected by the coupling to the environment, however the coherences are suppressed. We will assume a pure initial state for the system $\ket{\psi_\mathcal{S}} = \alpha \ket{\uparrow} +\beta \ket{\downarrow}$, where $\beta= \sqrt{1-\alpha^2}$  and for simplicity we restrict to $\alpha\in \mathbb{R}$. Additionally, all the environmental qubits are in $\ket{+} = \tfrac{1}{\sqrt{2}}\left(\ket{0}+\ket{1}\right)$, and therefore our overall state can be written
\begin{equation}
\ket{\psi_{\mathcal{S}\mathcal{E}}(0)} = \frac{\alpha}{(\sqrt{2})^N} \ket{\uparrow} \left( \ket{0} + \ket{1} \right)^{\otimes N} +  \frac{\beta}{(\sqrt{2})^N} \ket{\downarrow} \left( \ket{0} + \ket{1} \right)^{\otimes N}.
\end{equation}
The composite system then evolves according to the unitary $U(t)=\e{-i/\hbar\, H_I t}$. After a time $t$ our total system+environment state is
\begin{equation}
\begin{aligned}
\ket{\psi_{\mathcal{S}\mathcal{E}}(t)} &= U(t) \ket{\psi_{\mathcal{S}\mathcal{E}}(0)} \\
                           &= \frac{\alpha}{(\sqrt{2})^N} \ket{\uparrow} \left( e^{-i Jt} \ket{0} +  e^{iJt} \ket{1} \right)^{\otimes N} +  \frac{\beta}{(\sqrt{2})^N} \ket{\downarrow} \left( e^{i Jt}\ket{0} +  e^{iJt}\ket{1} \right)^{\otimes N} \label{eq:evolved}.
\end{aligned}
\end{equation}
Now examining the reduced state, written in its pointer basis $\{ \ket{\uparrow}, \ket{\downarrow} \}$ of the system at time $t$ we find
\begin{equation}
\rho_\mathcal{S}(t) = \left(
\begin{array}{cc}
 \alpha^2 & \alpha \beta \cos(2Jt)^N \\
\alpha \beta \cos(2Jt)^N & \beta^2 \\
\end{array}
\right),
\end{equation}
while any single environmental qubit, written in the computational basis, takes the form
\begin{equation}
\label{eq:spinstarE}
\rho_{\mathcal{E}_i}(t) = \left(
\begin{array}{cc}
 \frac{1}{2} &  \frac{1}{2} \left( \cos(2Jt) + i(1-2\alpha^2) \sin(2Jt) \right)\\
\frac{1}{2} \left( \cos(2Jt) - i(1-2\alpha^2) \sin(2Jt) \right) & \frac{1}{2} \\
\end{array}
\right),
\end{equation}
When $t=\pi/(4J)$ we find that the overall system+environment state, Eq.~\eqref{eq:evolved} has the same general form as given in Eq.~\eqref{eq:DarState} and furthermore that the system (and any environmental qubit) is fully decohered. We now consider the information that an observer can learn about the system by measuring some fraction, $f=n/N$, of the environment. It is a matter of direct calculation to find that regardless of the size of the environmental fragment the observer interrogates, the mutual information shared between the system and the environment will be $\bar{\mathcal{I}}_{\mathcal{S}\mathcal{E}_f} = S(\rho_\mathcal{S})$. In Fig.~\ref{figDarwin} we show the characteristic ``plateau" signalling the redundant encoding of information about the system throughout the environment.

\clearpage


\section{Checklist for ``Thermodynamics of Quantum Systems"}

\begin{enumerate}
\item Measuring low temperatures is not as simple as just sticking a thermometer into a quantum system.
\item Quantum systems with maximal energy variance, and therefore heat capacity, are the optimal thermometers.
\item Reversible Otto engines with quantum working fluids achieve the same efficiency as their classical counterparts.
\item Quantum working fluids allow to go beyond the Curzon-Ahlborn efficiency at maximum power for endoreversible Otto cycles.
\item Work may be extracted from multiple copies of a passive, but not completely passive, quantum battery by performing entangling operations.
\item Dynamical generation of entanglement leads to a significant boost in the charging power of arrays of quantum batteries.
\item Entropy production can be understood as the correlations established between the system and its environment.
\item Quantum Darwinism provides a framework to explain classical objectivity through `redundant encoding', i.e. imprinting the same system information onto multiple environment degrees of freedom.
\end{enumerate}

\section{Problems}

\subsection*{Quantum thermometry \ref{sec:qu_thermoetry}}

\begin{itemize}
\item[\textbf{[1]}] A spin-1 particle is found in a thermal state at temperature $T$. Compute the quantum Fisher information for the corresponding state. At optimal precision, can the temperature be estimated more accurately for spin-1/2 particles or spin-1 particles?
\item[\textbf{[2]}] Consider two quantum two-level systems (TLS). In which situation can the temperature be measured more precisely: (i) if the two TLS are non-interacting and independently prepared at temperature $T$; or (ii) if the two TLS interact as described by the quantum Ising model in the transverse field?
\end{itemize}

\subsection*{Quantum heat engines -- engines with atomic working fluids \ref{sec:HeatEngines}}

\begin{itemize}
\item[\textbf{[3]}] Consider a quantum Otto engine operating on a working medium with Hamiltonian,
$$H(t) =-\mu \vec{S}\cdot \vec{B}(t)\,,$$
where $\vec{S}=(S_x,S_y,S_z)$ describes a spin-1 particle, i.e.,
$$S_x=\frac{\hbar}{\sqrt{2}}\begin{pmatrix}
0&1&0\\1&0&1\\0&1&0
\end{pmatrix}, \quad
S_y=\frac{i\,\hbar}{\sqrt{2}}\begin{pmatrix}
0&-1&0\\1&0&-1\\0&1&0
\end{pmatrix}, \quad\mrm{and}\quad
S_z=\frac{\hbar}{\sqrt{2}}\begin{pmatrix}
1&0&0\\0&0&0\\0&0&-1
\end{pmatrix}\,.$$
Compute its efficiency for quasistatic variation of the magnetic field.
\item[\textbf{[4]}] The entropy of a semiclassical, ideal gas is given by the Sackur-Tetrode equation, 
$$S(E, V, N)= N k_B\,\left[\frac{5}{2}+\lo{\frac{V}{Nh^3}\left(\frac{4 \pi m E}{3 N}\right)^{(3/2)}}\right]\,. $$
What is the (endoreversible) efficiency at maximal power of an Otto engine operating with such a semiclassical gas as working medium? 
\end{itemize}

\subsection*{Work extraction from quantum systems \ref{sec:WorkExtraction}}

\begin{itemize}
\item[\textbf{[5]}] Consider again a spin-1 particle with Hamiltonian $$H =-\mu \vec{S}\cdot \vec{B}\,,$$
where  as before $\vec{S}=(S_x,S_y,S_z)$ is given by
$$S_x=\frac{\hbar}{\sqrt{2}}\begin{pmatrix}
0&1&0\\1&0&1\\0&1&0
\end{pmatrix}, \quad
S_y=\frac{i\,\hbar}{\sqrt{2}}\begin{pmatrix}
0&-1&0\\1&0&-1\\0&1&0
\end{pmatrix}, \quad\mrm{and}\quad
S_z=\frac{\hbar}{\sqrt{2}}\begin{pmatrix}
1&0&0\\0&0&0\\0&0&-1
\end{pmatrix}\,.$$
Determine the maximal ergotropy with respect to the completely passive state that can be stored in any quantum state $\rho$ as a function of temperature $T$.
\item[\textbf{[6]}] A specific quantum battery consists of an array of $n$ qubits. Compute the maximal ergotropy with respect to the completely passive state as a function of $T$ that can stored in such a battery.
\end{itemize}

\subsection*{Quantum decoherence and the tale of quantum Darwinism \ref{sec:Decoherence}}

\begin{itemize}
\item[\textbf{[7]}] A quantum system undergoes ``pure'' decoherence in the energy basis if the master equation takes the form 
$$\frac{d\rho}{dt}=-\frac{i}{\hbar}\com{H}{\rho}-\sum_{i\neq j} \gamma_{i,j} \ket{i}\bra{j}\,, $$
where $\ket{i}$ is an energy eigenstate and $\gamma_{i,j}$ are the coefficients of the coupling matrix. Show that this master equation can be brought into Lindblad form and that the resulting dynamics is unital. 
\item[\textbf{[8]}] Consider the reduced  two qubit state, $\rho_{SE_i}$, consisting of a system qubit, $S$, and any single environmental qubit, $E_i$, taken from a larger environment of the form
$$
\rho_{SE_i} = p \ketbrad{\phi} + (1-p)\ketbrad{\psi}
$$ 
where $\ket{\phi}=\sqrt{p}\ket{00}+\sqrt{1-p}\ket{11}$, $\ket{\psi}=\sqrt{p}\ket{01}+\sqrt{1-p}\ket{10}$ and $p\neq1/2$. Determine the marginal states, entropies, and the mutual information, and therefore show that the state of $S$ is objective according to quantum Darwinism. Is this state entangled? If so, what does this imply about the validity and applicability of defining classical objectivity through quantum Darwinism?
\end{itemize}

\addcontentsline{toc}{section}{References}
\bibliographystyle{plain}
\bibliography{book}

\clearpage


\chapter{\label{chap:info}Thermodynamics of Quantum Information}

The phrase \textit{quantum supremacy}  typically refers to situations in which information processing devices built on the principles of quantum physics solve computational problems that are not tractable by classical computers \cite{Savage2017}. The resulting quantum advantage is the ratio of classical resources, such as time or memory, to the associated quantum resources. Generally, the hardware requirements to achieve this computational supremacy can be summarized by three key properties \cite{Nielsen2010} (i) the quantum systems must initially be prepared in a well-defined state; (ii) arbitrary unitary operators must be available and controllable in order to launch an arbitrary entangled state; and (iii) measurements of the qubits must be performed with high quantum efficiency. 

As we have already discussed, however, all physical quantum systems are subject to decoherence and dissipation arising from their noisy interaction with the environment. Thus, thermodynamically speaking any realistic operation of quantum information processing devices will be accompanied by the production of irreversible entropy, and by the irretrievable loss of quantum information into the environment. 

In the present context, some questions appear immediate: how can the tools and techniques of quantum thermodynamics help to optimally operate quantum computers, i.e., how can we keep quantum computers in the deep quantum regime so that we can actually utilize their supremacy? In this chapter, we will quantify the thermodynamic cost of quantum information processing in Sec.~\ref{sec:qu_info}. Then, Sec.~\ref{sec:dwave} is dedicated to assessing the performance of adiabatic quantum computers by means of Quantum Stochastic Thermodynamics. This will turn our attention to thermodynamic properties of critical systems. Specifically, in Sec.~\ref{sec:KibbleZurek} we assess the Kibble-Zurek mechanism of defect formation and its relation with the irreversible entropy production. Such an analysis is particularly relevant in certain quantum computational platforms, particularly those that rely on annealing. We close the chapter with a brief outline on recent efforts in developing quantum error correcting schemes for adiabatic quantum computers in Sec.~\ref{sec:STA}.

\section{Quantum thermodynamics of  information \label{sec:qu_info}}

\subsection{Thermodynamics of classical information processing}
\label{sec:land1}
Information is physical. This is the conclusion established by Rolf Landauer in his landmark 1961 paper~\cite{Landauer1961}. The question posed was relatively simple: What are the physical limitations on information processing set by the laws of thermodynamics? As has been widely noted since, and even acknowledged by Landauer himself in the original paper, the notion of an energetic price to pay for processing information is not surprising, after all information is encoded in physical systems and therefore must be subject to the laws of thermodynamics. Indeed any process occurring at a finite rate will be accompanied by some dissipation. What was remarkable about Landauer's insight was that it established an absolute minimum cost, independent of any other constraints, that must be paid to {\it erase} information. Thus showing that, far from being an abstract concept, information truly is as physical as any other quantity.  Among the many consequences of Landauer's principle, probably the most famous is the exorcism of Maxwell's Demon~\cite{Bennett2003}. However, another notable consequence was the theoretical proposals for fully reversible models of computation~\cite{Bennett1982}. There are many subtleties in understanding Landauer's insight, therefore to begin we shall discuss the basic motivations that lead to the result.

\paragraph{Landauer, Bennett, and the famous $k_B T \ln 2$.}
To first gain an intuitive picture, consider the operation that restores a bit to a given state, say 1. If the state of the bit is initially known then there will be two possible procedures to achieve this: either the bit is already in 1 and we do nothing, or the bit is initially in 0 and we must change it to 1. Notice that regardless, in both of these settings the entropy of the bit is left unchanged since it always begins and ends in a definite state, thus has zero entropy. Both of these operations can be done, in principle, in a fully reversible manner and therefore correspond to no ``wasted" energy being dissipated. However, it is clear that this is not how a normal information processing device works, as such devices will operate on data that is independent of the particular process or computation it is performing. 

Therefore, Landauer asks the question whether there exists some process that can always perform the action restore to 1, regardless of the state that the bit is in, without dissipating heat. If we consider the two processes mentioned already, since they are both reversible we can imagine running them backwards in time. Now we see the problem: by definition the individual processes are fully deterministic, however in the time-reversed scenario we have a single initial condition, the bit in 1, but two possible final states. Therefore, there cannot be a conservative force that always restores the bit to 1 regardless of its initial state. This is the basic reasoning that Landauer followed to show that information erasure comes at an inescapable thermodynamic cost. 

Indeed, in the above reasoning we clearly see that two possible inputs lead to a single output, and therefore we can  define the notion of logical irreversibility as those processes which the output does not uniquely define the inputs. This form of irreversibility is common in most computing devices, the AND gate is an example of a logically irreversible gate, as its two inputs that lead to a single output. However, such irreversible operations can be avoided, either by saving the entire history of the process, or by embedding the usual irreversible gate operations, such as the AND, into more complex but reversible gate operations, e.g. using a Toffoli gate. This is a remarkable observation as it indicates that the processing of information has no intrinsic thermodynamic cost.

How then do we arrive at Landauer's principle? First we must examine a subtle difference between information copying and information erasure. For a single bit in a given state, we can faithfully copy this bit onto a blank bit in a fully reversible manner. This is clear as it is a one-to-one process. This can again be seen by considering the entropy before and after the copying procedure: both bits have definite states before and after, and therefore the entropy is constant and zero. The subtlety arises when we consider erasure of a bit of information. In this case we have two possible states being mapped to a single definite state. Landauer's principle states that: the entropy decrease of the information bearing degrees of freedom must be compensated by an equal or greater entropy increase in the environment.

For a two-state system containing a bit of information the initial Shannon entropy, $G=-\sum_i p_i\ln{p_i}$, which quantifies its information content, is $\ln 2$. After the erasure the entropy is zero. If our system bit, $\mathcal{S}$, is surrounded by a large thermal reservoir, $\mathcal{E}$, in equilibrium at temperature $T$, from the second law the total change in entropy of both $\mathcal{S}+\mathcal{E}$ must be positive, i.e.
\begin{equation}
\Delta S_\mrm{tot} =\Delta S_\mathcal{S} + \Delta S_\mathcal{E} \geq 0.
\end{equation}
where $\Delta S = S_\text{final} - S_\text{initial}$. Assuming the reservoir is large, and therefore always in equilibrium, we can then use the Clausius inequality to write down the heat flow into the reservoir as
\begin{equation}
Q_\mathcal{E} \geq -T \Delta S_\mathcal{S}
\end{equation}
We can see a link between the thermodynamic and the information entropy of the system, in particular $S=k_B G$. Given that the change in information entropy from the initial to final state of the system is $\Delta G =-\ln2$, it follows then that the heat into the reservoir is bounded by
\begin{equation}
\label{eq:Landauer_classical}
Q_\mc{E} \geq k_B T \Delta G = k_B T \ln 2.
\end{equation}

\begin{figure}
\begin{mdframed}[roundcorner=10pt]
\begin{minipage}[l]{.4\textwidth}
\begin{center}
\includegraphics[height=5cm]{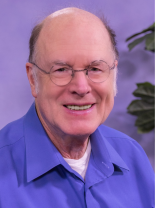}
\end{center}
\end{minipage}
\hfill
\begin{minipage}[r]{.58\textwidth}
Charles Bennett:\\ \emph{Computers may be thought of as engines for transforming free energy into waste heat and mathematical work  \cite{Bennett1982}.}
\end{minipage}
\end{mdframed}
\end{figure}

\begin{figure}[t]
\begin{centering}
\includegraphics[width=0.52\columnwidth]{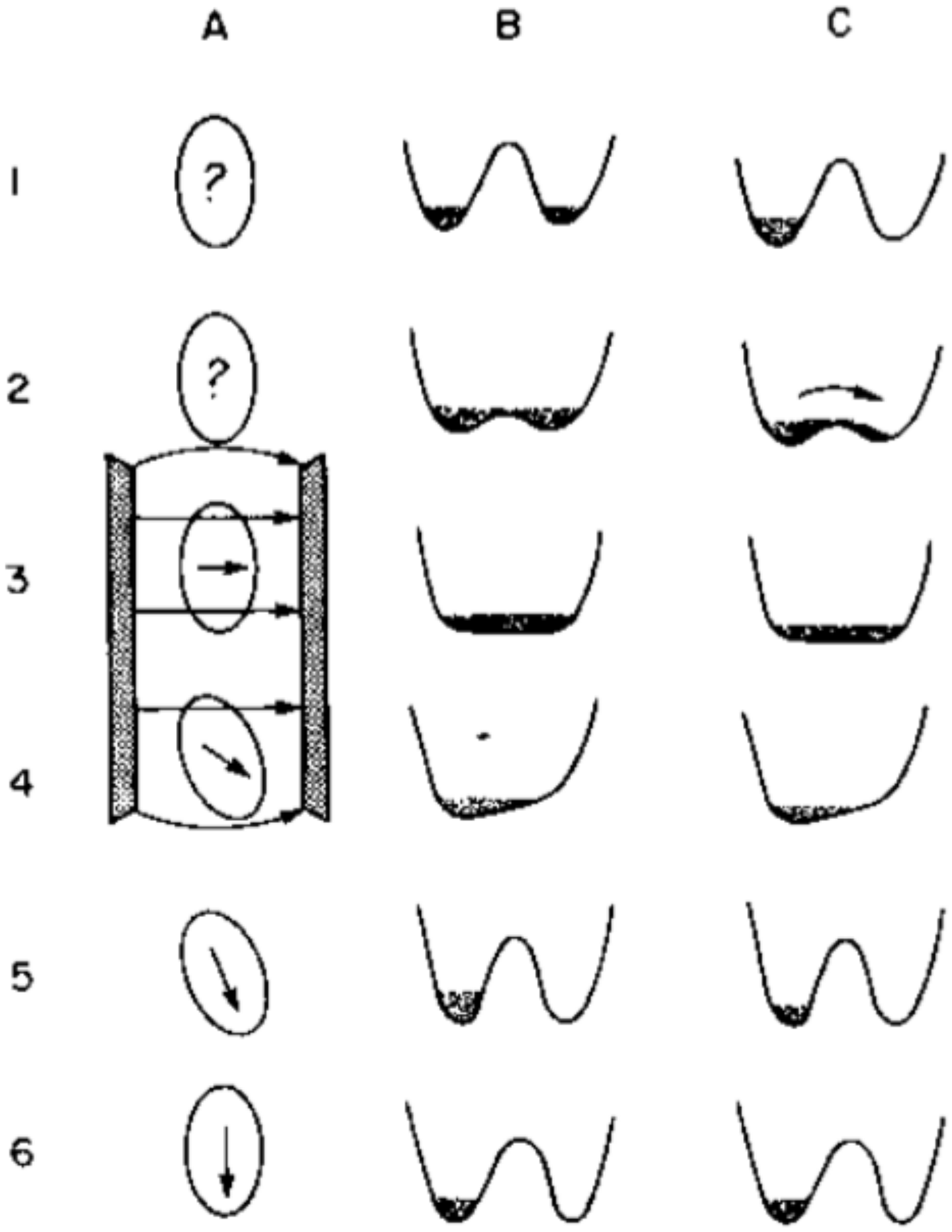}
\caption{Figure taken from Bennett's 1982 paper~\cite{Bennett1982}. Column A shows a possible implementation of Landauer's erasure: a bistable element which can be either in spin-up or spin-down is slowly moved into  magnetic field at step 3. At step 4 a slight bias is applied to abolish the symmetry before the bit is moved out of the field leaving it in a definite final state. Column B shows the evolution of the probability density when the bit is a random, unknown initial state. Column C corresponds to when the bit is in a definite state, as would be typical at the end of a computation. An irreversible entropy increase occurs at step 2.}
\label{fig:Erasure}
\end{centering}
\end{figure}

To highlight the subtleties further it is useful to revisit some considerations from Bennett~\cite{Bennett1982}. If our bit is encoded onto a particle trapped in a double well, such that we assign logical 0 as the particle in the left-well and logical 1 as the particle on the right-well, then we can consider the situations in Fig.~\ref{fig:Erasure}, where shaded regions represent the probability that the particle is found in either the left or right well. The basic ``erasure" protocol in both is as follows: given an initial state for the bit, the barrier in the double well is slowly ramped down until it is gone. Then a small perturbation is applied to break the symmetry such that the probability of the particle concentrates in the right well, and finally the barrier is slowly ramped back up. The total work required to perform this process is then $k_B T \ln2$.

Now let us consider how the initial state affects things. If the bit is random then this process is in fact fully reversible, running the process backwards results in precisely the random state we started with. What is important to note is the work done leads to the complementary entropy decrease of the bit, and therefore is thermodynamically reversible. On the other hand, when the bit has a definite state (one that we may not have yet measured) then the process is irreversible: running the process backwards will not result in the same initial state we began with. There is an irreversible entropy increase which occurs when the barrier is ramped down; at this point the information entropy of the bit has increased by $\ln2$. Thus, the $k_B T \ln2$ of work which is converted to heat into the environment while performing the task is not compensated by a corresponding entropy decrease in the bit. Remarkably this setting is almost exactly the one used in the first experimental confirmation of Landauer's erasure principle~\cite{BerutNature}.

\subsubsection{Maxwell's Demon and Szilard's Engine.}
Before moving to examining Landauer's principle for quantum systems, we would be remiss to not include a mention of Maxwell's Demon, upon which there have been many excellent discussions already~\cite{Bennett1982, Sagawa2015, Vedral2009}. Maxwell's original Gedankenexperiment envisaged an intelligent being, the Demon, that was able to observe an ideal gas in a container. By inserting a partition with a controllable, frictionless trap door in the middle of the container, the Demon is able to open the door as faster moving, hotter particles approach from one side while closing it when slower moving, colder particles approach from the other. In this way the Demon is able to sort the particles. and consequently reduce the entropy of the gas, in apparent violation of the second law.

Arguably the clearest formulation of the Maxwell Demon paradox is exemplified by the Szilard engine. In 1929 Szilard considered essentially the same setting but within the context of a single particle gas, in thermal equilibrium with a reservoir, in a box under going a cyclic process. The Demon, initially ignorant of where in the box the particle is, inserts a partition in the middle. It now measures which side of the partition the particle is on and, using this information, attaches a piston to the side of the partition not containing the particle. The expansion of the particle then leads to $k_B T \ln 2$ of work being extracted. At the end of the process the only record of where the particle was when the partition was introduced is in the Demon's memory. Thus, it was often believed that it was the act of measurement that accounted for the discrepancy. 

Once again, Bennett, using the insight from Landauer's principle, clarified that it is actually the erasure the Demon's memory that restores the second law. To see this, consider a Demon with a memory that can be in three possible states, left (L), right (R), and standard (S). Assuming it starts in S, after inserting the partition, the act of measurement is similar to copying the single bit information about where the particle is to the Demon's memory. As already discussed, the copying of information comes at no intrinsic thermodynamic cost. However, evidently to close the cycle the Demon's memory must be returned to the standard state, i.e. the information about which side of the partition the particle was on must be erased which, from Landauer's principle, leads to the corresponding dissipation of $k_B T \ln 2$ of heat. While the inclusion of the standard state for the Demon's memory is not crucial (a similar argument can be followed for a memory with only two possible states), it does greatly help in identifying erasure as the source of the apparent paradox and not the measurement process~\cite{Bennett1982}. 

\subsection{A quantum sharpening of Landauer's bound}
Interestingly Landauer's principle as considered above appeared to largely apply in the quantum domain, however only recently has the statement been made more mathematically rigorous. The statement of Landauer's principle can be made more concrete by first establishing a minimal set of conditions for which the notion of information erasure has a definite meaning. To this end, consider a system $\mc{S}$ whose information content we want to erase by making it interact with an environment $\mc{E}$. Following Landauer's own reasoning~\cite{Landauer1961}, Reeb and Wolf determined that the following constitute a minimal set of assumptions, which ensure the validity of Landauer's principle~\cite{ReebWolfNJP}, and apply to both classical and quantum settings:
\begin{enumerate}
\item Both $\mc{S}$ and $\mc{E}$ are quantum systems, living in Hilbert spaces $\mathscr{H}_\mc{S}$ and $\mathscr{H}_\mc{E}$ respectively;
\item The initial state of the composite system is factorized, i.e. $\rho_{\mc{S}\mc{E}}(0) = \rho_\mc{S}(0)\otimes\rho_\mc{E}(0)$, such that no initial correlations are present;
\item The environment is prepared in the thermal state $\rho_\mc{E}(0)=\rho_{\beta}=\e{-\beta H_\mc{E}}/Z_\mc{E}$ with $H_\mc{E}$ the Hamiltonian of the environment, which we spectrally decompose as $$H_\mc{E}= \sum_m E_m \ket{E_m}\bra{E_m} = \sum_m E_m \Pi_m.$$ Here, $\ket{E_m}$ is the $m^{\rm th}$ eigenstate of $H_\mc{E}$, associated with eigenvalue $E_m$. Finally, we have introduced the partition function $Z_\mc{E} = \ptr{\mc{E}}{\e{-\beta H_\mc{E}}}$;
\item System and environment interact via the overall unitary transformation $U(t) = \e{-i\, H t/\hbar}$ with $H = H_\mc{S} + H_\mc{E} + H_{\mc{S}\mc{E}}$ the total Hamiltonian. 
\end{enumerate}

Assumptions 1 and 3 are well motivated: one must be able to clearly partition our overall system so as to (initially) be able to unambiguously identify the system and environment. Furthermore, the requirement for an initially thermal state of the reservoir is also physically well motivated, recall from Sec.~\ref{sec:WorkExtraction} that thermal states are the only completely passive states. Thus, if the environment is in any other state it would be possible to violate Landauer's principle. Regarding assumption 2, as has been extensively used throughout this book, the initially factorized state is common in thermodynamics, and in fact is required for Landauer's principle to hold: if $\mathcal{S}$ and $\mathcal{E}$ share some initial correlations then it can be shown that the system entropy can be reduced without a corresponding increase in the environment. Finally, assumption 4 is required to ensure that no auxiliary environment can be involved in the evolution, and therefore all resources used during the process are fully taken into account.

Using only these assumptions, we can derive a rigorous equality version of Landauer's principle. First let us carefully set some notation. Landauer's principle is related to the change in entropy of the system, therefore we will denote an entropy change for the system/reservoir as the difference between their initial and final entropies\footnote{It is important to note that Reeb and Wolf's rigorous treatment relate the change in entropy between {\it initial} and {\it final} system states with the heat transferred to the reservoir, while the more `intuitive' derivation in Sec.~\ref{sec:land1} employed the entropy production.}
\begin{equation}
\label{eq:entchange}
\Delta S_{\mc{S}(\mc{E})} = S(\rho_{\mc{S}(\mc{E})}^0) - S(\rho_{\mc{S}(\mc{E})}^t).
\end{equation}
Notice that these entropy changes can be positive or negative. To establish Landauer's principle, we first consider the entropy production of the process, i.e. the difference between the {\it final} and {\it initial} entropies, which according to the second law must be positive
\begin{equation}
\begin{split}
\label{2ndLawLand}
&-\Delta S_\mc{S} - \Delta S_\mc{E} = S(\rho_\mc{S}^t)-S(\rho_\mc{S}^0) + S(\rho_\mc{E}^t)-S(\rho_\mc{E}^0) \\
                                       &\qquad = S(\rho_\mc{S}^t) + S(\rho^t_\mc{E}) - S(\rho_\mc{S}^0 \otimes \rho_\mc{E}^0) \\
                                       &\qquad = S(\rho_\mc{S}^t) + S(\rho_\mc{E}^t) - S(\rho_{\mc{S}\mc{E}}^t) \\
                                       & \qquad= I(\rho_{\mc{S}\mc{E}}^t) \geq 0,
\end{split}
\end{equation}
i.e. the mutual information shared between system and reservoir at the end of the process. In moving from the second to the third line we have used the invariance of the von Neumann entropy under unitary transformations. From the non-negativity of the mutual information, it follows that this is a statement of the second law.

Simply rearranging Eq.~\eqref{2ndLawLand} we have 
\begin{equation}
\begin{split}
&I(\rho_{\mc{S}\mc{E}}^t) + \Delta S_\mc{S} = - \Delta S_\mc{E} \\
                                           &\qquad= -\tr{\rho_\mc{E}^t \ln \rho_\mc{E}^t} + \tr{\rho_\mc{E}^0 \ln\left[ \frac{e^{-\beta H_\mc{E}}}{\tr{e^{-\beta H_\mc{E}}}}\right]} \\
                                           &\qquad= -\tr{\rho_\mc{E}^t \ln \rho_\mc{E}^t} - \beta\tr{ H_\mc{E} \rho_\mc{E}^0} - \ln \tr{ e^{-\beta H_\mc{E}} } + \beta \tr{ H_\mc{E} \rho_\mc{E}^t } - \beta \tr{ H_\mc{E} \rho_\mc{E}^t } \\
                                           &\qquad= \beta \tr{ H_\mc{E} \left( \rho_\mc{E}^t  - \rho_\mc{E}^0 \right)  } - \tr{ \rho_\mc{E}^t \ln \rho_\mc{E}^t } + \tr{\rho_\mc{E}^t \ln \rho_\mc{E}^0} \\
                                           &\qquad= \beta \mean{Q_\mc{E}} - S(\rho_\mc{E}^t \| \rho_\mc{E}^0)
\end{split}
\end{equation}
Therefore the following equality follows~\cite{EspositoNJP, ReebWolfNJP}
\begin{equation}
\label{eq:LanduaerSharp}
\beta \mean{Q_\mc{E}} = \Delta S_\mc{S} + I(\rho_{\mc{S}\mc{E}}^t) + S(\rho_\mc{E}^t \| \rho_\mc{E}^0),
\end{equation}
which is the equality version of Landauer's principle, valid for non-equilibrium settings. As both the relative entropy and the mutual information are non-negative functions, we can arrive at Landauer's bound by simply dropping them to give
\begin{equation} 
\label{LandauerBound}
\beta \mean{Q_\mc{E}} \geq \Delta S_\mc{S},
\end{equation}
which is the non-equilibrium (quantum) Landauer's principle. Notice that Eq.~\eqref{eq:LanduaerSharp} is in fact equivalent to the expression for entropy production derived in Sec.~\ref{chap2:semigroup} Eq.~\eqref{eq:EntropyChange}. Reeb and Wolf's result has a further interesting consequence: equality in Landauer's bound holds only for trivial processes, i.e. those that in essence do nothing~\cite{ReebWolfNJP}.

\subsection{New Landauer bounds for non-equilibrium quantum systems}
Beyond the clear conceptional milestone that Landauer's insight provided, recently some efforts have gone into exploring and extending the idea of Landauer's bound~\cite{GooldPRL, GiacomoNJP}. Indeed Landauer's principle, in its most basic reading, is simply a lower bound on the dissipated heat. In the previous section we have established that the change in information entropy provides a perfectly valid non-equilibrium lower bound to the dissipated heat. Here we examine an alternative method originally published in Ref.~\cite{GiacomoNJP} to lower bound the dissipated heat that, rather than relying on the information change, is instead derived by applying the two-time energy measurement protocol from Sec.~\ref{sec:quantum_Jarzynski}.

\paragraph{Full counting statistics approach to dissipated heat.}
We can use the full counting statistics of the heat dissipated by the system (which corresponds to the change in environmental energy~\cite{ReebWolfNJP, EspositoNJP}) to access its mean value. Consider the same two-time energy measurement protocol for determining the quantum work introduced in Sec.~\ref{sec:quantum_Jarzynski}. The heat probability distribution, $\mc{P}(Q_\mc{E})$, to record a transferred amount of heat $Q_\mc{E}$, can be formally defined in the same manner. In line with the minimal assumptions for Landauer's principle to be applicable, we assume $\mc{S}$ to be initially uncorrelated with $\mc{E}$, which is prepared in an equilibrium state. Therefore $\rho_{\mc{S}\mc{E}}(0) = \rho_\mc{S}(0)\otimes\rho_{\beta}$ with $\left[ H_\mc{E}, \rho_{\beta}\right]=0$. Now we apply the two-time energy measurement approach to the environment: A projection over one of the energy eigenstates of the environment at time $t=0$ is carried out, obtaining $E_n$ as an outcome. As a result, the total $\mc{S}$-$\mc{E}$ state is 
\begin{equation}
\rho'_{\mc{S}\mc{E}}(0)=\rho_\mc{S}(0)\otimes\Pi_n.
\end{equation}
Immediately after the measurement, the interaction between $\mc{S}$ and $\mc{E}$ is switched on and the overall system undergoes a joint evolution up to a generic time $t$, when the interaction is switched off and a second projective measurement of the environmental energy is performed, this time obtaining an outcome $E_m$. After the second measurement, we have
\begin{equation}
\rho''_{\mc{S}\mc{E}}(t) = \frac{\Pi_{m} U(t)\rho'_{\mc{S}\mc{E}}(0)  U(t)^{\dagger} \Pi_{m}}{\ptr{\mc{S}\mc{E}}{\Pi_{m} U(t)\rho'_{\mc{S}\mc{E}}(0)  U(t)^{\dagger}}} .
\end{equation}
It is worth stressing that the set of assumptions and steps used in the two-time measurement protocol are perfectly compatible with those required by the erasure process. The joint probability to have obtained the two stated outcomes at times $0$ and $t$ respectively is given by the Born rule
\begin{equation}
\label{joint}
p(\ket{n}\rightarrow \ket{m})= \tr{ \Pi_{m} U(t) \Pi_{n} \rho_\mc{S}(0)\otimes\rho_{\beta} \Pi_{n} U^{\dagger}(t) \Pi_{m}},
\end{equation}
from which the probability distribution $\mc{P}(Q_\mc{E})$ follows as
\begin{equation}\label{probQ}
\mc{P}(Q_\mc{E})=\sum_{n,m} \delta(Q_\mc{E} - (E_m-E_n)) p(\ket{n}\rightarrow \ket{m}).
\end{equation}
The cumulant generating function is defined as the Laplace transform of the probability distribution
\begin{equation}\label{ThetaStart}
\Theta(\eta,\beta,t) \equiv \ln \langle  e^{-\eta Q_\mc{E}} \rangle = \ln \int dQ_\mc{E}\, \mc{P}(Q_\mc{E}) e^{-\eta Q_\mc{E}}\, ,
\end{equation}
which can be seen as the Wick rotated version of the usual definition given by the Fourier transform of $\mc{P}(Q_\mc{E})$. The cumulant of $n$th order is simply obtained by differentiation with respect to the real parameter $\eta$ as
\begin{equation}\label{Cumulants}
\langle Q_\mc{E}^n \rangle = (-1)^n\frac{\partial^n}{\partial\eta^n}\Theta(\eta,\beta,t)|_{\eta=0}.
\end{equation}
Note that in the definition of the cumulant generating function we have explicitly written the dependence on the inverse temperature $\beta$ of the bath, which enters in the joint probability Eq.~\eqref{joint} through the initial environmental state $\rho_{\beta}$. The crucial point in using the full counting statistics approach is that the cumulant generating function introduced in Eq.~\eqref{ThetaStart} can be expressed as
\begin{equation}
\Theta(\eta,\beta,t) = \ln\big( \ptr{\mc{S}}{\rho_\mc{S}(\eta,\beta,t)} \big),
\end{equation}
where
\begin{equation}
\rho_\mc{S}(\eta,\beta,t) = \ptr{\mc{E}}{ U_{\eta/2}(t)\rho_\mc{S}(0)\otimes\rho_{\beta}U^{\dagger}_{\eta/2}(t)},
\end{equation}
with $U_{\eta/2}(t) \equiv e^{-(\eta/2)H_\mc{E}} U(t) e^{(\eta/2)H_\mc{E}}$. By invoking the same approximations and techniques used to derive a master equation for the density matrix of the system $\rho_\mc{S}(t)$~\cite{BreuerBook}, one can obtain a new equation for $\rho_\mc{S}(\eta,\beta,t)$. Solving this is a task with the same degree of complexity as accessing the dynamics of the reduced system. However we may circumvent this difficulty by deriving a family of lower bounds to $\left<Q_\mc{E}\right>$ using the counting statistics arising from the two-time measurement protocol.

\paragraph{Lower bounds on the mean dissipated heat.} 
In order to derive a lower bound for $\mean{Q_\mc{E}}$, we consider the cumulant generating function of its probability distribution. Having it defined as in Eq.~\eqref{ThetaStart}, we can apply H\"{o}lder's inequality to prove that $\Theta(\eta,\beta,t)$ is a convex function with respect to the counting parameter $\eta$. This condition can be equivalently expressed as
\begin{equation}\label{Convexity}
\Theta(\eta,\beta,t) \geq \eta \frac{\partial}{\partial\eta}\Theta(\eta,\beta,t)\big|_{\eta=0}.
\end{equation}
Combining Eq.~\eqref{Cumulants} and Eq.~\eqref{Convexity}, we obtain a one-parameter family of lower bounds for the mean dissipated heat $\mean{Q_\mc{E}}$ leading to
\begin{equation}
\label{bound}
\beta \mean{Q_\mc{E}} \geq -\frac{\beta}{\eta}\Theta(\eta,\beta,t) \equiv \mathcal{B}^{\eta}_{Q_\mc{E}}(t)\quad (\eta > 0).
\end{equation}
As with Eq.~\eqref{LandauerBound}, Eq.~\eqref{bound} is valid in the case of a generic erasure protocol. It is worth noting however that the two bounds arise from totally different underlying frameworks, the former rooted in an information approach while the latter takes a more thermodynamic take on the problem, as such their performance as meaningful bounds to the dissipated heat in a given non-equilibrium setting has been shown to be quite different~\cite{CampbellPRALand}.



\section{Performance diagnostics of quantum annealers \label{sec:dwave}}

In the preceding sections we have seen that purely quantum resources can lead to a modification of the statements of thermodynamics. This is due to the additional (quantum) informational contribution to the entropy production. The natural question arises whether these generalized statements can be used in a practically relevant situation to teach us something about the quantum information processing system that we would not have known otherwise. That this is, indeed, the case has been shown by using the generalized quantum fluctuation theorem \eqref{eq:GFT} to assess the performance of adiabatic quantum computers \cite{Gardas2018}.

\emph{Adiabatic quantum computing} is a distinct paradigm of quantum computing \cite{Farhi2000}, that relies on  \emph{quantum annealing} \cite{Kadowaki1998}. In quantum annealing a quantum system is initially prepared in the ground state of a simple and controllable Hamiltonian. Then, the Hamiltonian is slowly varied, such that the system remains in the instantaneous ground state at all times. According to the quantum adiabatic theorem ``slow'' means that the rate with the Hamiltonian changes is much smaller than one over the energy gap between instantaneous ground state and first excited state \cite{Messiah1966}.  The target is the ground state of a complicated many-body Hamiltonian that cannot be diagonalized efficiently by classical algorithms. The desired outcome of the computation is encoded in this final ground state.

\subsection{Fluctuation theorem for quantum annealers}

Above in Sec.~\ref{sec:gft} we already outlined the general framework for quantum fluctuation theorems for arbitrary observables. For an application to quantum annealers we now need to choose meaningful and experimentally accessible observables. To this end, we will assume for the remainder of the discussion  that the quantum system is described by the quantum Ising model in transverse field~\cite{Zurek2005a},
\begin{equation} 
\label{eq:ising}
H(t)/(2\pi\hbar) = -g(t)\sum_{n=1}^{L}\sigma_x^n-\Delta(t)\sum_{n=1}^{L-1}J_n\sigma_z^n\sigma_z^{n+1}\,.
\end{equation}
Although, the current generation of quantum annealers can implement more general many body systems~\cite{Lanting2014}, we focus on the one dimensional case for the sake of simplicity~\cite{Kadowaki1998}. Typically, the parameterization is chosen, such that $\Delta(0)=0$ and $g(\tau)=0$, where $\tau$ is the \emph{anneal time}, i.e., the length of the process.

Thus, the somewhat obvious choice for the observables is the (customary renormalized) Hamiltonian in the beginning and the end of the computation, 
\begin{equation}
\Omega^i=|H(0)|/[2\pi\hbar g(0)]-\id\quad\mrm{and}\quad\Omega^f=|H(\tau)|/[2\pi\hbar J\Delta(\tau)]\,.
\end{equation}
Consequently, we have
\begin{equation}
\label{s1}
\Omega^i = \sum_{n=1}^{L}\sigma_x^n - \id \quad \text{and} \quad 
\Omega^f = \sum_{n=1}^{L-1}\sigma_z^n\sigma_z^{n+1}\,,
\end{equation}
where we included $\id$ in the definition of $\Omega^i$ to guarantee for the quantum efficacy $\varepsilon=1$ for unital dynamics, cf. Sec.~\ref{sec:gft}.

For the ideal computation, the initial state, $\rho_0$, is chosen to be given by $\rho_0=\ket{\boldsymbol{\rightarrow}}\bra{\boldsymbol{\rightarrow}}$, where $\ket{\boldsymbol{\rightarrow}}:=\ket{\cdots \rightarrow\rightarrow\rightarrow\cdots}$ is a non-degenerate, paramagnetic state -- the ground state of $H(0)$ (and thus of $\Omega^i$), where all spins are aligned along the $x$-direction. Consequently,
\begin{equation}
\label{s2}
M^i[\rho_0]=\rho_0 \quad \text{and\,\,measurement\,\,outcome} \quad \omega^i=L-1,
\end{equation}
as $\Omega^i$ and $H(0)$ commute by construction.

Moreover, if the quantum annealer is ideal, then the dynamics is not only unitary, but also adiabatic. For adiabatic evolution, we can write $\mathbb{E}_{\tau}\left[\rho\right] = U_{\tau}\rho U_{\tau}^{\dagger}$, where  
\begin{equation}
U_{\tau} = \mathcal{T}_{>}\exp\left(-\frac{i}{\hbar} \int_0^{\tau} H(s)\, ds\right) 
\end{equation}  
and as a result $\mathbb{E}_{\tau}\left[\rho_0\right] = \ket{\boldsymbol{f}}\bra{\boldsymbol{f}}$, where $\ket{\boldsymbol{f}}$ is the final state, a defect-free state where all spins are aligned along the $z$-direction, \emph{i.e.} $\ket{\boldsymbol{\uparrow}}$ or $\ket{\boldsymbol{\downarrow}}$. Therefore, $\omega^f=\omega^i$.

In general, however, due to decoherence~\cite{ZurekRMP}, dissipation~\cite{Chenu2017} or other (hardware) issues that may occur~\cite{Young2013PRA}, the evolution may be neither unitary nor adiabatic. Nevertheless, for the annealer to perform a useful~\cite{Hastings2009} computation its evolution, $\mathbb{E}_{\tau}$, has to map $\ket{\boldsymbol{\rightarrow}}$ onto $\ket{\boldsymbol{f}}$. 
Therefore, the quantum efficacy~\eqref{eq:qu_efficacy} simply becomes
\begin{equation}
\label{sgamma}
\varepsilon = e^{-\Delta\omega} \braket{\boldsymbol{f}}{\boldsymbol{f}} = 1,
\end{equation}
that is, a process independent quantity.    
\begin{figure}
\centering
	\includegraphics[width=.95\textwidth]{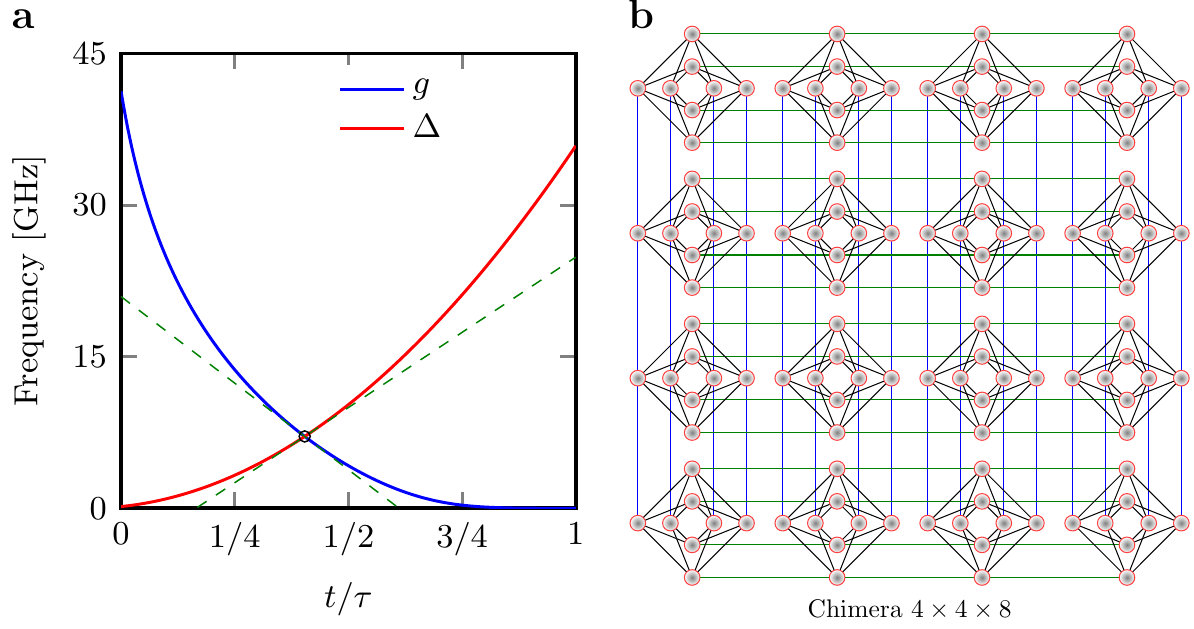}
	\caption{\label{fig:chimera} A typical annealing protocol for the quantum Ising chain implemented on the chimera graph (right panel). 
		The red lines are active couplings between qubits. The annealing time reads $\tau$.}
\end{figure}

Since the system starts, to very good approximation, from its ground state, $\ket{\boldsymbol{\rightarrow}}$, we can further write
\begin{equation}
p_{m\rightarrow n} = \delta_{0,m}\,p_{n|m} =  \delta_{0,m}\,p_{n|0}, 
\end{equation}
where $p_{n|0}$ is the probability of measuring $\omega_n^f$, conditioned on having first measured the ground state. Since we assume the latter event to be certain, $p_{n|0}\equiv p_n$ is just the probability of measuring the final outcome $\omega_n$ (we dropped the superscript). Therefore,
\begin{equation}
\langle e^{-\Delta\omega}\rangle = 
e^{-\Delta\omega}\,p_0 + \sum_{n\not = 0} e^{-\Delta\omega_n}\,p_n.
\end{equation}
Comparing this equation with  Eq.~\eqref{eq:GFT} we finally obtain a condition that is verifiable experimentally: 
\begin{equation}
\label{fdist}
p_n = \mathcal{P}(|\omega_n|) = 
\left\{
\begin{array}{l}
1 \quad \text{if} \quad |\omega_n| = L-1,  \\
0  \quad \text{otherwise.} 
\end{array}
\right.
\end{equation}

The probability density function $ \mathcal{P}(|\omega_n|) $  is characteristic for every process that transforms one ground state of the Ising Hamiltonian~\eqref{eq:ising} into another. It is important to note that the quantum fluctuation theorem~\eqref{eq:GFT} is valid for arbitrary duration $\tau$ -- any slow and fast processes. Therefore, even if a particular hardware does \emph{not} anneal the initial state adiabatically, but only unitally (which is not easy to verify experimentally) Eq.~(\ref{fdist}) still holds -- given that the computation 
starts and finishes in a ground state, as outlined above.

As an immediate consequence, every $\tau$-dependence of $\mathcal{P}$ must come from dissipation or decoherence. This is a clear indication that the hardware interacts with its environment in a way that cannot be neglected. 

\subsection{Experimental test on the D-Wave machine}


The above described framework was experimentally tested on a commercially available system -- the D-Wave machine \cite{Gardas2018}. An implementation of the Ising Hamiltonian \eqref{eq:ising} on the D-Wave machine is depicted in Fig.~\ref{fig:chimera}. On this platform, users can choose couplings $J_i$ and longitudinal magnetic field $h_i$, which were all set to zero \cite{Gardas2018}. In general, however, one can\emph{not} control the annealing process by manipulating $g(t)$ and $\Delta(t)$. In the ideal quantum annealer the quantum Ising chain~\eqref{eq:ising} undergoes unitary and adiabatic dynamics, while $\Delta(t)$ is varied from $\Delta(0)\approx 0$ to $\Delta(\tau) \gg 0$, and $g(t)$ from $g(0) \gg 0 $ to $g(\tau)\approx 0$ (cf. Fig.~\ref{fig:chimera}).

\begin{figure}
\centering
	\includegraphics[width=.8\textwidth]{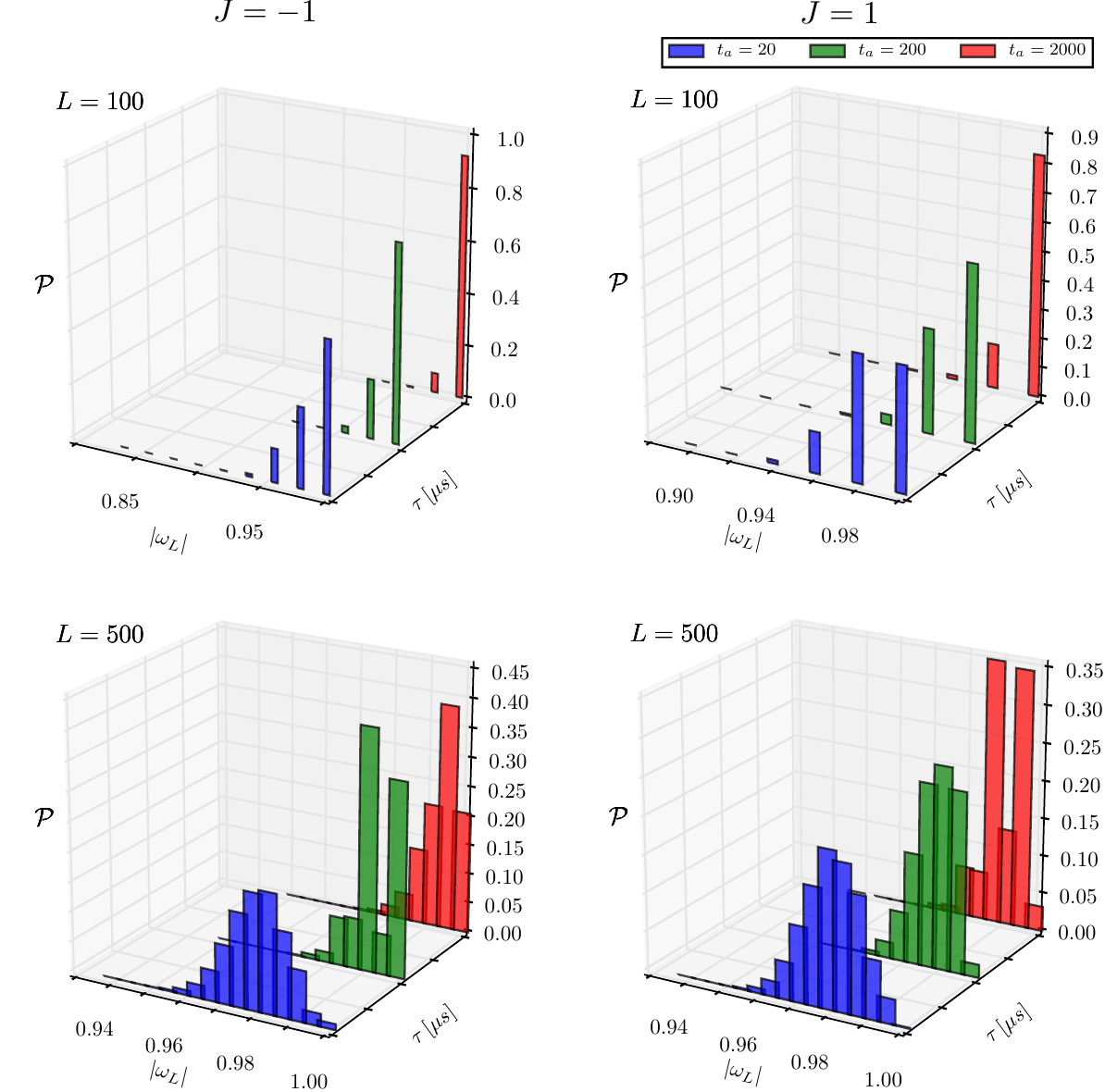}
	\caption{\label{fig:results_2000Q} Distribution $\mathcal{P}(\Delta\omega)$ for the quantum Ising chain~(\ref{eq:ising}) implemented on a D-Wave 2000Q annealer. Plot shows the final results for $J=-1$ (antiferromagnetic) and $J=1$ (ferromagnetic) cases, respectively. The renormalized energy is given by $\omega_L = \omega/(L-1)$, where $L$ is the length of a randomly chosen Ising chain.}
\end{figure}

\begin{figure}
\centering
	\includegraphics[width=.65\textwidth]{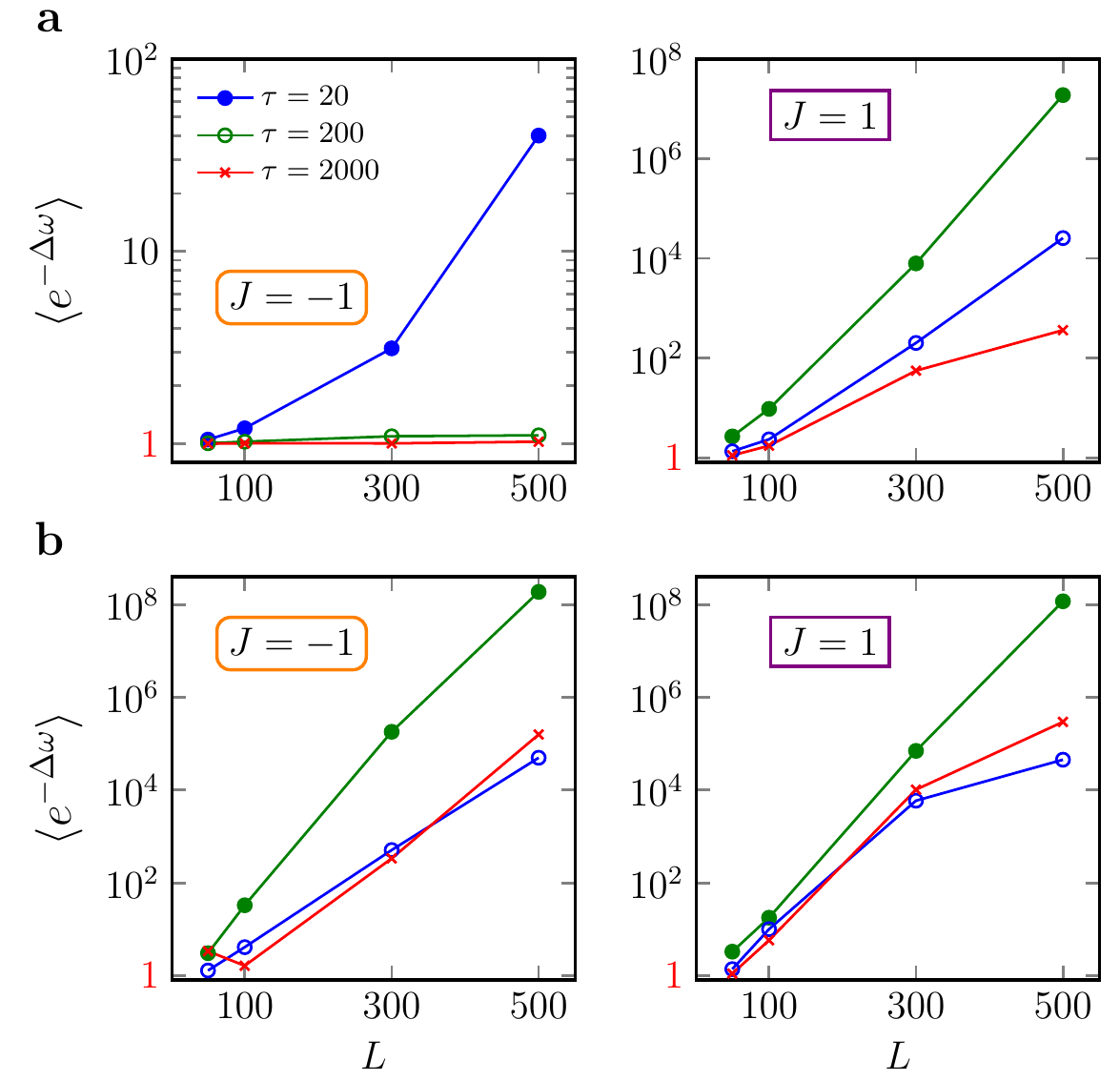}
	\caption{\label{fig:GFT_DWave} Exponential averages $\la\e{-\omega}\ra$ \eqref{sgamma} for experiments run on 2X (\textbf{a}) and 2000Q (\textbf{b}), which for unital dynamics should be identical to 1.}
\end{figure}

Ref.~\cite{Gardas2018} reported the experimental implementation of the above described protocol. To this end, several work distributions $\mathcal{P}(|\omega_n|) $  were generated  through ``annealing'' on two generations of the D-Wave machine ($2$X and $2000$Q), which implemented an Ising chain as encoded in Hamiltonian~(\ref{eq:ising}). All connections on the chimera graph were chosen \emph{randomly}. A typical example is shown in Fig.~\ref{fig:chimera}, where the red lines indicate nonzero $zz$-interactions between qubits \cite{Gardas2018}. The experiment was conducted $N=10^6$ times. Fig.~\ref{fig:results_2000Q} shows and example of the results obtained for different chain lengths $L$, couplings between qubits $J_i$ and annealing times $\tau$ on $2000$Q. The current D-Wave solver reports the final state energy which is computed classically from the measured eigenstates of the individual qubits. Fig.~\ref{fig:GFT_DWave} depicts the resulting exponential averages, $\la \e{-\Delta\omega}\ra$ for 2X and 2000Q \cite{Gardas2018}.

It was observed, that in the vast majority of all tested situations  $\mathcal{P}(|\omega_n|) $ is far from the theoretical prediction~\eqref{fdist} and the dynamics is clearly not even unital. Importantly, $\mathcal{P}$ clearly depends on $\tau$ indicating a large amount of computational errors are generated during the annealing. Similar conclusions were also obtained by other authors \cite{Albash2013,Boixo2013experimental,Albash2015,Albash2015a,Albash2012}.

Finally, it is interesting to realize that any departure from the  ideal distribution $\mathcal{P}$ (\ref{fdist}) for the Ising model indicates that the final state carries ``kinks'' (topological defects). Counting the exact number of such imperfections allows to determine by \emph{how much} the annealer misses the true ground state. Thus, one would expect that the excitations could be described by the  Kibble-Zurek  mechanism~\cite{Kibble1976,Zurek1985}, which we will discuss in the following section\footnote{Note, however, that a careful analysis \cite{Gardas2018KZM} revealed that the occurrence of computational errors in the D-Wave machine is more frequent as the phenomenological approach would predict.}.

\section{Kibble-Zurek Scaling of Irreversible Entropy \label{sec:KZM}}
\label{sec:KibbleZurek}

If the Universe started with a Big Bang during which all mass and energy was concentrated in an infinitely small volume, how come that nowadays matter is so sparsely distributed? Realizing that the early Universe must have undergone a phase transtion, Kibble noted that relativistic causality alone makes the creation of topological defects and the existence of finite domain sizes inevitable \cite{Kibble1976}. In laboratory phase transitions, however, relativistic causality does not lead to useful insights \cite{Zurek1985}. 

In thermodynamics  second order phase transitions can be classified into \emph{universality classes} \cite{Callen1985}. At the critical point thermodynamic response functions, such as the magnetic susceptibility, diverge, $\chi\sim |T-T_c|^{-\gamma}$, where $T$ is the temperature and $\gamma$ is called critical exponent. Typically,  $\gamma$ only depends on symmetries and not on microscopic details, and thus the values of $\gamma$ are \emph{universal} for classes of systems \cite{Fisher1974}. This divergence of response functions at the critical point can be understood as a ``freezing out'' of all dynamics. It is exactly this critical slowing down in the vicinity of the critical point that allows for the prediction of the density of defects, the size of typical domains, and their excitations \cite{Zurek1985}. The \emph{Kibble-Zurek mechanism} (KZM) has been very successfully tested in thermodynamic phase transitions, in trapped ions, in Bose-Einstein condensates in inhomogeneous systems, quantum phase transitions, and biochemical networks\footnote{For a comprehensive review of the current state-of-the-art we refer to the literature \cite{DelCampo2013c,DelCampo2014}}.

\subsection{Fundamentals of the Kibble-Zurek mechanism}

We begin by briefly reviewing the main notions of the KZM and establish notations. Close to the critical point both the correlation length, $\xi$, as well as the correlation time, $\tau_c$, diverge. Renormalization group theory predicts \cite{Fisher1974} that
\begin{equation}
\label{eq02}
\xi(\epsilon) = \xi_0\,\left|\epsilon\right|^{-\nu} \quad\mathrm{and}\quad\tau_c(\epsilon)=\tau_0\,\left|\epsilon\right|^{-z\nu}\,,
\end{equation}
where $\epsilon$ is a dimensionless parameter measuring the distance from the critical point, $\nu$ is the spatial and $z$ the dynamical critical exponent. In thermodynamic phase transtions $\epsilon$ is the relative temperature \cite{Zurek1985}, whereas in quantum phase transitions $\epsilon$ is a relative external field \cite{Zurek2005a}.

For the sake of simplicity we will assume that the system is driven through its phase transition by a linear ``quench''
\begin{equation}
\label{eq03}
\epsilon(t)=t/\tau_Q\,,
\end{equation}
and thus the constant quench rate $\dot{\epsilon}(t)$ is given by one over the quench time $\tau_Q$. 

\begin{figure}
\centering
\includegraphics[width=.8\textwidth]{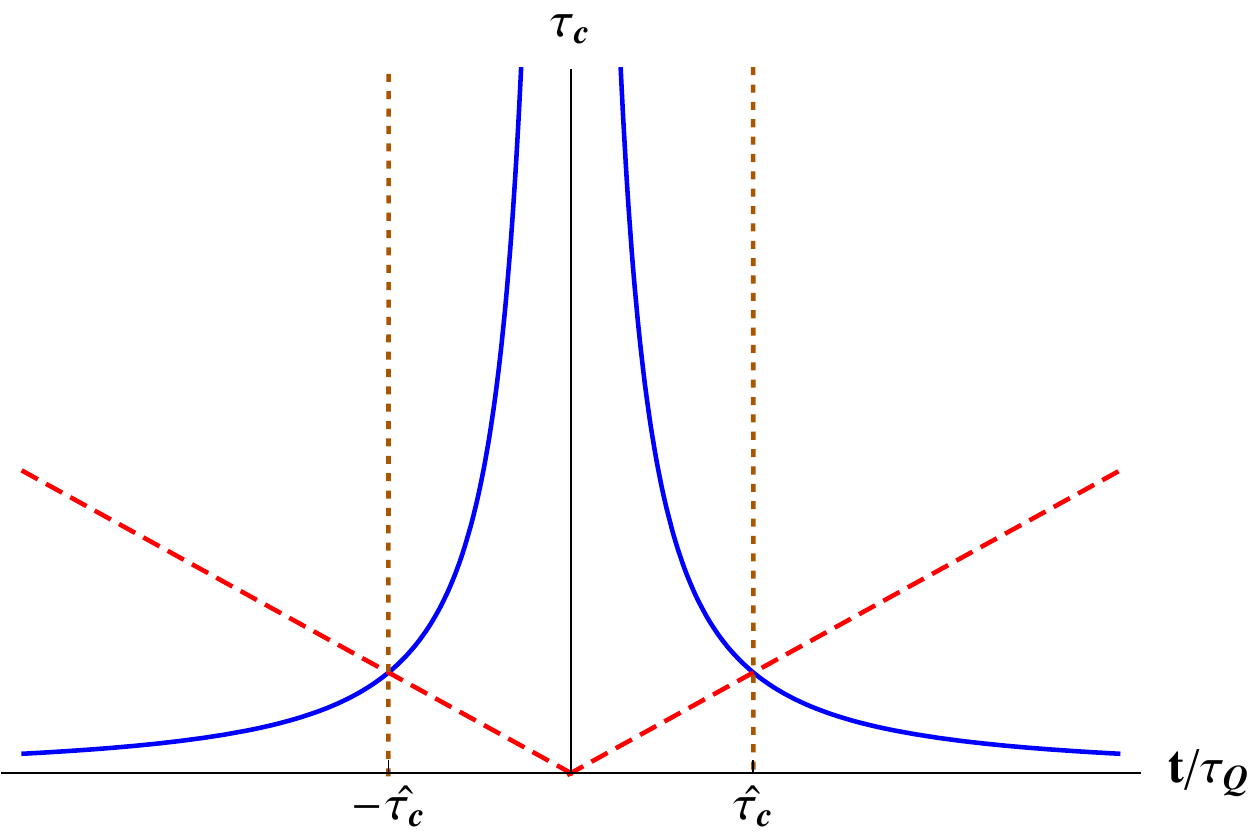}
\caption{\label{fig:plot_KZ}Relaxation time $\tau_c(t)$ \eqref{eq02} (blue, solid line) and rate of driving $|\dot{\epsilon}/\epsilon|$ (red, dashed line) for $\nu=1$ and $z=3/2$. The vertical lines illustrate the separation of the thermodynamic behavior into adiabatic and impulse regimes \cite{Zurek1996}.}
\end{figure}

For slow-enough driving and far from the critical point, $\tau_c\ll t$, the dynamics of the system is essentially adiabatic. This means, in particular, that all nonequilibrium excitations and defects equilibrate much faster than they are created. Close to the critical point, $\tau_c\simeq t$ the situation dramatically changes, since the response freezes out and defects and excitations cannot ``heal'' any longer.  This change of thermodynamic behavior, from adiabatic to ``impulse'' \cite{Zurek1996},  happens when the rate of driving becomes equal to the rate of relaxation, or more formally at
\begin{equation}
\label{eq04}
\hat{\tau}_c(\hat{t})=\hat{t}\quad\mrm{with}\quad\hat{\tau}_c=\left(\tau_0\,\tau_Q^{z \nu}\right)^\frac{1}{z\nu+1}\,.
\end{equation}
This insight is illustrated in Fig.~\ref{fig:plot_KZ}.

Accordingly the typical domain size is determined by the correlation length at $\hat{t}$, which can be written as,
\begin{equation}
\label{eq05}
\hat{\xi}=\xi(\hat{t})=\xi_0\,\left(\tau_Q/\tau_0\right)^\frac{\nu}{z\nu+1}\,.
\end{equation}
In many situations it is useful to introduce the density of defects $\rho_d$, which is given by the ratio $\hat{\xi}^d/\hat{\xi}^D$. Here $d$ and $D$ are the dimensions of defects and the space they live in, respectively. Thus, we can write,
\begin{equation}
\label{eq06}
\rho_d=\hat{\xi}^{(d-D)}\sim\tau_Q^{-\frac{(d-D)\,\nu}{z\nu+1}}\,,
\end{equation}
which sometimes is also called \emph{KZ-scaling}. It is important to emphasize that Eq.~\eqref{eq06} quantifies an effect of finite-rate, \emph{nonequilbirum} driving entirely in terms of the \emph{equilibrium} critical exponents. Note that in the original formulation of the KZM topological defects were considered since they constitute  robust signatures of the quench that can be easily counted. If, however, even correlation functions are accessible the scaling of the correlations length \eqref{eq05} can be directly measured.

\subsection{Example: the Landau-Zener model}

That the Kibble-Zurek mechanism might also apply to quantum phase transitions was first proposed in Ref.~\cite{Zurek2005a}. There it was numerically shown that final state of a quantum Ising chain driven at finite rate through its phase transitions is properly characterized by Eq.~\eqref{eq06}.

\begin{figure}
\centering
\includegraphics[width=.75\textwidth]{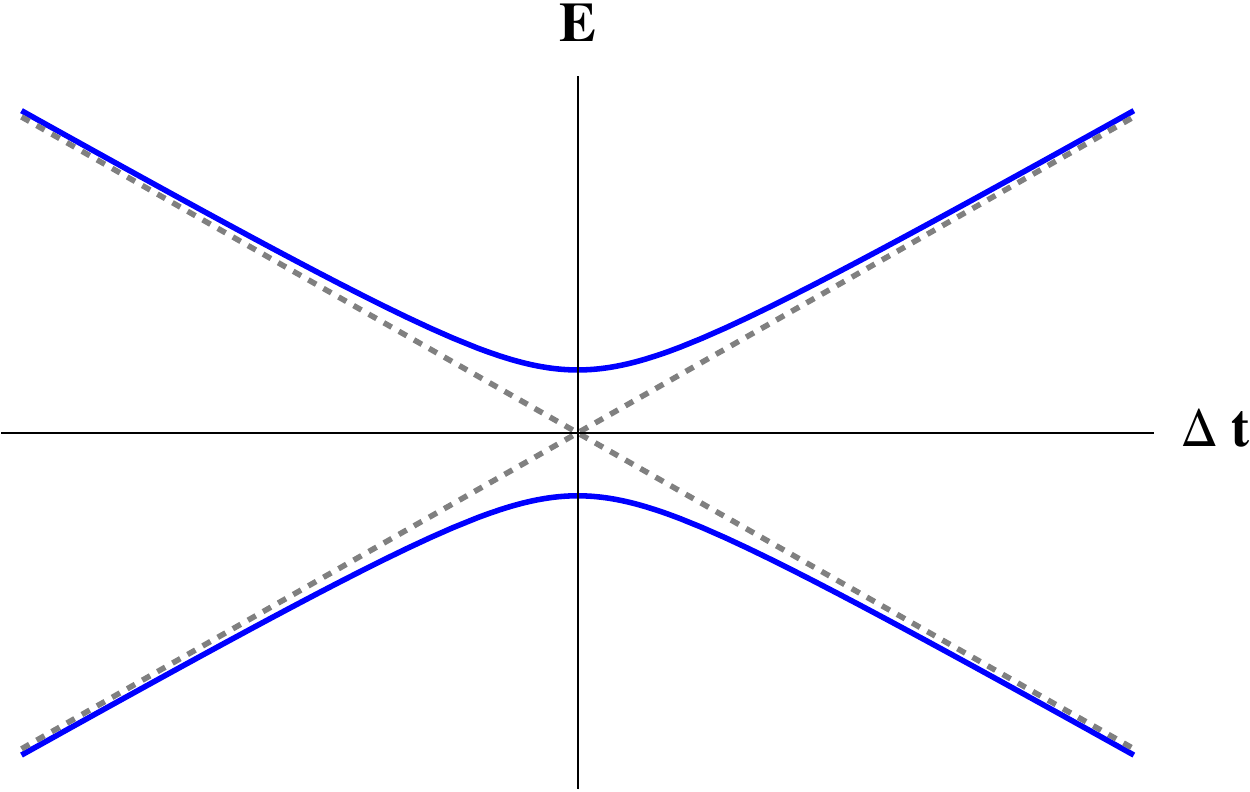}
\caption{\label{fig:LZ} Energy spectrum of the Landau-Zener Hamiltonian \eqref{eq:H_LZ} as a function of time for $\nu_0=1$ (dashed lines are for $\nu_0=0$).}
\end{figure}

Almost simultaneously it was recognized that the dynamics of the Landau-Zener model can provide illustrative insight into underpinnings of the phenomenological framework \cite{Damski2005}. Let us again consider a two-level system (TLS), for which we now write the time-dependent Hamiltonian as
\begin{equation}
\label{eq:H_LZ}
H(t)=\frac{1}{2}\,\begin{pmatrix}\Delta\, t&\nu_0\\\nu_0&-\Delta\, t\end{pmatrix}
\end{equation}
with eigenvalues $E_{1,2}=\pm \sqrt{\nu_0^2+(\Delta\,t)^2}$. In Fig.~\ref{fig:LZ} we plot the energy eigenvalues as a function of $t$, which clearly exhibits a so-called \emph{avoided crossing}.

For $t\ll0$ the ground state is given by $\ket{\uparrow}$, whereas for $t\gg0$ the ground state becomes $\ket{\downarrow}$. Therefore, at $t=0$ the TLS undergoes behavior that is reminiscent of a second order phase transition in mean-field theory \cite{Callen1985}. An analog of ``topological defects'' can then be introduced by considering the expectation value of the angular momentum. For perfectly adiabatic dynamics the angular momentum would be pointing downwards. However, if the system is driven at finite rate some population is excited, which results in precession of the angular momentum vector. Thus, the ``density of defects'' can be quantified by the fidelity with respect to the ground state fidelity,
\begin{equation}
\label{eq:dens}
\mc{D}(t)\equiv\left|\braket{\psi(t)}{\uparrow}\right|^2\,,
\end{equation}
where $\ket{\psi(t)}$ is a solution of the corresponding time-dependent Schr\"odinger equation.

In the following it will be instructive to further identify
\begin{equation}
\label{eq:tau}
\tau_c(t)=\frac{\tau_0}{\sqrt{1+\epsilon(t)^2}}\quad\mrm{and}\quad\epsilon(t)=\frac{\Delta\,t}{\nu_0}\equiv\frac{t}{\tau_Q}\,,
\end{equation}
where $\tau_Q\equiv\nu_0/\Delta$. Thus, to compare the solution of the exact dynamics we only need to determine the instant $\hat{t}$. Solving Eq.~\eqref{eq:tau} for $\hat{\tau}_c(\hat{t})=\hat{t}$ we obtain,
\begin{equation}
\label{eq:tau_hat}
\hat{t}=\pm\sqrt{\frac{\tau_Q}{2}}\,\sqrt{\sqrt{4 \tau_0^2+\tau_Q^2}-\tau_Q}\,.
\end{equation}

We now have all the ingredients to compute the density of defects $\mc{D}$ \eqref{eq:dens}. It is a straight forward exercise to show that for systems that initially started in $\ket{\uparrow}$ at $t_i=-\infty$ we have
\begin{equation}
\lim_{t\rightarrow\infty}\mc{D}(t)=\left|\braket{\uparrow(\hat{t})} {\downarrow(\hat{t})}\right|^2=\frac{\hat{\epsilon}^2}{1+\hat{\epsilon}^2}\,,
\end{equation}
where $\hat{\epsilon}=\epsilon(\hat{t})$ and $\ket{\uparrow(\hat{t})}$ and $\ket{\downarrow(\hat{t})}$ are the eigenstates of $H(t)$ \eqref{eq:H_LZ} at $t=\hat{t}$. Therefore, the density of defects behaves in leading order like,
\begin{equation}
\lim_{t\rightarrow\infty}\mc{D}(t)\sim\tau_Q^2
\end{equation}
Ref.~\cite{Damski2005} compared this phenomenological prediction with a numerical solution of the dynamics, and almost perfect agreement was found.

\subsection{Kibble-Zurek mechanism and entropy production}

More generally, a natural question is whether the irreversible entropy production, $\la\Sigma\ra$, exhibits a similar behavior. Naively one would expect that per excitation the system is accompanied by a characteristic amount of entropy, $\sigma$,
\begin{equation}
\label{eq07}
\la \Sigma\ra \sim \rho_d\,\cdot\,\sigma\sim \tau_Q^{-\frac{d\,\nu}{z\nu+1}}\,.
\end{equation}
Remarkably this naive expectation is not entirely correct.

\paragraph*{Maximum available work theorem.}

As we have discussed several times, the only processes that can be fully described by means of conventional thermodynamics are infinitely slow, equilibrium, a.k.a. \emph{quasistatic} processes \cite{Callen1985}. Nonequilibrium processes are characterized by the \emph{maximum available work theorem} \cite{Schlogl1989}. Consider a general thermodynamic system which supplies work to a work reservoir, and which is in contact, but \emph{not in equilibrium} with a heat reservoir, $\mc{B}$. Then the first law of thermodynamics can be written as,
\begin{equation}
\label{app:eq01}
\Delta E+\Delta E_\mc{B}=\la W\ra\,,
\end{equation}
where $\Delta E$ is the change of internal energy of the system, $\Delta E_\mc{B}$ is the energy exchanged with $\mc{B}$, and as before $\la W\ra$ denotes the average work. Accordingly the second law of thermodynamics states,
\begin{equation}
\label{app:eq02}
\Delta S+\Delta S_\mc{B}\geq 0\,,
\end{equation}
where $\Delta S$ is the change of thermodynamic entropy of the system, $\Delta S_\mc{B}$ is the change of entropy in $\mc{B}$, and where we used that the entropy of the work reservoir is negligible \cite{Callen1985}. Since the heat reservoir is so large that it is always in equilibrium at inverse temperature $\beta$ we immediately can write $\beta \Delta E_\mc{B}=\Delta S_\mc{B}$, and hence we always have
\begin{equation}
\label{app:eq03}
\la W\ra\geq \Delta E-\Delta S/\beta\equiv \Delta \mc{E}\,.
\end{equation}
The thermodynamic quantity $\mc{E}$ is called exergy or availability \cite{Schlogl1989}, since it quantifies the maximal available work in any thermodynamic process.

\paragraph*{KZ-scaling of the excess work.}

The \emph{maximal available work theorem} \cite{Schlogl1989} can be re-written in terms of the excess work, $\la W_\mrm{ex}\ra$, which is given by the total work, $\la W\ra$, minus the quasistatic contribution, i.e., the availability $\Delta\mc{E}$, 
\begin{equation}
\label{eq08}
\la W_\mrm{ex}\ra=\la W\ra-\Delta \mc{E}\,.
\end{equation}
At constant temperature we can write $\Delta \mc{E}\equiv\la W\ra-\Delta E+\Delta S/\beta$ \cite{Schlogl1989}, where $\Delta E$ is the change of internal energy in a quastistatic process, $\Delta S$ denotes the change of entropy, and $\beta$ is the inverse temperature. For open equilibrium systems and isothermal processes the availability further reduces to the difference in Helmholtz free energy, $\Delta\mc{E}=\Delta F$, why we can also write $\la \Sigma\ra=\beta \la W_\mrm{ex}\ra$. However, more generally $\Delta\mc{E}$ is the work performed during any quasistatic process, and thus $\la W_\mrm{ex}\ra$ quantifies the nonequilibrium excitations arising from finite time driving -- in isothermal as well as in more general processes, and in open as well as in isolated systems. 

Motivated by insights from finite-time thermodynamics \cite{Andresen1984} it has recently become clear  that for sufficiently slow processes $\la W_\mrm{ex}\ra$ can be  expressed as quadratic form  \cite{Sivak2012a,Bonanca2014a},
\begin{equation}
\label{eq09}
\la W_\mrm{ex}\ra=\int dt\,\frac{d \mb\lambda^\dagger}{dt}\,\tau_c(t)\,\mc{I}(t)\,\frac{d \mb\lambda}{dt}\,,
\end{equation}
where $\mb{\lambda}=(T,V,H,\dots)$ is the vector of all intensive parameters varied during the process, such as temperature $T$, volume $V$, magnetic field $H$, etc., and the integral is taken over the whole process. Furthermore, $\mc{I}(t)$ is the Fisher information matrix, which  for a $d$ dimensional system close to the critical point and for only two intensive parameters such as $T$ and $H$ can be written as \cite{Prokopenko2011},
\begin{equation}
\label{eq10}
\mc{I}(t)\sim\begin{pmatrix}|\epsilon(t)|^{-\alpha}&|\epsilon(t)|^{b-1}\\|\epsilon(t)|^{b-1}&|\epsilon(t)|^{-\gamma}\end{pmatrix}
\end{equation}
where $\gamma=d \nu-2 b$, and $\alpha$ is the critical exponent corresponding to changes in temperature.

For the sake of simplicity we will now assume that only one intensive parameter, $\lambda(t)$, is varied. Thus, we can express the $(1\times 1)$-dimensional Fisher information matrix in terms of the general susceptibility $\mc{X}(t)$,
\begin{equation}
\label{eq11}
\mc{I}(t)=\mc{X}(t)=\mc{X}_0\,|\epsilon(t)|^{-\Lambda}
 \end{equation}
where $\Lambda$ is the critical exponent corresponding to the varied control parameter, e.g., for varied magnetic fields we have $\Lambda=\gamma$, and for processes with time-dependent temperatures $\Lambda=\alpha$.

The Kibble-Zurek hypothesis predicts that far from the critical point, $|t|\gg \hat{\tau}_c$, the dynamics is essentially adiabatic, and hence $\la W_\mrm{ex}\ra$ has non-vanishing contributions only in the impulse regime, $|t|\leq\hat{\tau}_c$, cf. Fig.~\ref{fig:plot_KZ}. Therefore, we can write,
\begin{equation}
\label{eq12}
\la W_\mrm{ex}\ra\simeq\lambda_c^2\,\int_{-n \hat{\tau}_c}^{n \hat{\tau}_c}dt\,|\dot{\epsilon}(t)|^2\,\tau_c(t)\,\mc{X}(t)
\end{equation}
where $\lambda(t)=\lambda_c (1-\epsilon(t))$ and $n>1$ is a small, real constant \footnote{We included the small, real constant $n>1$ to guarantee that no non-negligible contributions to the excess work are neglected.}. Employing Eqs.~\eqref{eq02} and \eqref{eq11} it is then a simple exercise to show that
\begin{equation}
\label{eq13}
\la W_\mrm{ex}\ra=\frac{2\lambda_c\,\mc{X}_0\,n^{-z\nu-\Lambda+1}}{z\nu+\Lambda-1}\,\tau_0^{\frac{2-\Lambda}{z\nu+1}}\,\tau_Q^{\frac{\Lambda-2}{z\nu+1}}\,.
\end{equation}
Thus, we have shown that for systems that are driven at constant rate through a critical point the excess work, $\la W_\mrm{ex}\ra$, universally scales like \cite{KibbleZurekPRE},
\begin{equation}
\label{eq14}
\la W_\mrm{ex}\ra\sim\tau_Q^{\frac{\Lambda-2}{z\nu+1}}\,,
\end{equation}
which explicitly depends on the critical exponent $\Lambda$ corresponding to how the system is driven. This behavior is in full agreement with thermodynamics, since thermodynamic work is a \emph{process dependent} quantity \cite{Callen1985}. In other words, Eq.~\eqref{eq14} expresses the fact that the excess work depends on \emph{how} the system is driven through the critical point, whereas the typical domain size $\hat{\xi}$ \eqref{eq05} is independent on the choice of the intensive control parameter.

\paragraph{Quantum Ising model.}

Before we proceed we briefly comment on the consistency of our conceptual arguments with an analytically solvable model. The Kibble-Zurek mechanism has been extensively studied for the quantum Ising chain \eqref{eq:ising}. In the limit of infinitely many spins, $N\to\infty$, and at $\Delta\, J=1$ the Ising has two critical points at $g_c =\pm 1$, with a ferromagnetic phase for $|g| < 1$ and two paramagnetic phases for  $|g| > 1$. The critical exponents are given by $z=1$ and $\nu=1$.

A recent and very thorough study of the KZM in this model \cite{Francuz2015} revealed that the excess work scales like $\la W_\mrm{ex}\ra\sim \tau_Q^{-1}$. In Ref.~\cite{Francuz2015} this behavior was explained by noting that close to the critical point the dispersion relation is no longer flat, but rather $\la W_\mrm{ex}\ra\sim\hat{\xi}^{-2}$.  This finding is fully consistent with our general result \eqref{eq14}. Note that $\Lambda=0$ \cite{Herbut2007a} and hence Eq.~\eqref{eq14} immediately  predicts $\la W_\mrm{ex}\ra\sim \tau_Q^{-1}$.

\section{Error correction in adiabatic quantum computers \label{sec:STA}}

Not only for adiabatic quantum computers, but actually for \emph{any} quantum computer it has been recognized that the implementation of \emph{quantum error correcting algorithms} \cite{Nielsen2010} is a necessity. Loosely speaking, any such algorithm works by encoding logical quantum states in several physical states that can be controlled separately and in parallel. In this way, the logical quantum states can be made resilient against the effects of noise, such as decoherence and dissipation. However, due to the delicate nature of entanglement and decoherence, quantum error correction is a little more involved than correcting errors in classical computers.

\paragraph{Classical error correction.}

The basic principle is most easily demonstrated by a standard communication problem. Imagine we wish to send one bit through a noisy classical channel. The effect of the noise can be described by a probability $p$, with which a bit flip occurs. Thus, with probability $1-p$ the bit is transmitted correctly. Then, the bit can be protected by sending several independent copies, and by ``taking majority'' votes at the receiving end. Suppose a \emph{logical bit} is encoded in three \emph{physical bits}
\begin{equation}
0\rightarrow 000 \quad\mrm{and}\quad 1\rightarrow 111,
\end{equation}
and at the receiving end we obtain $010$. Then with probability $1-(1-p)^2 p$ the logical bit $0$ was transmitted, and only one physical bit was affected by the noise. It should be noted that such \emph{majority voting} fails if two or more of the physical bits were flipped, which however is very unlikely as long as $p$ is not too large.

This kind of code is known as \emph{repetition code}, which closely resembles error correction in every day conversations. To make sure to be understood, the same message is repeated several times. Unfortunately, generic features of quantum physics make error correction in quantum computers more involved.

\paragraph{Quantum error correction.}

Naively, one would hope that quantum error correction could be facilitated by similar principles. However, the intricacies of quantum information pose significant challenges. These can be summarized under the following three issues \cite{Nielsen2010}:
\begin{enumerate}
\item Measurement back action: Standard quantum measurements are rather invasive, and they typically ``destroy'' the quantum state. Thus recovery of quantum information after observation is not possible.
\item Continuous quantum noise: Quantum noise is not restricted to only discrete bit flips, but rather continuous and cumulative errors can occur affecting the phase, or  resulting in loss of coherence and entanglement.
\item Quantum states cannot be copied \cite{Wootters1982}: A hallmark result of quantum information theory is the \emph{no cloning theorem} \cite{Wootters1982}. Thus, it is not immediately clear how one would implement a repetition code.
\end{enumerate}
Fortunately, none of these complications are debilitating enough to make quantum computing an impossibility. Rather, it has been shown that quantum bit flips as well as quantum phase flips can be corrected. A seminal result is the \emph{Shor code} \cite{Shor1995} that protects a single qubit against any arbitrary error. In this scheme,
\begin{equation}
\ket{0}\rightarrow\frac{\left(\ket{000}+\ket{111}\right)^3}{2\sqrt{2}}\quad\mrm{and}\quad \ket{1}\rightarrow\frac{\left(\ket{000}-\ket{111}\right)^3}{2\sqrt{2}}\,.
\end{equation}
Rather remarkably, the utility of the Shor code has been demonstrated  in several experiments, see for instance Ref.~\cite{Briegel2000,Reed2012,Bell2014}. 

However, most quantum error correcting codes have in common that they have been developed for \emph{gate based quantum computation}. In this paradigm, a quantum algorithm is constructed as a sequence of unitary maps acting upon a set of logical qubits.

\subsection{Quantum error correction in quantum annealers}

Naturally, any real quantum annealer will also be subject to effects of environmental noise, such as decoherence and dissipation. However, in contrast to gate based quantum computers, for quantum annealers computational errors fall into two independent categories \cite{Young2013}: (i) \emph{fundamentally correctable errors} that are effects of environmental noise and (ii) \emph{fundamentally non-correctable errors} that arise from excitations away from the ground state manifold due to finite time driving.

\paragraph{Fundamentally correctable errors -- quantum annealing correction.} 

For the first type of errors a successful and experimentally tested error correction scheme was devised in Ref.~\cite{Pudenz2014}. To this end, consider the general Ising Hamiltonian
\begin{equation}
H_\mrm{Ising}=\sum_{i=1}^N h_i \,\sigma^i_z+\sum_{i<j}^N J_{ij}\,\sigma^i_z\sigma^j_z\,,
\end{equation}
which allows to encode many hard and important optimization problems. Similarly to above, cf. Sec.~\ref{sec:dwave}, the solution of the optimization problem is found by letting the initial quantum state evolve under the time-dependent Hamiltonian
\begin{equation}
H(t)=A(t)\,H_X+B(t)\, H_\mrm{Ising}\quad\mrm{for}\quad 0\leq t\leq \tau\,.
\end{equation}
Here, $H_X=\sum_{i=1}^N \sigma^i_x$, and $A(t)$ and $B(t)$ are time-dependent functions satisfying $A(\tau)=B(0)=0$.

\emph{Quantum annealing correction} is a combined strategy comprising an energy penalty (EP) together with encoding and error correction. To this end, $H_\mrm{Ising}$ is ``encoded'' by replacing each $\sigma^i_z$ by its encoded counterpart $\la\sigma^i_z\ra=\sum_{\ell=1}^n \sigma^{i,\ell}_z$ and each $\sigma^i_z\sigma^j_z$ by $\la\sigma^i_z\sigma^j_z\ra=\sum_{\ell=1}^n \sigma^{i,\ell}_z\sigma^{j,\ell}_z$, where $\ell$ is an index counting the \emph{physical} qubits encoding a single \emph{logical} qubit. Therefore, the encoded Hamiltonian can be written as
\begin{equation}
\la{H}_\mrm{Ising}\ra=\sum_{i=1}^{\la N\ra} h_i \,\la \sigma^i_z\ra + \sum_{i<j}^{\la N\ra} J_{ij}\,\la \sigma^i_z \sigma^j_z\ra
\end{equation}
where $\la N\ra$ is the number of logical qubits. It is important to note that $H_X$ \emph{cannot} be encoded in this simple manner, as this would require $n$-body interactions.

Additional protection is provided by introducing a ferromagnetic penalty term
\begin{equation}
H_P=-\sum_{i=1}^{\la N\ra}\sum_{\ell=1}^n \sigma_z^{i,\ell}\,\sigma_z^{iP}\,,
\end{equation}
which is a sum of stabilizer generators for the $n+1$ qubit repetition code. In effect, $H_P$ detects and energetically penalizes all physical bit-flip errors, but not a bit-flip of the logical qubit.

Combining both encoding and energy penalty the total, encoded Hamiltonian becomes,
\begin{equation}
\label{eq:H_encoded}
\la H(t)\ra=A(t)\,H_X+B(t)\,\la H_\mrm{Ising}(\nu,\mu)\ra\,,
\end{equation}
where $\la H_\mrm{Ising}(\nu,\mu)\ra\equiv \nu\,\la H_\mrm{Ising}\ra+\mu\, H_P$ and $\nu$ is the ``problem scale'' and $\mu$ the penalty scale. 

This \emph{quantum annealing correction} was successfully tested for $n=3$ on the second generation of the D-Wave machine \cite{Pudenz2014,Pudenz2015}. It was demonstrated that environmental noise can be efficiently corrected for. However, such a quantum error correction code cannot circumvent fundamentally non-correctable errors.

\subsection{Adiabatic quantum computing -- A case for shortcuts to adiabaticity}

Finally, we will briefly outline a possible way of avoiding \emph{fundamentally non-correctable errors} from happening in the first place. As we have already seen above, excitations that will naturally occur in non-equilibrium processes can be avoided by driving the systems sufficiently slowly. However, this requires a high-degree of control over the systems and still limits the utility of such devices as the time scales required for their operation will tend to grow as the size of the system grows. Therefore, in recent years a great deal of theoretical and experimental research has been dedicated to mathematical tools and practical schemes to suppress these excitations in finite-time, nonequilibrium processes. To this end, a variety of techniques has been developed, such as the use of dynamical invariants, the inversion of scaling laws, the fast-forward technique, transitionless quantum driving, local counterdiabatic driving, optimal protocols from optimal control theory, optimal driving from properties of the quantum work statistics, ``environment'' assisted methods, using the properties of Lie algebras, and approximate methods such as linear response theory and fast quasistatic dynamics\footnote{See Ref.~\cite{Torrontegui2013} and references therein for an extensive review of these techniques.}. Among this plethora of different approaches, \emph{transitionless quantum driving} stands out, since it is the only method that suppresses excitations away from the adiabatic manifold at all instants.

\paragraph{Transitionless quantum driving.}

In the paradigm of transitionless quantum driving~\cite{Demirplak2003, Demirplak2005, Berry2009} one considers a time-dependent Hamiltonian $H_0(t)$ with instantaneous eigenvalues $\{\epsilon_n(t)\}$ and eigenstates $\{\ket{n(t)}\}$. In the limit of infinitely slow variation of $H_0(t)$ a solution of the dynamics is given by
\begin{equation}
\label{eq:adiabatic_approximation}
\ket{\psi_n(t)}=\e{-\frac{i}{\hbar}\int_0^t\td s\,\epsilon_n(s)-\int_0^t\td s\la n|\partial_{s}n\ra }\ket{n(t)}. 
\end{equation}
In this adiabatic limit no transitions between eigenstates occur~\cite{Messiah1966}, and each eigenstate acquires a time-dependent phase that can be separated into a dynamical and a geometric contribution~\cite{Berry1984}, represented by the two terms inside the exponential in the above expression.

Now, a corresponding Hamiltonian $H(t)$ is constructed, such that the adiabatic approximation associated with $H_0(t)$~\eqref{eq:adiabatic_approximation} is an exact solution of the dynamics generated by $H(t)$ under the time-dependent Schr\" odinger equation. Writing the time-evolution operator as $U(t)=\sum_n \ket{\psi_n(t)}\bra{n(0)}$, one arrives at an explicit expression for $H(t)$ \cite{Demirplak2003, Demirplak2005, Berry2009}:
\begin{equation}
\label{q01}
H= H_0+H_1=H_0+ i\hbar\sum_n\left(\ket{\partial_tn}\bra{n}-\braket{n}{\partial_t n}\ket{n}\bra{ n}\right)\,.
\end{equation}
Here, the {\it auxiliary Hamiltonian} $H_1(t)$ enforces evolution along the adiabatic manifold of $H_0(t)$: if a system is prepared in an eigenstate $\ket{n(0)}$ of $H_0(0)$ and subsequently evolves under $H(t)$, then the term $H_1(t)$ effectively suppresses the non-adiabatic transitions out of $\ket{n(t)}$ that would arise in the absence of this term. Through a little manipulation an equivalent expression for $H_1(t)$ can be found~\cite{Berry2009}
\begin{equation}
\label{H1berry}
H_1(t) = i\hbar \sum_{m\neq n} \sum \frac{\ketbrad{m} \partial_t H_0 \ketbrad{n}}{E_n-E_m}.
\end{equation}

From the set-up, transitionless quantum driving appears to be uniquely suited to suppress excitations from finite time driving and thereby bypass fundamentally non-correctable errors in quantum annealers.

\subsection{Counterdiabatic Hamiltonian for scale-invariant driving\label{sec:xp}}

A major obstacle arises from the fact that it is rarely feasible to find closed-form expressions for the counterdiabatic field \eqref{H1berry}, i.e., expressions that do not depend on the full spectral decomposition of $H_0(t)$. However, there is a reasonably broad class of systems in which the situation greatly simplifies \cite{Deffner2014}.

\paragraph{Example: Transitionless quantum driving of the harmonic oscillator.}

As a instructive example consider the parametric harmonic oscillator
\begin{equation}
\label{eq:para_harm}
H_0(t) = \frac{\hbar \omega(t)}{2} \left(a^\dagger a + \frac{1}{2}\right)
\end{equation}
where $a^\dagger$ and $a$ are the creation and annihilation operators, respectively. Using Eq.~\eqref{H1berry} we see that in order to determine the auxiliary Hamiltonian we require the time derivative of the bare Hamiltonian. Using the definitions of $a^\dagger$ and $a$ in terms of position and momentum operators, $x$ and $p$
\begin{equation}
\label{adaggerdef}
\begin{aligned}
a^\dagger &=& \sqrt{\frac{m \omega(t)}{2 \hbar}} \left( x - \frac{i}{m \omega(t)} p  \right) \\
a &=& \sqrt{\frac{m \omega(t)}{2 \hbar}} \left( x + \frac{i}{m \omega(t)} p  \right)
\end{aligned}
\end{equation}
it is evident that they are time-dependent operators. Taking the derivative of $H_0$ we find
\begin{equation}
\partial_t H_0 = \frac{\hbar \dot{\omega}}{2} \left( a^\dagger a + \frac{1}{2} \right) + \frac{\hbar \omega}{2}\left( \dot{a}^\dagger a + a^\dagger \dot{a} \right).
\end{equation}
where we have dropped the explicit time dependence for brevity and where `$\dot{~}$' denotes the time derivative. From Eq.~\eqref{adaggerdef} we can readily determine the derivatives of the creation and annihiliation operators
\begin{equation}
\dot{a}^\dagger = \frac{\dot{\omega}}{2\omega} a \qquad\qquad \dot{a} = \frac{\dot{\omega}}{2\omega} a^\dagger
\end{equation}
Therefore we have the derivative of the bare Hamiltonian is given by
\begin{equation}
\partial_t H_0 = \frac{\hbar \dot{\omega}}{2} \left( a^\dagger a + \frac{1}{2} \right) + \frac{\hbar \dot{\omega}}{4}\left( \dot{a}^{\dagger^2} + a^2 \right).
\end{equation}
To determine the auxiliary Hamiltonian we use Eq.~\eqref{H1berry}. Since in the sums $m\neq n $, it is clear that the first term will not contribute. After some manipulation we finally arrive at the concise expression
\begin{equation}
\label{TQDHO}
H_1(t) = i\hbar \frac{\dot{\omega}}{4\omega} \left( a^2 - a^{\dagger^2} \right).
\end{equation}
Notice that no assumptions or constraints have been put on the form of time-dependence. Thus, the transitionless driving approach allows for an arbitrary ramp to be applied and for the driving to occur, at least in principle, in arbitrarily short times~\cite{CampbellPRL2017, FunoPRL2017}.

\paragraph{General case.}

The parametric harmonic oscillator \eqref{eq:para_harm} belongs to the broader class of so-called scale-invariantly driven systems. Scale-invariant driving refers to transformations of the Hamiltonian which can be absorbed by scaling of coordinates, time, energy, and possibly other variables to rewrite the transformed Hamiltonian in its original form up to a multiplicative factor. If only the potential term $V(q,\mb{\lambda}(t))$ is modulated, its overall shape does not change under $\mb{\lambda}(0)\rightarrow\mb{\lambda}(t)$. 

In the simplest case, consider a quantum system with a single degree of freedom,
\begin{equation}
\label{q02}
H_0(t)=\frac{p^2}{2m}+V(q,\mb{\lambda}(t))=\frac{p^2}{2m}+\frac{1}{\gamma^2}\,V_0\left(\frac{q-f}{\gamma}\right)\,,
\end{equation}
where $\mb{\lambda}=(\gamma,f)$ and $V_0(q)=V(q,\mb{\lambda}(0))$. Note that generally $\gamma=\gamma(t)$ and $f=f(t)$ are both allowed to be time-dependent, but we assume that they are independent of each other. This time-dependence encompasses transport processes ($\gamma(t)=1$), dilations (such as an expansion or compression, with $f(t)=0$) and combined dynamics.

It can now be shown \cite{Deffner2014} that  the auxiliary term $H_1(t)$ \eqref{q01} can be brought into a form that does not rely on the spectral decomposition of $H_0(t)$. To see this, consider that of $\psi_n^0(q)=\braket{n}{q}$ is an eigenfunction of the Hamiltonian $H_0(\gamma=1,f=0)$, then $\psi_n(q,\gamma,f)=\alpha(\gamma)\,\psi_n^0\left((q-f)/\gamma\right)$ is an eigenfunction of $H_0(\gamma,f)$, where $\alpha(\gamma)=1/\sqrt{\gamma}$ is a normalization constant. 

Now, we want to use this symmetry to simplify $H_1(t)$ in Eq.~\eqref{q01}. We can write,
\begin{equation}
\label{q04}
H_1(t)=i\hbar \dot{\mb{\lambda}}\cdot \,\sum\limits_m\,\left(\ket{\nabla_{\mb{\lambda}}\,m}\bra{m}-\braket{m}{\nabla_{\mb{\lambda}}\,m}\ket{m}\bra{m}\right)\,,
\end{equation}
which reads in space representation
\begin{equation}
\label{q05}
\begin{split}
H_1(t)=i\hbar\dot{\mb{\lambda}}\cdot\,\sum\limits_m\,\int\td q\,\ket{q}\,\nabla_{\mb\lambda}\,\psi_m(q,\mb{\lambda})\,\bra{m}-i\hbar\dot{\mb{\lambda}}\cdot\,\sum\limits_m\,\int\td q\,\braket{m}{q}\,\nabla_{\mb\lambda} \psi_m(q,\mb{\lambda}) \ket{m}\bra{m}\,. 
\end{split}
\end{equation}
To simplify this expression,  we note that
\begin{equation}
\label{q07}
\begin{split}
\nabla_{\mb{\lambda}}\psi_n(q,\mb{\lambda})=\left(\frac{\alpha'(\gamma)}{\alpha(\gamma)}\,\psi_n(q,\mb{\gamma})-\frac{q-f}{\gamma}\,\pd_q\psi_n(q,\gamma),\,\,-\pd_q\psi_n(q,\mb{\gamma})\right)\,.
\end{split}
\end{equation}
For the sake of clarity, let us treat both terms of $H_1(t)$ in \eqref{q05} separately. We obtain for the first term
\begin{equation}
\label{q08}
\begin{split}
i\hbar\dot{\mb{\lambda}}\cdot\,\sum\limits_m\,\int\td q\,\ket{q}\,\nabla_{\mb\lambda}\,\psi_m(q,\mb{\lambda})\,\bra{m}=\frac{\dot{\gamma}}{\gamma}\,\left(q-f\right)p+i\hbar\dot{\gamma}\,\frac{\alpha'(\gamma)}{\alpha(\gamma)}+\dot{f} \,p\,,
\end{split}
\end{equation}
while the second term reduces to
\begin{equation}
\label{q09}
\begin{split}
-i\hbar\dot{\mb{\lambda}}\cdot\,\sum\limits_m\,\int\td q\,\braket{m}{q}\,\nabla_{\mb\lambda} \psi_m(q,\mb{\lambda}) \ket{m}\bra{m}=-\frac{i\hbar\dot{\gamma}}{2\gamma}-i\hbar\dot{\gamma}\,\frac{\alpha'(\gamma)}{\alpha(\gamma)}\,.
\end{split}
\end{equation}
Note that the second component of $\nabla_{\mb{\lambda}}\psi_n(q,\mb{\lambda})$ does not contribute, since the wavefunction vanishes at infinity due to normalizability. In conclusion, we obtain the explicit expression of  the auxiliary CD Hamiltonian,
\begin{equation}
\label{q10}
H_1(t)=\frac{\dot{\gamma}}{2\gamma}\left[\left(q-f\right)\,p+p\,\left(q-f\right)\right]+\dot{f}\,p\,,
\end{equation}
where we used $\com{q-f}{p}=i\hbar$. Notice that $H_1(t)$ in Eq.~\eqref{q10} is of the general form $\hat{H}_1\propto (qp+pq)$ which is identical to \eqref{TQDHO}.

Equation~\eqref{q10} is a remarkable result. For all driving protocols under which the original Hamiltonian $H_0(t)$ is scale-invariant, i.e., where the time-dependent potential is of the form \eqref{q02}, the auxiliary term $H_1(t)$ takes the closed form \eqref{q10}. In particular, $H_1(t)$ is independent of the explicit energy eigenfunctions, and only depends on the anticommutator, $H_1\propto \{q,p\}=qp+pq$, the generator of dilations. 

As a result, CD applies not only to single eigenstates, but also to non-stationary quantum superpositions and mixed states.
However, the expression \eqref{q10} is still not particularly practical as non-local  Hamiltonians \footnote{Hamiltonians that include products of space and momentum operator, $q$ and $p$, are \textit{non-local}, whereas \textit{local} Hamiltonians contain only terms that depend on at most sums of $q$ and $p$.} are hard to realize in the laboratory. 

\paragraph{Transitionless quantum driving and quantum annealing.}

We have seen above, cf. Sec.~\ref{sec:dwave}, that quantum spin chains, such as the Ising chain \eqref{eq:ising}, offer a promising architecture for realizing quantum computational models. With this in mind, let us examine an alternative potential spin-system given by the ferromagnetic Lipkin-Meshkov-Glick (LMG) model, which allows us to exploit the previous results for transitionless driving of the harmonic oscillator. The LMG model is described by the Hamiltonian~\cite{LMG0, LMG1, LMG2, Campbell2014}.
\begin{equation}
H_0(t)=-\frac{1}{N}\left( \sum_{i<j} \sigma_x^i \otimes \sigma_x^j+\chi\sigma_y^i \otimes \sigma_y^j \right) - h(t) \sum_{i} \sigma_z^i 
\end{equation}
where $\sigma_{x,y,z}$ are again the Pauli spin-operators, $h(t)$ is the time-dependent magnetic field strength, and $\chi$ is here the anisotropy parameter. By considering the collective spin operators $S_j=\sum_i \sigma_{j}^i/2$ with $j=\{x,y,z\}$, the model can be written as
\begin{equation}
\label{collspinLMG}
H_0=-\frac{2}{N}\left( S_x^2 + \chi S_y^2 \right) - 2h S_z + \frac{1+\chi}{2}.
\end{equation}
For $N\to\infty$, the model can be solved through the Holstein-Primakoff (HP) transformation that allows us to map the spin model to an equivalent harmonic oscillator. Similarly to the quantum Ising model, the LMG model exhibits a quantum phase transition in its ground state when $h\!=\!1$. Depending on the phase that one is considering, the HP transformation must be taken along the direction that the classical angular momentum, 
\begin{equation}
\text{{\bf S}}=\frac{N}{2}(\sin\varphi\cos\phi,\sin\varphi\sin\phi,\cos\varphi),
\end{equation}
points. For $h>1$ we find this is always along the $z$-axis. Neglecting terms higher than $O(N)$ the HP transformation in this limit is
\begin{equation}
\label{HP1}
S_+=\sqrt{N} a, ~~~~~S_-=\sqrt{N} a^\dagger,~~~~~S_z=\frac{N}{2}-a^\dagger a,
\end{equation}
with
\begin{equation}
\label{HP2}
S_x=\frac{1}{2}(S_+ + S_-)~~~~~\text{and}~~~~S_y=\frac{1}{2i}(S_+ - S_-).
\end{equation}
This results in the mapped Hamiltonian in terms of bosonic creation and annihilation operators
\begin{equation}
H_b=-\frac{1-\chi}{2} \left( a^2 + a^{\dagger~2} \right) + (2h-1-\chi) a^\dagger a - hN,
\end{equation}
which can then be written in diagonal form by performing the following Bogoliubov transformation
\begin{eqnarray}
a&=\sinh\left(\frac{\alpha}{2}\right) b^\dagger + \cosh\left(\frac{\alpha}{2}\right) b,\\
a^\dagger&=\sinh\left(\frac{\alpha}{2}\right) b + \cosh\left(\frac{\alpha}{2}\right) b^\dagger,
\end{eqnarray}
and taking 
\begin{equation}
\tanh \alpha = \frac{1-\chi}{2h-1-\chi},
\end{equation}
we finally obtain the harmonic oscillator equivalent for our Eq.~\eqref{collspinLMG}
\begin{equation}
\label{mappedLMGlargeh}
H_\text{ho}= 2\sqrt{(h-1)(h-\chi)}  \left( b^\dagger b +\frac{1}{2}  \right) - h ( N+1 ) + \frac{1+\chi}{2}.
\end{equation}

For $0<h<1$ this classical vector moves as the field, $h$, is varied. Therefore before performing the HP transformation the Hamiltonian must be rotated to be inline with the direction of the classical angular momentum, or equivalently, we must take the HP transformation along the direction this vector points for a given value of $h$. We shall take the latter approach. For clarity, let us look at the slightly simpler case of $\chi=0$. In this case the classical vector moves between pointing along the $x$-axis ($h=0$) and pointing along the $z$-axis ($h=1$) according to $\varphi=\arccos h$. Therefore, we take the HP transformation along this new direction
\begin{equation}
\begin{aligned}
H^\varphi &= -\left(\frac{2}{N} \right) \left(S_x^{\varphi}\right)^2 -2h S_z^\varphi +\frac{1+\chi}{2},\\
S_x^\varphi &=  S_x \cos\varphi  - S_z \sin\varphi ,\\
S_z^\varphi  &=  S_z \cos\varphi  + S_x\sin\varphi .
\end{aligned}
\end{equation}
We now use the same operators as in Eqs.~\eqref{HP1} and \eqref{HP2} and therefore we have no need to perform any inverse rotations after the mapping is complete. Doing this results in a different bosonic representation,
\begin{equation}
H_b=-\frac{1}{2}h^2 \left(a^2+a^{\dagger~2}\right)+\left(2-h^2\right) a^{\dagger}a- \left( \frac{h^2N}{2}+\frac{h}{2} +\frac{N}{2}  \right)+\frac{1}{2}.
\end{equation}
Taking
\begin{equation}
\tanh \alpha = \frac{h^2}{2-h^2},
\end{equation}
in the Bogoliubov operators, we obtain the harmonic oscillator equivalent for $\chi=0$ 
\begin{equation}
H_\text{ho}=2\sqrt{(1-h^2)}  \left( b^\dagger b +\frac{1}{2}  \right) - \frac{1+h^2}{2} N - \frac{1}{2}.
\end{equation}
When considering arbitrary $\chi$, the calculation is slightly more involved. However the final form achieved is
\begin{equation}
\label{mappedLMGsmallh}
H_\text{ho}=2\sqrt{(1-h^2)(1-\chi)}  \left( b^\dagger b +\frac{1}{2}  \right) - \frac{1+h^2}{2} N - \frac{1-\chi}{2}.
\end{equation}

Using these mappings we can then use Eq.~\eqref{TQDHO} to determine the corresponding auxiliary Hamiltonian. For our purposes, working in units of $\hbar=1$, we take the effective frequency term appearing in Eqs.~\eqref{mappedLMGlargeh} and \eqref{mappedLMGsmallh} as $\omega$, i.e.
\begin{equation}
\omega=\begin{cases}
                   2\sqrt{(h-1)(h-\chi)}, ~~~&h>1,\\
                   2\sqrt{(1-h^2)(1-\chi)},~~~&0<h<1,
                   \end{cases}
\end{equation}
recalling that $h$ is time-dependent. Given that $\left( a^2 - a^{\dagger^2}  \right) = \left( b^2 - b^{\dagger^2}  \right)$, and returning to the collective spin operators, we find
\begin{equation}
\label{TQDLMG}
H_{1}=\begin{cases}
                               \frac{2h-1-\chi}{4N(h-1)(h-\chi)}\left( S_xS_y + S_yS_x  \right),~~~h>1,\\
                               \frac{2h(\chi-1)}{4N(1-h^2)(1-\chi)}\left( S_xS_y + S_yS_x  \right),~~~0<h<1.
                               \end{cases}
\end{equation}
Which are the exact correction terms required to achieve perfect finite-time adiabatic dynamics for $N\to\infty$. What is immediately apparent is that the auxiliary Hamiltonian is not well-defined at the critical point. Furthermore, examining Eqs.~\eqref{TQDLMG} we see that $H_1(t)$ contains complex and highly non-local terms, meaning their experimental implementation is extremely challenging. A further issue regarding the LMG model is that, while analytically exact in the thermodynamic limit, the correction Hamiltonians do not recover exactly adiabatic dynamics when applied {\it verbatim} to finite $N$ systems~\cite{Campbell2014}.  While we focused the analysis here on the harmonic oscillator and the LMG model, similar results can be achieved in the case of the quantum Ising model~\cite{delCampo2012LZ} by exploiting a concatenation of Landau-Zener models. Also in that case the corresponding auxiliary Hamiltonian was shown to be highly non-local. 

\paragraph{Inadequacy of transitionless quantum driving for  computing.}
Furthermore, despite its appealing set-up and potentially powerful applications, transitionless quantum driving is to date \emph{not very useful} from a computational point of view. With the exception of scale-invariant processes, to compute the auxiliary Hamiltonian the instantaneous eigenstates have to be known \eqref{q01}. Since in adiabatic quantum computation the outcome is encoded in the final ground state, one actually needs more information to implement $H_1(t)$  \eqref{q01} then is necessary to perform the computation exactly, i.e., without any errors. 

While this paints a somewhat bleak picture, we should realize that such analyses provide much information regarding what must be done in order to achieve the requisite level of control. Clearly, critical spin systems are promising prototype platforms, but evidently for practical utility we will require alternative methods for coherently controlling them. Such a realization has led to an encouraging line of research attempting to circumvent the most restrictive aspects of transitionless quantum driving~\cite{Campbell2014, Saberi2014}.  

Nevertheless further research has analyzed transitionless quantum driving in the context of universal quantum computation~\cite{SarandySciRep2015} and gate teleportation~\cite{SarandyPRA2016}. Furthermore, transitionless quantum driving stands as one of the most promising control techniques for coherently manipulating small scale quantum systems, and in particular when applied to facilitating the adiabatic strokes of quantum heat cycles and even in enhancing metrological protocols~\cite{TQDMetrology}. Thus, we anticipate the coming years to experience dedicated research efforts to adapt and generalize the framework of shortcuts to adiababiticty to develop novel tools and techniques tailored for quantum error correction.

\clearpage


\section{Checklist for ``Thermodynamics of Quantum Information"}

\begin{enumerate}
\item Information is physical and processing -- writing as well as erasing -- information ``costs'' thermodynamic energy.
\item Purely quantum effects, such as coherence and entanglement can be utilized as additional thermodynamic resources.
\item Quantum Stochastic Thermodynamics allows to assess the performance of quantum computers.
\item Quantum entropy production quantifies the amount of computational errors in quantum annealers.
\item The Kibble-Zurek mechanism phenomenlogically predicts the occurrence of computational errors.
\item Quantum entropy production exhibits Kibble-Zurek scaling.
\item Quantum error correction is essential and schemes exist for any computational paradigm.
\item Shortcuts to adiabaticity suppress \emph{fundamentally non-correctable errors} in quantum annealers.
\end{enumerate}

\section{Problems}

\subsection*{Quantum thermodynamics of  information \ref{sec:qu_info}}

\begin{itemize}
\item[\textbf{[1]}] An exactly soluble double-well potential is given by $$V(x)=\frac{1}{8} \cosh{\left(4 x\right)}-\alpha \cosh{\left(2 x\right)}-\frac{1}{8}\,, $$ where $\alpha>1/2$. Assume that the system was initially prepared in a corresponding Maxwell-Boltzmann distribution at temperature $T$, and that we seek to reset the system into right well with accuracy $\delta$. To this end, assume that the final distribution is a narrow Gaussian centered in the right well such that the probability to find the system in the left well is smaller than $\delta$. For this situation, verify Landauer's principle \eqref{eq:Landauer_classical}.
\item[\textbf{[2]}] Two qubits, $\mc{A}$ and $\mc{B}$, are found in a quantum state, $\rho$, that is an even mixture of the Bell states $$\ket{\Phi^\pm}=\frac{1}{\sqrt{2}}\left(\ket{0}_\mc{A}\otimes\ket{0}_\mc{B}\pm\ket{1}_\mc{A}\otimes\ket{1}_\mc{B}\right)\quad\mrm{and}\quad\ket{\Psi^\pm}=\frac{1}{\sqrt{2}}\left(\ket{0}_\mc{A}\otimes\ket{1}_\mc{B}\pm\ket{1}_\mc{A}\otimes\ket{0}_\mc{B}\right)\,.$$
Compute the amount of heat that is dissipated during a complete erasure of the stored quantum information, i.e., $\mc{A}$ and $\mc{B}$ are returned to $$\rho_T=\ket{0}_\mc{A}\otimes\ket{0}_\mc{B}\bra{0}_\mc{A}\otimes\bra{0}_\mc{B}\,.$$ How much of this heat is due to the erasure of classical information, and how much of this heat originates in destroying quantum correlations?
\end{itemize}

\subsection*{Performance diagnostics of quantum annealers \ref{sec:dwave}}
\begin{itemize}
\item[\textbf{[3]}] The dynamics of a TLS weakly coupled to thermal noise is given by the Lindblad master equation
$$\frac{d\rho}{dt}=-\frac{i}{\hbar}\,\com{H}{\rho}+\gamma\left(2\sigma_-\rho\sigma_+-\sigma_+\sigma_-\rho-\sigma_+\rho\sigma_-\right)\,, $$
where $H=-\hbar\omega\, \sigma_z/2$. Compute the quantum efficacy $\varepsilon$ \eqref{eq:qu_efficacy} for $A^i=\sigma_x$, $A^f=\sigma_x$, and $\rho_0=\e{-\beta H}/Z$.
\item[\textbf{[4]}]The strong coupling limit master equation with a $\sigma_z$ system-bath operator reads, 
$$\frac{d\rho}{dt}=-\frac{i}{\hbar}\,\com{H}{\rho}+\gamma\sum_i \left(\sigma_z^i \rho \sigma_z^i-\rho\right)\,, $$
where again $H=-\hbar\omega\, \sigma_z/2$. Show that the dynamics under this master equation is unital.
\end{itemize}

\subsection*{Kibble-Zurek Scaling of Irreversible Entropy \ref{sec:KibbleZurek}}
\begin{itemize}
\item[\textbf{[5]}] Consider a TLS described by the Landau-Zener Hamiltonian \eqref{eq:H_LZ}. Assuming that the system was initially prepared in state $\ket{\uparrow}$ at $t_i=-\hat{t}$ compute the quantum work distribution \eqref{eq:P_quwork} for processes that end at $t_f=\hat{t}$. Show that the excess work exhibits Kibble-Zurek scaling.
\item[\textbf{[6]}]  In mean-field theory the Landau free energy of a critical system is given by 
$$E(m)=\frac{h^2}{8}\,\ch{2 m}- h \,\ch{m}- \frac{h^2}{8}\,.$$
Identify the critical point and determine the critical exponents. Predict the behavior of the irreversible entropy production if the system is driven through the critical point at constant, but finite rate.
\end{itemize}

\subsection*{Error correction in adiabatic quantum computers \ref{sec:STA}}

\begin{itemize}
\item[\textbf{[7]}]Consider two coupled qubits, which are described by the Ising Hamiltonian in transverse field
$$H(t)= -g(t)\,\sigma_x^1-g(t)\,\sigma_x^2-\Delta(t)\, \sigma_z^1\sigma_z^{2}\,.$$
Compute the auxiliary Hamiltonian $H_1(t)$ \eqref{q01} for transitionless quantum driving.
\item[\textbf{[8]}] Consider the encoded Hamiltonian $\la H(t)\ra$ in Eq.~\eqref{eq:H_encoded} for $n=2$ and $\la N\ra=3$. Compute the corresponding auxiliary Hamiltonian $H_1(t)$ \eqref{q01} that would suppress finite-time excitations in quantum annealing correction. Would it be possible to implement the total Hamiltonian $H_\mrm{tot}(t)=\la H(t)\ra+H_1(t)$ on a quantum annealer such as the D-Wave machine?
\end{itemize}

\addcontentsline{toc}{section}{References}
\bibliographystyle{plain}
\bibliography{book}

\clearpage


\addcontentsline{toc}{chapter}{Epilogue}

\chapter*{Epilogue}
In this book we have attempted to, concisely, explore several facets of modern thermodynamics -- from its axiomatic origins through to the development of \emph{Stochastic Thermodynamics} and right up to the most recent advances in its quantum formulation. Indeed, as a physical theory thermodynamics is imposing in both its range of applicability and the deep insights into the workings of the universe it provides. For instance, as we have seen in Chapter~\ref{chap:termo}, the role of entanglement in providing a unique means of deriving canonical concepts in statistical mechanics enhances the special place that the seemingly counterintuitive notions of quantum mechanics play in dictating how the world around us emerges. Naturally, we have seen that a consistent quantum formulation of the core tenants of thermodynamics -- quantum work and heat -- is a delicate issue. Nevertheless, as established throughout Chapters~\ref{chap:devices} and \ref{chap:info}, as technological progress marches (and miniaturizes) on understanding the thermodynamics in this regime is crucial. It is therefore our hope that the material in this book has provided the necessary tools to handle the exciting challenges ahead.

Of course there is a whole host of interesting topics that we simply could not cover in the limited space available, one particular field being so-called resource theories. As the field of quantum information reached maturity, a greater focus was given to understanding the manipulation of quantum systems from a resource theoretic viewpoint. Indeed, it is clear that quantum features, in particular entanglement and other quantum correlations, are quantifiable resources for information processing and other tasks. Such an approach is fruitful when applied to understanding \emph{Quantum Thermodynamics}. The resource theory of Quantum Thermodynamics has shed light into what constitutes thermally free states and operations, thus providing insight into the thermodynamic cost of quantum information. Other exciting work has gone into exploring thermodynamic principles in cold atomic systems, where theoretical and experimental tools in this arena are progressing in tandem, and quantum biology, which studies the impact of genuine quantum effects on biological processes.

We close with some aspirations for the future. As new quantum technologies develop the understanding of their thermodynamic working principles is key to ensuring practical, energy efficient devices. The topics covered in Chapter~\ref{chap:info} gave a snap-shot of some of the more recent developments in this regard, however as mentioned, a great deal of work still needs done before the full promise of quantum technologies can be realized. Nevertheless, the great pace at which the young community continues to drive the field leaves us with no doubt that, as with the incredible advances that classical thermodynamics provided little over a century and a half ago,  Quantum Thermodynamics has many more remarkable insights yet to come.

\vspace{1em}
\begin{flushright}
 \textit{If one cannot enjoy reading a book over and over again, there is no use in reading it at all.}\\ \vspace{0.3em}\footnotesize(Oscar Wilde)
\end{flushright}


\bibliographystyle{plain}
\bibliography{book}
\clearpage







\addcontentsline{toc}{chapter}{Acknowledgments}

\chapter*{Acknowledgments}

{\bf Sebastian Deffner.}--
Every academic strives to live up to his mentors. In my case, this is an almost inconceivable challenge, since I have been very fortunate to have learned from the best. In particular, I would like to thank my \emph{Doktorvater} and friend Eric Lutz for making me his first student. Without his vision and foresight I would never have started to work in Quantum Thermodynamics. I will also forever be indebted to Chris Jarzynski for putting up with me during my early postdoctoral phase. His kindness and unwavering support paired with his unmatched understanding of Thermodynamics allowed me to grow into the physicist I am today. Finally, I will never forget the lessons I was taught by Wojciech H. Zurek. Being one of the most influential theoretical physicists he opened my eyes to the insurmountable variety of questions that can be addressed with the tools of Quantum Thermodynamics. His dedication to and his joy in unlocking the mysteries of the Universe, while at the same time remaining grounded in what really counts in life, remind me almost every day why I became a theoretical physicist and what kind of man I want to be.

I would also like to thank my dear friends and collaborators, who helped me hone my thinking and whose work contributed to this book. In particular, I am grateful to Marcus Bonan\c{c}a, Bart\l omeij Gardas, Frederico Brito, Haitao Quan, and Obinna Abah. I am looking forward to all the exciting research we will be tackling in the years to come.

Finally, I would like to thank my family, my parents, Alfred and Isabella, and my brother, Christoph, for accepting me for who I am and reminding me to never give up on my dreams. Last but not least, I am lacking words to express the importance of my partner in crime, my closest confidante, and mother of my children, Catherine. Thank you for always reminding me to keep fighting, for making me a better man, and for never giving up on me.

Maximillian and Alexander, like everything I do, this is for you!
\bigskip

\noindent
{\bf Steve Campbell.}--
I have been fortunate to have enjoyed a menagerie of collaborators over my relatively short research career so far. They all, in their own way, have contributed to how my interests have developed over the years which ultimately led to this work, and for this I am forever grateful. I am particularly indebted to those friends whose work formed the basis for some parts of this book: Marco Genoni, Gian Luca Giorgi, John Goold, Giacomo Guarnieri, Simon Pigeon, Maria Popovic, and Bassano Vacchini. I am also eternally grateful to Tony Apollaro, Bar\i \c{s} \c{C}akmak, Gabriele De Chiara, Mossy Fogarty, and Massimo Palma for the many years of stimulating discussions, punctuated with great refreshments, may they long continue.

I am lucky to have gained much of my scientific training from two world-leading physicists, Mauro Paternostro and Thomas Busch. The lessons learned from their expert guidance is woven throughout this book. I feel privileged to have benefitted from their friendship for so many years.

My parents, Larry and Shirley, and brother, Jaymz, I am thankful for all they continue to do for me. Finally, to my loves Flora and Qubit (the cat). Your support and encouragement to undertake and complete this book is the only reason I made it through. For putting up with the life of an early career academic trying to find his place, I owe you everything.


\end{document}